\documentclass[english,aps,preprint]{revtex4-2}
\usepackage[latin9]{inputenc}
\usepackage{mathrsfs}
\usepackage{mathtools}
\usepackage{amsmath}
\usepackage{amssymb}
\PassOptionsToPackage{normalem}{ulem}
\usepackage{ulem}

\makeatletter
\usepackage[english]{babel}
\usepackage{graphicx}
\usepackage{bm}
\usepackage{hyperref}
\usepackage{dcolumn}
\usepackage{xcolor}
\usepackage{bbm}
\usepackage{soul}
\usepackage{verbatim}

\DeclareMathOperator*{\sumint}{%
\mathchoice%
  {\ooalign{$\displaystyle\sum$\cr\hidewidth$\displaystyle\int$\hidewidth\cr}}
  {\ooalign{\raisebox{.14\height}{\scalebox{.7}{$\textstyle\sum$}}\cr\hidewidth$\textstyle\int$\hidewidth\cr}}
  {\ooalign{\raisebox{.2\height}{\scalebox{.6}{$\scriptstyle\sum$}}\cr$\scriptstyle\int$\cr}}
  {\ooalign{\raisebox{.2\height}{\scalebox{.6}{$\scriptstyle\sum$}}\cr$\scriptstyle\int$\cr}}
}

\DeclareMathSymbol{\mhyphen}{\mathord}{AMSa}{"39}

\def\gcoups{g_S}
\def\gcoupr{g_{V}}
\def\gcoupw{g_{V_0}}
\def\gcoupa{g_{A_0}}

\def\cc{\alpha}
\def\gmatrix{{\cal G}}
\def\jmatrix{{\cal J}}
\def\gnomatrix{{G}}
\def\jnomatrix{{J}}

\def\trmin{{\rm tr}}
\def\mf{{\mbox{\tiny MF}}}
\def\calF{{\cal F}_{\scriptscriptstyle{Q}}}

\def\xi{v}

\def\taua{{  \tau_b}}
\def\sigmaa{{  \sigma_b}}
\def\aamu{{  a_b^\mu}}

\def\sigmac{{  \sigma_0}}
\def\pia{{  \pi_b}}
\def\pic{{  \pi_0}}
\def\rhocmu{{  \rho_0^\mu}}
\def\rhocnu{{  \rho_0^\nu}}
\def\rhocmud{{  \rho_{0\mu}}}
\def\rhoc{{  \rho_{0}}}
\def\acmud{{  a_{0\mu}}}
\def\acmu{{  a_0^\mu}}
\def\acnu{{  a_0^\nu}}
\def\ac{{  a_{0}}}
\def\sigmat{{  \sigma_3}}
\def\pit{{  \pi_3}}
\def\rhotmu{{  \rho_3^\mu}}
\def\rhotnu{{  \rho_3^\nu}}
\def\rhot{{  \rho_3}}
\def\atmu{{  a_3^\mu}}
\def\atnu{{  a_3^\nu}}
\def\at{{  a_3}}
\def\pib{{  \pi_{b}}}
\def\rhoamu{{  \rho_b^\mu}}

\usepackage{babel}

\begin{document}

\global\long\def\thefootnote{\arabic{footnote}}
\setcounter{footnote}{0}

\title{\sc\Large{\vspace*{-0.5cm} Masses of magnetized pseudoscalar and vector mesons
in an extended NJL model: the role of axial vector mesons} \vspace*{0.5cm}}

\author{M. Coppola$^1$, D. Gomez Dumm$^2$, S. Noguera$^3$ and N.~N.\ Scoccola$^{1,4}$ \vspace*{0.5cm}}

\affiliation{\small $^{1}$ Departamento de F\'isica, Comisi\'{o}n Nacional de Energ\'{\i}a At\'{o}mica, \\
Avenida del Libertador 8250, 1429 Buenos Aires, Argentina}
\affiliation{$^{2}$ IFLP, CONICET $-$ Departamento de F\'{\i}sica, Facultad de Ciencias Exactas,
Universidad Nacional de La Plata, C.C. 67, 1900 La Plata, Argentina}
\affiliation{$^{3}$ Departamento de F\'isica Te\'orica and IFIC, Centro Mixto \\
Universidad de Valencia-CSIC, E-46100 Burjassot (Valencia), Spain}
\affiliation{$^{4}$ CONICET, Rivadavia 1917, 1033 Buenos Aires, Argentina \vspace*{\fill}}

\begin{abstract}
We study the mass spectrum of light pseudoscalar and vector mesons in the
presence of an external uniform magnetic field $\vec B$, considering the
effects of the mixing with the axial vector meson sector. The analysis is
performed within a two-flavor NJL-like model which includes isoscalar and
isovector couplings together with a flavor mixing 't Hooft-like term. The
effect of the magnetic field on charged particles is taken into account by
retaining the Schwinger phases carried by quark propagators, and expanding
the corresponding meson fields in proper Ritus-like bases. The
spin-isospin and spin-flavor decomposition of meson mass states is also
analyzed. For neutral pion masses it is shown that the mixing with axial
vector mesons improves previous theoretical results, leading to a monotonic
decreasing behavior with $B$ that is in good qualitative agreement with LQCD
calculations, both for the case of constant or $B$-dependent couplings.
Regarding charged pions, it is seen that the mixing softens the enhancement
of their mass with $B$. As a consequence, the energy becomes lower than the
one corresponding to a pointlike pion, improving the agreement with LQCD
results. The agreement is also improved for the magnetic behavior of the 
lowest $\rho^+$ energy state, which does not vanish for the considered range 
of values of $B$ --- a fact that can be relevant in connection with the 
occurrence of meson condensation for strong magnetic fields.
\end{abstract}

\maketitle
\newpage

\section{Introduction}
\label{intro}

The effects caused by magnetic fields larger than $|eB| \sim
\Lambda_{QCD}^2$ on the properties of strong-interacting matter have
attracted a lot of attention along the last
decades~\cite{Kharzeev:2012ph,Andersen:2014xxa,Miransky:2015ava}. In part,
this is motivated by the realization that such magnetic fields might play an
important role in the study of the early
Universe~\cite{Vachaspati:1991nm,Grasso:2000wj}, in the analysis of high
energy noncentral heavy ion collisions~\cite{Skokov:2009qp,Voronyuk:2011jd}
and in the description of compact stellar objects like the
magnetars~\cite{Duncan:1992hi,Kouveliotou:1998ze}. In addition to this
phenomenological relevance, from the theoretical point of view, external
magnetic fields can be used to probe QCD dynamics, allowing for a
confrontation of theoretical results obtained through different approaches
to nonperturbative QCD. In this sense, several interesting phenomena have
been predicted to be induced by the presence of strong magnetic fields. They
include the chiral magnetic
effect~\cite{Kharzeev:2007jp,Fukushima:2008xe,Kharzeev:2015znc}, the
enhancement of the QCD quark-antiquark condensate (magnetic
catalysis)~\cite{Gusynin:1995nb}, the decrease of critical temperatures for
chiral restoration and deconfinement QCD transitions (inverse magnetic
catalysis)~\cite{Bali:2011qj,Andersen:2021lnk}, etc.

In this context, the understanding of the way in which the properties of
light hadrons are modified by the presence of an intense magnetic field
becomes a very relevant task. Clearly, this is a nontrivial problem, since
first-principle theoretical calculations require to deal in general with QCD
in a low energy nonperturbative regime. As a consequence, the corresponding
theoretical analyses have been carried out using a variety of approaches.
The effect of intense external magnetic fields on $\pi$ meson properties has
been studied e.g.~in the framework of Nambu-Jona-Lasinio (NJL)-like
models~\cite{Fayazbakhsh:2012vr,Fayazbakhsh:2013cha,Liu:2014uwa,Avancini:2015ady,
Zhang:2016qrl,Avancini:2016fgq,Mao:2017wmq,GomezDumm:2017jij,Wang:2017vtn,Liu:2018zag,
Coppola:2018vkw,Mao:2018dqe,Avancini:2018svs,Coppola:2019uyr,Cao:2019res,Cao:2021rwx,Sheng:2021evj,
Avancini:2021pmi,Xu:2020yag,Lin:2022ied}, quark-meson
models~\cite{Kamikado:2013pya,Ayala:2018zat,Ayala:2020dxs,Das:2019ehv,Wen:2023qcz},
chiral perturbation theory
(ChPT)~\cite{Andersen:2012zc,Agasian:2001ym,Colucci:2013zoa}, path integral
Hamiltonians~\cite{Orlovsky:2013gha,Andreichikov:2016ayj}, effective chiral
confinement Lagrangians~\cite{Simonov:2015xta,Andreichikov:2018wrc} and QCD
sum rules~\cite{Dominguez:2018njv}. In addition, several results for the
$\pi$ meson spectrum in the presence of background magnetic fields have been
obtained from lattice QCD (LQCD)
calculations~\cite{Bali:2011qj,Luschevskaya:2015bea,Luschevskaya:2014lga,Bali:2017ian,Ding:2020hxw,Ding:2022tqn}.
Regarding the $\rho$ meson sector, studies of magnetized $\rho$ meson masses
in the framework of effective models and LQCD can be found in
Refs.~\cite{Xu:2020yag,Chernodub:2011mc,Andreichikov:2016ayj,Zhang:2016qrl,Liu:2018zag,
Cao:2019res,Kawaguchi:2015gpt,Ghosh:2016evc,Ghosh:2020qvg,Avancini:2022qcp}
and
Refs.~\cite{Luschevskaya:2012xd,Luschevskaya:2015bea,Hidaka:2012mz,Luschevskaya:2014lga,Luschevskaya:2016epp,
Bali:2017ian}, respectively. The effect of an external magnetic field on
nucleon masses has also been considered in several
works~\cite{Tiburzi:2008ma,Deshmukh:2017ciw,Taya:2014nha,Haber:2014ula,Mukherjee:2018ebw,
He:2016oqk,Dominguez:2020sdf,Endrodi:2019whh,Coppola:2020mon}.

In most of the existing model calculations of meson masses the mixing
between states of different spin/isospin has been neglected. Although such
mixing contributions are usually forbidden by isospin and/or angular
momentum conservation, they can be nonzero (and may become important) in the
presence of the external magnetic field. Effects of this kind have
been studied recently by some of the authors of the present work, for both
neutral~\cite{Carlomagno:2022inu} and charged
mesons~\cite{Carlomagno:2022arc}. Those analyses have been performed in the
framework of an extended NJL-like model, where, for simplicity, possible
axial vector interactions have been neglected. The aim of the present work
is to study how those previous results get modified when the presence of
axial vector mesons is explicitly taken into account. In fact, due to
symmetry reasons, in the context of the NJL model and its
extensions~\cite{Vogl:1991qt,Klevansky:1992qe,Hatsuda:1994pi} vector and
axial vector interactions are expected to be considered on the same footing
(see e.g.\ Refs.~\cite{Klimt:1989pm,Bijnens:1995ww}). This, in turn, implies
the existence of the so-called ``$\pi$-a$_1$ mixing'' even at vanishing
external magnetic field. Such a mixing has to be properly taken into account
in order to correctly identify the pion mass states. Thus, the inclusion of
the axial interactions is expected to be particularly relevant for the
analysis of lowest meson masses.

Regarding the explicit calculation, as shown in previous
works~\cite{Coppola:2018vkw,Coppola:2019uyr,Carlomagno:2022arc,
Coppola:2018ygv,Coppola:2019wvh}, one has to deal with the meson
wavefunctions that arise as solutions of the equations of motion in the
presence of the external magnetic field (which we assume to be static and
uniform). In particular, in the case of charged mesons, it is seen that
one-loop level calculations involve the presence of Schwinger phases that
induce a breakdown of translational invariance in quark
propagators~\cite{GomezDumm:2023owj}. 
As a consequence, the corresponding meson polarization functions are
not diagonal for the standard plane wave states. One should describe
meson states in terms of wavefunctions characterized by a set of quantum
numbers that include the Landau level $\ell$, which is associated
to the quantization of momentum in the plane perpendicular to the
magnetic field. It is worth mentioning that although we consider a
magnetic field that extends over all space, in a realistic scenario
---such as the core of a neutron star, or a heavy ion collision--- the
existence of a large magnetic field will be limited to a confined region. 
In fact, if a charged meson is to be tracked by some detector,
the latter will be in general located away from the zone affected by the
magnetic field; thus, the theoretical analysis would require the
projection onto a proper basis determined by the particular features of
the experiment.

As for the model specifications, it is important to care about the regularization of
ultraviolet divergences, since the presence of the external magnetic field
can lead to spurious results, such as unphysical oscillations of physical
observables~\cite{Allen:2015paa,Avancini:2019wed}. As in previous
works~\cite{Carlomagno:2022inu,Carlomagno:2022arc}, we use the so-called
magnetic field independent regularization (MFIR)
scheme~\cite{Menezes:2008qt,Avancini:2015ady,Avancini:2016fgq,Coppola:2018vkw},
which has been shown to be free from these oscillations; moreover, it is
seen that within this scheme the results are less dependent on model
parameters~\cite{Avancini:2019wed}. Concerning the effective coupling
constants of the model, we consider both the case in which these
parameters are kept constant and the case in which they show some explicit
dependence on the external magnetic field. This last possibility, inspired
by the magnetic screening of the strong coupling constant occurring for a
large magnetic field~\cite{Miransky:2002rp}, has been previously explored in
effective
models~\cite{Ayala:2014iba,Farias:2014eca,Ferreira:2014kpa,Endrodi:2019whh,
Sheng:2021evj} in order to reproduce the inverse magnetic catalysis effect
observed at finite temperature LQCD calculations.

The paper is organized as follows. In Sec.~\ref{sect2a} we introduce the magnetized
extended NJL-like lagrangian to be used in our calculations, as well as the
expressions of the relevant mean field quantities to be evaluated, such as
quark masses and chiral condensates. In Sec.~\ref{sec:M0} and~\ref{chargedmes} we present the
formalisms used to obtain neutral and charged meson masses, respectively, in
the presence the magnetic field. In Sec.~\ref{sec:Num} we present and discuss our
numerical results, while in Sec.~\ref{sec:conclu} we provide a summary of our work,
together with our main conclusions. We also include several appendices to
provide some technical details of our calculations.


\section{Effective Lagrangian and mean field quantities}

\label{sect2a}

Let us start by considering the Lagrangian density for an extended
NJL two-flavor model in the presence of an electromagnetic field.
We have, in Minkowski space,
\begin{eqnarray}
{\cal L} & = & \bar{\psi}(x)\left(i\,\rlap/\!D-m_{c}\right)\psi(x)+\gcoups\sum_{a=0}^{3}\Big[\left(\bar{\psi}(x)\taua\psi(x)\right)^{2}+\left(\bar{\psi}(x)\,i\gamma_{5}\taua\psi(x)\right)^{2}\Big]\nonumber \\
 &  & -\,\gcoupr\left[\left(\bar{\psi}(x)\,\gamma_{\mu}\vec{\tau}\,\psi(x)\right)^{2}+\left(\bar{\psi}(x)\,\gamma_{\mu}\,\gamma_{5}\,\vec{\tau}\,\psi(x)\right)^{2}\right]\nonumber \\
 &  & -\,\gcoupw\left(\bar{\psi}(x)\,\gamma_{\mu}\,\psi(x)\right)^{2}-\,\gcoupa\left(\bar{\psi}(x)\,\gamma_{\mu}\,\gamma_{5}\,\psi(x)\right)^{2}+\,2g_{D}\,\left(d_{+}+d_{-}\right)\ ,\label{lagrangian}
\end{eqnarray}
where $\psi=(u\ d)^{T}$, $\taua=(\mathbbm{1},\vec{\tau})$, $\vec{\tau}$
being the usual Pauli-matrix vector, and $m_{c}$ is the current quark
mass, which is assumed to be equal for $u$ and $d$ quarks. The model
includes isoscalar/isovector vector and axial vector couplings, as
well as a 't Hooft-like flavor-mixing term, where we have defined
$d_{\pm}={\rm det}[\bar{\psi}(x)(1\pm\gamma_{5})\psi(x)]$. The interaction
between the fermions and the electromagnetic field ${\cal A}_{\mu}$
is driven by the covariant derivative
\begin{equation}
D_{\mu}\ =\ \partial_{\mu}+i\,\hat{Q}\mathcal{A}_{\mu}\ ,\label{covdev}
\end{equation}
where $\hat{Q}=\mbox{diag}(Q_{u},Q_{d})$, with $Q_{u}=2e/3$ and
$Q_{d}=-e/3$, $e$ being the proton electric charge. A summary of
the notation and conventions used throughout this work can be found
in App.~\ref{conventions}.

We consider here the particular case in which one has a homogenous
stationary magnetic field $\vec{B}$ orientated along the axis 3,
or $z$. Now, to write down the explicit form of $\mathcal{A}^{\mu}$
one has to choose a specific gauge. Some commonly used gauges are
the symmetric gauge (SG) in which ${\cal A}^{\mu}(x)=(0,-B\,x^{2}/2,B\,x^{1}/2,0)$,
the Landau gauge 1 (LG1) in which ${\cal A}^{\mu}(x)=\left(0,-B\,x^{2},0,0\right)$
and the Landau gauge 2 (LG2), in which ${\cal A}^{\mu}(x)=\left(0,0,B\,x^{1},0\right)$.
In what follows we refer to them as ``standard gauges''. To test
the gauge independence of our results, all these gauges will be considered
in our analysis.

Since we are interested in studying meson properties, it is convenient
to bosonize the fermionic theory, introducing scalar, pseudoscalar,
vector and axial vector fields $\sigmaa(x)$, $\pia(x)$, $\rhoamu(x)$,
$\aamu$, with $b=0,1,2,3$, and integrating out the fermion fields.
The bosonized action can be written as
\begin{eqnarray}
S_{\mathrm{bos}} & = & -i\ln\det\!\big(i\mathcal{D}\big)-\frac{1}{4g}\int d^{4}x\ \Big[\sigmac(x)\,\sigmac(x)+\vec{\pi}(x)\cdot\vec{\pi}(x)\Big]\nonumber \\
 &  & -\,\frac{1}{4g(1-2\cc)}\int d^{4}x\ \Big[\vec{\sigma}(x)\cdot\vec{\sigma}(x)+\pic(x)\,\pic(x)\Big]\nonumber \\
 &  & +\,\frac{1}{4\gcoupr}\int d^{4}x\ \left[\vec{\rho}_{\mu}(x)\cdot\vec{\rho}^{\,\mu}(x)+\vec{a}_{\mu}(x)\cdot\vec{a}^{\,\mu}(x)\right]\nonumber \\
 &  & +\frac{1}{4\gcoupw}\int d^{4}x\ \rhocmud(x)\,\rhocmu(x)+\frac{1}{4\gcoupa}\int d^{4}x\ \acmud(x)\,\acmu(x)\ ,\label{sbos}
\end{eqnarray}
with
\begin{equation}
i\mathcal{D}_{x,x'}\ =\ \delta^{(4)}(x-x')\,\big[i\,\rlap/\!D-m_{0}-\taua\left(\sigmaa(x)+i\,\gamma_{5}\,\pia(x)+\gamma_{\mu}\,\rhoamu(x)+\gamma_{\mu}\gamma_{5}\,\aamu(x)\right)\big]\ ,\label{dxx}
\end{equation}
where a direct product to an identity matrix in color space is understood.
For convenience we have introduced the combinations
\begin{eqnarray}
g = \gcoups + g_{D}\ ,\qquad\qquad\cc=\frac{g_{D}}{\gcoups+g_{D}}\ ,
\end{eqnarray}
so that the flavor mixing in the scalar-pseudoscalar sector is regulated
by the constant $\cc$. For $\cc=0$ quark flavors $u$ and $d$ get
decoupled, while for $\cc=0.5$ one has maximum flavor mixing, as
in the case of the standard version of the NJL model.

We proceed by expanding the bosonized action in powers of the fluctuations
of the bosonic fields around the corresponding mean field (MF) values.
We assume that the fields $\sigma_b(x)$ have nontrivial translational
invariant MF values given by $\tau_b\,\bar{\sigma}_b=\mbox{diag}(\bar{\sigma}_{u},\bar{\sigma}_{d})$,
while vacuum expectation values of other bosonic fields are zero;
thus, we write
\begin{equation}
\mathcal{D}_{x,x'}\ =\ \mathcal{D}_{x,x'}^{\mbox{\tiny MF}}+\delta\mathcal{D}_{x,x'}\ .\label{dxxp}
\end{equation}
The MF piece is diagonal in flavor space. One has
\begin{equation}
\mathcal{D}_{x,x'}^{\mf}\ =\ {\rm diag}\big(\mathcal{D}_{x,x'}^{\mf,\,u}\,,\,\mathcal{D}_{x,x'}^{\mf,\,d}\big)\ ,
\end{equation}
where
\begin{equation}
\mathcal{D}_{x,x'}^{\mf,\,f}\ =\ -i\,\delta^{(4)}(x-x')\left(i\rlap/\partial+Q_{f}\,B\,x^{1}\,\gamma^{2}-M_{f}\right)\ ,
\end{equation}
with $f=u,d$. Here $M_{f}=m_{c}+\bar{\sigma}_{f}$ is the quark effective
mass for each flavor $f$.

The MF action per unit volume is given by
\begin{equation}
\frac{S_{\mathrm{bos}}^{\mbox{\tiny MF}}}{V^{(4)}}\ =-\ \frac{(1-\cc)(\bar{\sigma}_{u}^{2}+
\bar{\sigma}_{d}^{2})-2\,\cc\,\bar{\sigma}_{u}\bar{\sigma}_{d}}{8g(1-2\,\cc)}-\frac{iN_{c}}{V^{(4)}}\sum_{f=u,d}\int d^{4}x\,d^{4}x'
\ \trmin_{D}\,\ln\left(\mathcal{S}_{x,x'}^{\mf,\,f}\right)^{-1}\ ,
\label{seff}
\end{equation}
where $\trmin_{D}$ stands for the trace over Dirac space, and $\mathcal{S}_{x,x'}^{\mf,\,f}=\big(i\mathcal{D}_{x,x'}^{\mf,\,f}\big)^{-1}$
is the MF quark propagator in the presence of the magnetic field.
Its explicit expression can be written as
\begin{equation}
\mathcal{S}_{x,y}^{\mf,\,f}=e^{i\Phi_{Q_{f}}(x,y)}\int\frac{d^{4}p}{(2\pi)^{4}}\ e^{-i\,p(x-y)}
\,\bar{S}^{f}(p_{\parallel},p_{\perp})\ ,
\label{uno}
\end{equation}
where
\begin{eqnarray}
\bar{S}^{f}(p_{\parallel},p_{\perp}) & = & -\,i\int_{0}^{\infty}d\sigma\ \exp\!\bigg[\!-i\sigma\Big(M_{f}^{2}-p_{\parallel}^{2}+\vec{p}_{\perp}^{\ 2}\,\dfrac{\tan(\sigma B_{f})}{\sigma B_{f}}-i\epsilon\Big)\bigg]\nonumber \\
 &  &
\times\left[\left(p_{\parallel}\cdot\gamma_{\parallel}+M_{f}\right)(1-s_f\,\gamma^{1}\gamma^{2}
\tan(\sigma B_{f}))-\frac{\vec{p}_{\perp}\cdot\vec{\gamma}_{\perp}}{\cos^{2}(\sigma B_{f})}\right]\ ,
\label{sfp_schw}
\end{eqnarray}
with $B_f = |B Q_f|$ and $s_f={\rm sign}(B Q_f)$. Here we have defined the
``parallel'' and ``perpendicular'' four-vectors
\begin{equation}
p_{\parallel}^{\mu}=(p^{0},0,0,p^{3})\ , \qquad\qquad
p_{\perp}^{\mu}=(0,p^{1},p^{2},0)\ ,
\end{equation}
and equivalent definitions have been used for $\gamma_\parallel$,
$\gamma_\perp$. The function $\Phi_{Q}(x,y)$ in Eq.~(\ref{uno}) is the
so-called Schwinger phase, which is shown to be a gauge dependent quantity.
For the standard gauges one has
\begin{align}
 & \mbox{SG:} &  & \hspace{-1.5cm}\Phi_{Q}(x,y)=-\frac{QB}{2}(x^{1}y^{2}-y^{1}x^{2})\ ,\hspace{1.5cm}\nonumber \\
 & \mbox{LG1:} &  & \hspace{-1.5cm}\Phi_{Q}(x,y)=-\frac{QB}{2}(x^{2}+y^{2})(x^{1}-y^{1})\ ,\hspace{1.5cm}\nonumber \\
 & \mbox{LG2:} &  & \hspace{-1.5cm}\Phi_{Q}(x,y)=\frac{QB}{2}(x^{1}+y^{1})(x^{2}-y^{2})\ .\hspace{1.5cm}\label{sphase}
\end{align}

Let us consider the quark-antiquark condensates $\phi_{f}\equiv\langle\bar{\psi}_{f}\psi_{f}\rangle$.
For each flavor $f=u,d$ we have
\begin{eqnarray}
\phi_{f}=iN_{c}\int\frac{d^{4}p}{(2\pi)^{4}}\ \trmin_{D}\,\bar{S}^{f}(p_{\parallel},p_{\perp})\ .
\end{eqnarray}
The integral in this expression is divergent and has to be properly
regularized. As stated in the Introduction, we use here the magnetic field
independent regularization (MFIR) scheme: for a given unregularized
quantity, the corresponding (divergent) $B\to0$ limit is subtracted and then
it is added in a regularized form. Thus, the quantities can be separated
into a (finite) ``$B=0$'' part and a ``magnetic'' piece. Notice that, in
general, the ``$B=0$" part still depends implicitly on $B$ (e.g.\ through
the values of the dressed quark masses $M_{f}$); hence, it should not be
confused with the value of the studied quantity at vanishing external field.
To deal with the divergent ``$B=0$'' terms we use here a proper time (PT)
regularization scheme. Thus, we obtain
\begin{equation}
\phi_{f}^{{\rm reg}}\ =\ \phi_{f}^{0,\,{\rm reg}}\,+\,\phi_{f}^{{\rm mag}}\ ,
\end{equation}
where
\begin{equation}
\phi_{f}^{0,{\rm reg}}=-N_{c}\,M_{f}\,I_{1f}\ ,\qquad\qquad\phi_{f}^{{\rm mag}}=-N_{c}\,M_{f}\,I_{1f}^{{\rm mag}}\ .\label{phif}
\end{equation}
The expression of $I_{1f}$ obtained from the PT regularization,
$I_{1f}^{\rm reg}$, is given in Eq.~(\ref{I1freg}) in App.~\ref{b0loops},
while the ``magnetic'' piece $I_{1f}^{{\rm mag}}$ reads
\begin{equation}
I_{1f}^{{\rm mag}}=\dfrac{B_{f}}{2\pi^{2}}\left[\ln\Gamma(x_{f})-\left(x_{f}-\dfrac{1}{2}\right)\ln x_{f}+x_{f}-\dfrac{\ln{2\pi}}{2}\right]\,,\label{i1}
\end{equation}
where $x_{f}=M_{f}^{2}/(2B_{f})$. The corresponding gap equations,
obtained from $\partial S_{\mathrm{bos}}^{\mbox{\tiny MF}}/\partial\bar{\sigma}_{f}=0$,
can be written as
\begin{eqnarray}
M_{u} & = & m_{c}-4g\left[(1-\cc)\,\phi_{u}^{{\rm reg}}+\cc\,\phi_{d}^{{\rm reg}}\right]\ ,\nonumber \\
M_{d} & = & m_{c}-4g\left[(1-\cc)\,\phi_{d}^{{\rm reg}}+\cc\,\phi_{u}^{{\rm reg}}\right]\ .\label{gapeqs}
\end{eqnarray}
As anticipated, for $\alpha=0$ these equations get decoupled. For
$\alpha=0.5$ the right hand sides become identical, thus one has
in that case $M_{u}=M_{d}$.

\section{The neutral meson sector}
\label{sec:M0}

To determine the meson masses we have to consider the terms in the
bosonic action that are quadratic in meson fluctuations. As expected
from charge conservation, it is easy to see that the terms corresponding
to charged and neutral mesons decouple from each other. In this section
we concentrate on the neutral meson sector; the charged meson sector
will be considered in Sec.~\ref{chargedmes}.

\subsection{Neutral meson polarization functions}
\label{neutralmesons}

For notational convenience we will denote isospin states by
$M=\sigmac,\pic,\rhocmu,\acmu,\sigmat,\pit,\rhotmu,\atmu$. Here, $\sigmac$,
$\pic$, $\rhoc$ and $\ac$ correspond to the isoscalar states $\sigma$,
$\eta$, $\omega$ and $f_{1}$, while $\sigmat$, $\pit$, $\rhot$ and $\at$
stand for the neutral components of the isovector triplets $\vec{\rm
a}_{0}$, $\vec{\pi}$, $\vec{\rho}$ and $\vec{\rm a}_{1}$, respectively.
Thus, the corresponding quadratic piece of the bosonized action can be
written as
\begin{equation}
S_{\mathrm{bos}}^{{\rm quad,\,neutral}}\ =-\ \frac{1}{2}\int d^{4}x\;d^{4}x'\sum_{M,M'}\delta M(x)^{\dagger}
\ {\gmatrix}_{MM'}(x,x')\ \delta M'(x')\ .
\end{equation}
Notice that the meson indices $M,M'$, as well as the functions
${\gmatrix}_{MM'}$, include Lorentz indices in the case of vector mesons.
This also holds for the functions $\delta_{MM'}$, ${\cal J}_{MM'}$,
$\Sigma_{MM'}^{f}$, $G_{MM'}$, etc., introduced below. In the corresponding
expressions, a contraction of Lorentz indices is understood when
appropriate. In particular, the functions ${\gmatrix}_{MM'}(x,x')$ can be
separated in two terms, namely
\begin{equation}
{\gmatrix}_{MM'}(x,x')\ =\ \frac{1}{2g_{M}}\ \delta_{MM'}\ \delta^{(4)}(x-x')-\,{\jmatrix}_{MM'}(x,x')\ ,
\end{equation}
where
\begin{equation}
\frac{1}{g_{M}}\ \delta_{MM'}\ =\ \left\{ \begin{array}{cl}
1/g & \ \ \ \ \mbox{for}\ \ M=M'=\sigmac,\pit\\
1/[g(1-2\cc)] & \ \ \ \ \mbox{for}\ \ M=M'=\sigmat,\pic\\
-\eta^{\mu\nu}/\gcoupr & \ \ \ \ \mbox{for}\ \ MM'=\rhotmu\rhotnu,\atmu\atnu\\
-\eta^{\mu\nu}/\gcoupw & \ \ \ \ \mbox{for}\ \ MM'=\rhocmu\rhocnu\\
-\eta^{\mu\nu}/\gcoupa & \ \ \ \ \mbox{for}\ \ MM'=\acmu\acnu
\end{array}\right.\qquad,\label{valgm}
\end{equation}
and $\delta_{MM'}=0$ otherwise. Here $\eta^{\mu\nu}$ is the Minkowski metric
tensor, which can be decomposed as $\eta^{\mu\nu} =
\eta_{\parallel}^{\mu\nu} +\eta_\perp^{\mu\nu}$, with
$\eta_\parallel^{\mu\nu} = \mbox{diag}(1,0,0,-1)$,
$\eta_{\perp}^{\mu\nu}=\mbox{diag}(0,-1,-1,0)$ (see App.~\ref{conventions}).
In turn, the polarization functions ${\jmatrix}_{MM'}(x,x')$ can be
separated into $u$ and $d$ quark pieces,
\begin{equation}
{\jmatrix}_{MM'}(x,x')\ =\ \mathit{\Sigma}_{MM'}^{u}(x,x')\,+\,\varepsilon_{M}\,\varepsilon_{M'}\,\mathit{\Sigma}_{MM'}^{d}(x,x')\ .
\label{jotas}
\end{equation}
Here $\varepsilon_{M}=1$ for the isoscalars $M=\sigmac,\pic,\rhocmu,\acmu$
and $\varepsilon_{M}=-1$ for $M=\sigmat,\pit,\rhotmu,\atmu$, while
the functions $\mathit{\Sigma}_{MM'}^{f}(x,x')$ are given by
\begin{eqnarray}
\mathit{\Sigma}_{MM'}^{f}(x,x')\ =-i\,N_{c}\ \trmin_{D}\bigg[i\,\mathcal{S}_{x,x'}^{\mf,\,f}\,\Gamma^{M'}i\,\mathcal{S}_{x',x}^{\mf,\,f}\,\Gamma^{M}\,\bigg]\ ,
\end{eqnarray}
with
\begin{equation}
\Gamma^{M}\ =\ \left\{ \begin{array}{cl}
\mathbbm{1} & \ \ \mbox{for}\ \ M=\sigmac,\sigmat\\
i\gamma_{5} & \ \ \mbox{for}\ \ M=\pic,\pit\\
\gamma^{\mu} & \ \ \mbox{for}\ \ M=\rhocmu,\rhotmu\\
\gamma^{\mu}\gamma_{5} & \ \ \mbox{for}\ \ M=\acmu,\atmu
\end{array}\right.\qquad.
\end{equation}

As stated, since we are dealing with neutral mesons, the contributions
of Schwinger phases associated with the quark propagators in Eq.~(\ref{uno})
cancel out, and the polarization functions depend only on the difference
$x-x'$, i.e., they are translationally invariant. After a Fourier
transformation, the conservation of momentum implies that the polarization
functions turn out to be diagonal in the momentum basis. Thus, in
this basis the neutral meson contribution to the quadratic action
can be written as
\begin{equation}
S_{\mathrm{bos}}^{{\rm quad,\,neutral}}\ =-\dfrac{1}{2}\sum_{M,M'}\int\dfrac{d^{4}q}{(2\pi)^{4}}\ \delta M (-q)^{\dagger}
\ {\gnomatrix}_{MM'}(q)\ \delta M'(q)\ .
\label{eqnneutral}
\end{equation}
We have
\begin{equation}
{\gnomatrix}_{MM'}(q)\ =\ \frac{1}{2g_{M}}\;\delta_{MM'}\,-\,{\jnomatrix}_{MM'}(q)\ ,\label{gmm}
\end{equation}
and the associated polarization functions can be written as
\begin{equation}
{\jnomatrix}_{MM'}(q)\ =\ \Sigma_{MM'}^{u}(q)+\varepsilon_{M}\varepsilon_{M'}\,\Sigma_{MM'}^{d}(q)\ .
\label{jotas2}
\end{equation}
Here the functions $\Sigma_{MM'}^{f}(q)$ read
\begin{equation}
\Sigma_{MM'}^{f}(q)\ =-i\,N_{c}\,\int\frac{d^{4}p}{(2\pi)^{4}}\ \trmin_{D}\left[i\,\bar{S}^{f}(p_{\parallel}^{+},p_{\perp}^{+})\,\Gamma^{M'}i\,\bar{S}^{f}(p_{\parallel}^{-},p_{\perp}^{-})\,\Gamma^{M}\right]\ ,\label{eqn26}
\end{equation}
where we have defined $p_{a}^{\pm}=p_{a}\pm q_{a}/2$, with $a=\parallel,\perp$,
and the quark propagators $\bar{S}^{f}(p_{\parallel},p_{\perp})$
in the presence of the magnetic field are those given by Eq.~(\ref{sfp_schw}).
The explicit expressions of the non-vanishing functions $\Sigma_{MM'}^{f}(q)$
for arbitrary four-momentum $q^{\mu}$ are given in App.~\ref{polneutral}.

Since we are interested in the determination of meson masses, let us focus
on the particular situation in which mesons are at rest, i.e.\
$q^{\mu}=(m,0,0,0)$, where $m$ is the corresponding meson mass. We
denote by $\hat{J}_{MM'}$ the resulting polarization functions. It can be
shown that some of these functions vanish, while the nonvanishing ones are
in general divergent. As we have done at the MF level, we consider the
magnetic field independent regularization scheme, in which we subtract the
corresponding unregularized ``$B=0$'' contributions and then we add them in
a regularized form. Thus, for a given polarization function
$\hat{\jnomatrix}_{MM'}$ we have
\begin{equation}
\hat{\jnomatrix}_{MM'}^{{\rm reg}}\ =\ \hat{\jnomatrix}_{MM'}^{{0,{\rm reg}}}\,+\,\hat{\jnomatrix}_{MM'}^{{\rm mag}}\ .\label{jregjmag}
\end{equation}

The ``$B=0$'' pieces of the polarization functions are quoted in
App.~\ref{b0loops}, considering arbitrary four-momentum $q^{\mu}$. In that
appendix we give the expressions for the unregularized functions
${\jnomatrix}_{MM'}^{0,{\rm unreg}}$, and use a proper time regularization
scheme to get the regularized expressions ${\jnomatrix}_{MM'}^{0,{\rm
reg}}$. The terms $\hat{\jnomatrix}_{MM'}^{{0,{\rm reg}}}$ in
Eq.~(\ref{jregjmag}) are then obtained from these expressions by taking
$q^{2}=m^{2}$. In the case of the ``magnetic'' contributions
$\hat{\jnomatrix}_{MM'}^{{\rm mag}}$, we proceed as follows: we take the
full expressions for the polarization functions ${\jnomatrix}_{MM'}(q)$
given in App.~\ref{polneutral}, and subtract the unregularized pieces
${\jnomatrix}_{MM'}^{0,{\rm unreg}}$; next, we take $q^{\mu}=(m,0,0,0)$ and
make use of the relations in App.~\ref{useful}, performing some integration
by parts when convenient. After a rather long calculation, it is found that
$\hat{\jnomatrix}_{MM'}^{{\rm mag}}$ can be expressed in the form given by
Eq.~(\ref{jotas2}), viz.
\begin{equation}
\hat{\jnomatrix}_{MM'}^{{\rm mag}}\ =\ \hat{\Sigma}_{MM'}^{u,\,{\rm mag}}+\varepsilon_{M}\varepsilon_{M'}\,\hat{\Sigma}_{MM'}^{d,\,{\rm mag}}\ ,
\end{equation}
where the functions $\hat{\Sigma}_{MM'}^{f,\,{\rm mag}}$ are given by
\begin{eqnarray}
\hat{\Sigma}_{\pi_{b}\pi_{b'}}^{f,\,{\rm mag}} & = & N_{c}\,\Big(I_{1f}^{{\rm mag}}-m^{2}I_{2f}^{{\rm mag}}\Big)\ ,\\[2mm]
\hat{\Sigma}_{\rho_{b}^\mu\rho_{b'}^\nu}^{f,\,{\rm mag}\,\mu\nu} & = &
N_{c}\,\Big(I_{4f}^{{\rm mag}}\,\eta_{\perp}^{\mu\nu}-m^{2}I_{5f}^{{\rm mag}}\,\delta_{\,3}^{\mu}\delta_{\,3}^{\nu}\Big)\ ,\\[2mm]
\hat{\Sigma}_{a_{b}^\mu a_{b'}^\nu}^{f,\,{\rm mag}\,\mu\nu} & = &
-N_{c}\,\Big[4M_{f}^{2}I_{2f}^{{\rm mag}}\,\delta_{\,0}^{\mu}\delta_{\,0}^{\nu}\,
-\left(I_{4f}^{{\rm mag}}+2M_{f}^{2}\,I_{7f}^{{\rm mag}}\right)\eta_{\perp}^{\mu\nu}\nonumber \\
 &  & +\left(m^{2}I_{5f}^{{\rm mag}}-4M_{f}^{2}\,I_{2f}^{{\rm mag}}\right)\delta_{\,3}^{\mu}\delta_{\,3}^{\nu}\Big]\ ,\\[2mm]
\hat{\Sigma}_{\pi_{b}\rho_{b'}^\mu}^{f,\,{\rm mag}\,\mu} & = & -\hat{\Sigma}_{\rho_{b}^\mu\pi_{b'}}^{f,\,{\rm mag}\,\mu}=-i\,s_{f}N_{c}\,I_{3f}^{{\rm mag}}\,\delta_{\,3}^{\mu}\ ,\\[2mm]
\hat{\Sigma}_{\pi_{b}a_{b'}^\mu}^{f,\,{\rm mag}\,\mu} & = & -\hat{\Sigma}_{a_{b}^\mu\pi_{b'}}^{f,\,{\rm mag}\,\mu}=2i\,N_{c}\,m\,M_{f}\,I_{2f}^{{\rm mag}}\,\delta_{\,0}^{\mu}\ ,\\[2mm]
\hat{\Sigma}_{a_{b}^\mu\rho_{b'}^\nu}^{f,\,{\rm mag}\,\mu\nu} & = &
\hat{\Sigma}_{\rho_{b}^\nu a_{b'}^\mu}^{f,\,{\rm mag}\,\nu\mu}
=s_{f}\,N_{c}\,\Big[\left(I_{6f}^{{\rm mag}}+M_{f}/m\,I_{3f}^{{\rm mag}}\right)\delta_{\,0}^{\mu}\delta_{\,3}^{\nu}\nonumber \\
 &  & \qquad\qquad\quad+\left(I_{6f}^{{\rm mag}}-M_{f}/m\,I_{3f}^{{\rm mag}}\right)\delta_{\,3}^{\mu}\delta_{\,0}^{\nu}\Big]\ .
\label{cfmag}
\end{eqnarray}
The expression for $I_{1f}^{{\rm mag}}$ has been given in Eq.~(\ref{i1}),
whereas the integrals $I_{nf}^{{\rm mag}}$ for $n=2,\dots,7$ read
\begin{eqnarray}
I_{2f}^{{\rm mag}} & = & \frac{1}{8\pi^{2}}\int_{0}^{1}dv\left[\psi(\bar{x}_{f})+\frac{1}{2\bar{x}_{f}}-\ln\bar{x}_{f}\right]\ ,\\
I_{3f}^{{\rm mag}} & = & \frac{M_{f}m}{8\pi^{2}}\int_{0}^{1}dv\;\frac{1}{\bar{x}_{f}}\ ,\\
I_{4f}^{{\rm mag}} & = & -\,I_{1f}^{{\rm mag}}-\frac{B_{f}}{4\pi^{2}}\sum_{s=\pm1}\int_{0}^{1}dv\ \bigg(\bar{x}_{f}
+ \frac{m^2}{4B_f} + \frac{sv}{2}\bigg)\Bigg[\ln\bar{x}_{f}-\psi\left(\bar{x}_{f}+\frac{1+sv}{2}\right)\Bigg]\ ,\\
I_{5f}^{{\rm mag}} & = & \frac{1}{8\pi^{2}}\int_{0}^{1}dv\ (1-v^{2})\left[\psi(\bar{x}_{f})+\frac{1}{2\bar{x}_{f}}-\ln\bar{x}_{f}\right]\ ,\\
I_{6f}^{{\rm mag}} & = & \frac{m^{2}}{32\pi^{2}}\int_{0}^{1}dv\; \frac{(1-v^{2})}{\bar{x}_{f}}\ ,\\
I_{7f}^{{\rm mag}} & = & \frac{1}{8\pi^{2}}\sum_{s=\pm1}\int_{0}^{1}dv\ \Big[\ln\bar{x}_{f}-\psi\left(\bar{x}_{f}+\frac{1+sv}{2}\right)\Big]\ .\label{inf}
\end{eqnarray}
Here we have defined
$\bar{x}_{f}=\left[M_{f}^{2}-(1-v^{2})m^{2}/4\right]/(2B_{f})$. For
$m<2M_{f}$, the integrals in the above expressions are well defined, while
for $m\geq 2M_{f}$ (i.e., beyond the $q\bar{q}$ production threshold) they
are divergent. Still, if this is the case one can obtain finite results by
performing analytic extensions~\cite{Carlomagno:2022inu}.

\subsection{Box structure of the neutral meson mass matrix}
\label{subs:neutralbox}

The quadratic piece of the bosonized action in Eq.~(\ref{eqnneutral})
involves 20 meson states. However, it can be seen that some of these
states do not get mixed, i.e., the $20\times20$ mass matrix can be
separated into several blocks, or ``boxes''.

The vector fields $\rhocmu$ and $\rhotmu$, as well as the axial vector
fields $\acmu$ and $\atmu$, can be written in a polarization vector basis.
Since the magnetic field defines a privileged direction in space, to exploit
the symmetries of the physical system it is convenient to choose one of the
polarization vectors $\epsilon^{\mu}$ in such a way that the spatial part
$\vec{\epsilon}\,$ is parallel to $\vec{B}$. A possible choice of a
polarization vector set satisfying this condition is introduced in
App.~\ref{pol_vec}: the polarization vector denoted by
$\epsilon^{\mu}(\vec{q},2)$ is such that $\vec{\epsilon}\,(\vec{q},2)$ is
parallel to the magnetic field, regardless of the three-momentum $\vec{q}$.
Now, as explained in App.~\ref{pbsymmetry}, the system has an invariance
related to the reflection on the plane perpendicular to the magnetic field
axis. If we associate to this transformation an operator $\mathcal{P}_{3}$,
the pseudoscalar and scalar particle states transform under
$\mathcal{P}_{3}$ by getting phases $\eta_{\mathcal{P}_{3}}^{\pi_{b}}=-1$
and $\eta_{\mathcal{P}_{3}}^{\sigma_{b}}=1$, respectively (here $b=0,3$). In
general, the transformation of the vector and axial vector states is more
complicated, depending on their polarizations. However, the choice of
$\epsilon^{\mu}(\vec{q},2)$ as one of the (orthogonal) polarization vectors
guarantees a well definite behavior of vector particle states; indeed,
considering the remaining polarization vectors in App.~\ref{pol_vec}, which
are denoted by $\epsilon^{\mu}(\vec{q},c)$ with $c=1$, $3$, $L$, one has
$\eta_{\mathcal{P}_{3}}^{\rho_{b,2}}=\eta_{\mathcal{P}_{3}}^{a_{b,1}}=
\eta_{\mathcal{P}_{3}}^{a_{b,3}}=\eta_{\mathcal{P}_{3}}^{a_{b,L}}=-1$ and
$\eta_{\mathcal{P}_{3}}^{a_{b,2}}=\eta_{\mathcal{P}_{3}}^{\rho_{b,1}}=
\eta_{\mathcal{P}_{3}}^{\rho_{b,3}}=\eta_{\mathcal{P}_{3}}^{\rho_{b,L}}=1$.
Here we have introduced the notation $\rho_{b,c}\,$, $a_{b,c}$, where $b=0$
and $b=3$ correspond to isoscalar an isovector states respectively, and the
index $c$ ($=1,2,3,L$) indicates the polarization state.

To get rid of the Lorentz indices, it is convenient to deal with a mass
matrix $\mathbf{G}$ in which the vector and axial vector meson entries are
given by the corresponding projections onto the polarization vector states.
Taking into account the matrix $G_{MM'}$ in Eq.~(\ref{eqnneutral}), and
using the above mentioned polarization basis, we have
\begin{eqnarray}
{\mathbf{G}}_{s_{b}s'_{b'}} & = & {G}_{s_{b}s'_{b'}}\ ,\nonumber \\
{\mathbf{G}}_{s_{b}v_{b'\!,c}} & = & {G}_{s_{b}v_{b'}^\mu}^{\,\mu}\,\epsilon_{\mu}(\vec{q},c)\ ,\nonumber \\
{\mathbf{G}}_{v_{b,c} s_{b'}} & = & \epsilon_{\mu}(\vec{q},c)^{\ast}\,{G}_{v_{b}^\mu s_{b'}}^{\,\mu}\ ,\nonumber \\
{\mathbf{G}}_{v_{b,c}{v'\!}_{b'\!,c'}} & = & \epsilon_{\mu}(\vec{q},c)^{\ast}
\,{G}_{v_{b}^\mu {v'}_{b'}^\nu}^{\,\mu\nu}\,\epsilon_{\nu}(\vec{q},c')\ ,
\end{eqnarray}
where $c,c'=1,2,3,L$. Here $s$ and $s'$ stand for the scalar or pseudoscalar
states $\pi,\sigma$, while $v$ and $v'$ stand for the vector or axial vector
states $\rho,a$. Now, as shown in App.~\ref{pbsymmetry}, the fact that the
system is invariant under the reflection in the plane perpendicular to the
magnetic field implies that particles with different parity phases
$\eta_{\mathcal{P}_{3}}^{M}$ cannot mix; therefore, the $20\times20$ matrix
$\mathbf{G}$ turns out to be separated into two $10\times10$ blocks. It can
be written as
\begin{equation}
\mathbf{G}\ =\ \mathbf{G}^{(-)}\oplus\,\mathbf{G}^{(+)}\ ,
\end{equation}
where the corresponding meson subspaces are
\begin{eqnarray}
\mathbf{G}^{(-)} & ,\  & {\rm states}\ \ \pi_{b},\ \rho_{b,2},\ a_{b,1},\ a_{b,3},\ a_{b,L}\ , \quad b=0,3\ ; \\
\mathbf{G}^{(+)} & ,\  & {\rm states}\ \ \sigma_{b},\ \rho_{b,1},\ \rho_{b,3},\
\rho_{b,L},\ a_{b,2}\ , \quad b=0,3\ .
\end{eqnarray}

There are more symmetry properties that can still be taken into account.
Notice that, according to its definition, the polarization vector
$\epsilon^{\mu}(\vec{q},2)$ is invariant under rotations around the axis 3,
which implies that it is an eigenvector of the operator
$S_{3}^{\mu\nu}=i\left(\delta_{1}^{\mu}\,\delta_{2}^{\nu}-\delta_{2}^{\mu}\,\delta_{1}^{\nu}\right)$
with eigenvalue $s_{3}=0$. Moreover, the whole physical system is invariant
under rotations around the axis 3, and consequently the third component of
total angular momentum, $J_{3}=\left(\vec{x}\times\vec{q}\right)_{3}+S_{3}$,
has to be a good quantum number. Thus, if we let $\vec{q}_{\perp}=\vec{0}$,
the quantum number $S_{3}$ will be a good one to characterize the meson
states.

Let us consider the polarization vectors defined in App.~\ref{pol_vec}. As
stated, $\epsilon^{\mu}(\vec{q},2)$ is an eigenvector of $S_{3}^{\mu\nu}$,
while $\epsilon^{\mu}(\vec{q},L)$ is defined as a ``longitudinal'' vector,
in the sense that its spatial part is parallel to $\vec{q}$. The
remaining polarization vectors, $\epsilon^{\mu}(\vec{q},1)$ and
$\epsilon^{\mu}(\vec{q},3)$, do not have in general a simple interpretation.
Now, if we let $\vec{q}_{\perp}=0$, they reduce to
\begin{equation}
\epsilon^{\mu}(\vec{q}_{\parallel},1)=\frac{1}{\sqrt{2}}\left(0,1,i,0\right)\ ,\qquad\epsilon^{\mu}(\vec{q}_{\parallel},3)=\frac{1}{\sqrt{2}}\left(0,1,-i,0\right)\ ,\label{polneu_a}
\end{equation}
where $\vec{q}_{\parallel}=(0,0,q^{3})$. Thus, it is seen that
$\vec{\epsilon}(\vec{q}_{\parallel},1)$ and
$\vec{\epsilon}(\vec{q}_{\parallel},3)$ lie in the plane perpendicular to
the magnetic field, and meson states with polarizations
$\epsilon^{\mu}(\vec{q}_{\parallel},1)$ and
$\epsilon^{\mu}(\vec{q}_{\parallel},3)$ are states of definite third
component of the spin, with eigenvalues $s_{3}=+1$ and $s_{3}=-1$,
respectively. The states with polarizations
$\epsilon^{\mu}(\vec{q}_{\parallel},2)$ and
$\epsilon^{\mu}(\vec{q}_{\parallel},L)$ are also eigenstates of $S_{3}$,
with eigenvalue $s_{3}=0$. As stated, in this case $S_{3}$ is a good quantum
number; this supports our choice of using for vector and axial vector states
the polarization basis $\rho_{b,c}$, $a_{b,c}$.

If mesons are taken to be at rest, i.e.\ if we take $\vec{q}=0$,
we can identify the mesons with polarizations $\epsilon^{\mu}(\vec{0},L)$
as spin zero states, and those with polarizations $\epsilon^{\mu}(\vec{0},2)$
as spin one ($s_{3}=0$) states. In this case one has simply
\begin{equation}
\epsilon^{\mu}(\vec{0},2)=\left(0,0,0,1\right)\ ,\qquad\epsilon^{\mu}(\vec{0},L)=\left(1,0,0,0\right)\ .\label{polneu_b}
\end{equation}
We notice, however, that our physical system is not fully isotropic, but
only invariant under rotations around the axis 3. Thus, $|\vec{S}|^{\,2}$ is
not a conserved quantum number, and in general the states with polarizations
$L$ and 2 will get mixed.

For clarification, we find it convenient to distinguish between the
polarization three-vectors $\vec{\epsilon}(\vec{0},c)$, $c=1,2,3$, and the
spin vectors of the $S=1$ vector and axial vector states. We define the spin
vector as the expected value
\begin{equation}
\langle\vec{S}\rangle_{c} \ = \ \frac{\epsilon_{\mu}(\vec{0},c)^{\ast}
\left(S_{1}^{\mu\nu},S_{2}^{\mu\nu},S_{3}^{\mu\nu}\right)\epsilon_{\nu}(\vec{0},c)}
{\epsilon_{\alpha}(\vec{0},c)^{\ast}\epsilon^{\alpha}(\vec{0},c)}\ ,
\end{equation}
with $S_{j}^{\mu\nu}=i\,\epsilon_{jkl}\,\delta_{k}^{\mu}\delta_{l}^{\nu}$. A
simple calculation leads to $\langle\vec{S}\rangle_{1}=\left(0,0,1\right)$,
$\langle\vec{S}\rangle_{3}=\left(0,0,-1\right)$ and
$\langle\vec{S}\rangle_{2}=\left(0,0,0\right)$, showing that for the
polarization vectors $\epsilon^{\mu}(\vec{0},1)$ and
$\epsilon^{\mu}(\vec{0},3)$ the spin is parallel or antiparallel to the
magnetic field, whereas for the polarization vector
$\epsilon^{\mu}(\vec{0},2)$ the spin has no preferred direction. Notice that
in Ref.~\cite{Carlomagno:2022inu} the $\rho^\mu$ states with polarizations
$\epsilon^{\mu}(\vec{0},2)$ and $\epsilon^{\mu}(\vec{0},c)$, $c=1,3$ were
denoted as ``perpendicular'' ($\rho_\perp$) and ``parallel''
($\rho_\parallel$), respectively.

Let us turn back to the mass matrix $\mathbf{G}$. From the regularized
polarization functions in Eq.~(\ref{jregjmag}) we can obtain a regularized
matrix $\hat{\mathbf{G}}(m^{2})$, where we have taken $q^{\mu}=(m,0,0,0)$.
Notice that the regularization procedure does not modify our previous
analysis about the symmetries of the problem. Thus, according to the above
discussion, we can conclude that ---for neutral mesons--- each one of the
$10\times10$ submatrices of $\hat{\mathbf{G}}(m^{2})$ gets further
decomposed as a direct sum of a subspace of $s_{3}=0$ states (that includes
vector and axial vector mesons with polarization states $c=2,L$), a subspace
of $s_{3}=+1$ states (polarization states $c=1$) and a subspace of
$s_{3}=-1$ states (polarization states $c=3$). In this way, the $20\times20$
matrix $\hat{\mathbf{G}}(m^{2})$ can be decomposed in ``boxes'' as
\begin{equation}
\hat{\mathbf{G}}=\hat{\mathbf{G}}^{(0,-)}\oplus\hat{\mathbf{G}}^{(1,-)}
\oplus\hat{\mathbf{G}}^{(-1,-)}\oplus\hat{\mathbf{G}}^{(0,+)}
\oplus\hat{\mathbf{G}}^{(1,+)}\oplus\hat{\mathbf{G}}^{(-1,+)}\ ,
\label{gboxes}
\end{equation}
where the superindices indicate the quantum numbers $(s_{3},\eta_{\mathcal{P}_{3}})$.
The meson subspaces corresponding to each box are the following:
\begin{equation}
\begin{array}{lcl}
\hat{\mathbf{G}}^{(0,-)}\ , & \  & {\rm states}\ \ \pi_{0},\pi_{3},\rho_{0,2},\rho_{3,2},a_{0,L},a_{3,L}\ ;\\
\hat{\mathbf{G}}^{(1,-)}\ , & \  & {\rm states}\ \ a_{0,1},a_{3,1}\ ;\\
\hat{\mathbf{G}}^{(-1,-)}\ , & \  & {\rm states}\ \ a_{0,3},a_{3,3}\ ;\\
\hat{\mathbf{G}}^{(0,+)}\ , & \  & {\rm states}\ \ \sigma_{0},\sigma_{3},\rho_{0,L},\rho_{3,L},a_{0,2},a_{3,2}\ ;\\
\hat{\mathbf{G}}^{(1,+)}\ , & \  & {\rm states}\ \ \rho_{0,1},\rho_{3,1}\ ;\\
\hat{\mathbf{G}}^{(-1,+)}\ , & \  & {\rm states}\ \ \rho_{0,3},\rho_{3,3}\ .
\end{array}
\end{equation}

Finally, it can also be seen that at the considered level of perturbation
theory the sigma mesons $\sigma_{b}$ get decoupled from other states. Thus,
the matrix $\hat{\mathbf{G}}^{(0,+)}$ can still be decomposed as
\begin{equation}
\hat{\mathbf{G}}^{(0,+)} \ = \ \hat{\mathbf{G}}_{S}^{(0,+)} \,\oplus\,
\hat{\mathbf{G}}_{V}^{(0,+)}\ .
\end{equation}
The submatrices in the right hand side correspond to the scalar meson
subspace $\sigma_{b}$, with $b=0,3$, and the meson subspace
$\rho_{b,L},a_{b,2}$, with $b=0,3$, respectively.

\subsection{Neutral meson masses and wave-functions}
\label{neumass}

{}From the expressions in the previous subsections one can
obtain the model predictions for meson masses and wave-functions.
Let us concentrate on the lightest pseudoscalar and vector meson states,
which can be identified with the physical $\pi^{0}$, $\eta$, $\rho^{0}$
and $\omega$ mesons. The pole masses of the neutral pion, the $\eta$,
and the $S_{z}=0$ neutral $\rho$ and $\omega$ mesons are given
by the solutions of
\begin{equation}
\mbox{det}\,\hat{\mathbf{G}}^{(0,-)}\ =\ 0\ ,\label{detgpar}
\end{equation}
while the pole masses of $S_{z}=\pm1$ vector meson states can be
obtained from
\begin{equation}
\mbox{det}\,\hat{\mathbf{G}}^{(\pm1,+)}\ =\ 0\ .\label{detgperp}
\end{equation}
Clearly, the symmetry under rotations around the axis 3, or $z$, implies
that the masses of $S_{z}=1$ and $S_{z}=-1$ states will be degenerate.

Once the mass eigenvalues are determined for each box, the spin-isospin
composition of the physical meson states can be obtained through the
corresponding eigenvectors. In the $S_z = 0$ sector, the physical
neutral pion state $\tilde{\pi}^0$ can be written as
\begin{equation}
|\tilde{\pi}^0\rangle\ =\ c_{\pi_3}^{\tilde{\pi}^0}\,|\pi_3\rangle\, +\,
c_{\pi_0}^{\tilde{\pi}^0}\,|\pi_0\rangle\, +
\,i\,c_{\rho_{3,2}}^{\tilde{\pi}^0}\,|\rho_{3,2}\rangle\, +
\,i\,c_{\rho_{0,2}}^{\tilde{\pi}^0}\,|\rho_{0,2}\rangle\, +
\,c_{a_{3,L}}^{\tilde{\pi}^0}\,|a_{3,L}\rangle\, +\,
c_{a_{0,L}}^{\tilde{\pi}^0}\,|a_{0,L}\rangle\ ,
\label{physmes}
\end{equation}
and in a similar way one can define coefficients $c_{M}^{\tilde M}$ for
other physical states $\tilde M$. On the other hand, in the $S_z =\pm
1$ sector it is convenient to write isospin states in terms of the flavor
basis $(\rho_{u,c},\rho_{d,c})$ for $c=1,3$, viz.
\begin{equation}
| \rho_{0,c} \rangle = \frac{1}{\sqrt2}\left(| \rho_{u,c} \rangle + | \rho_{d,c}\rangle\right) \ ,
\qquad\ \
| \rho_{3,c} \rangle  = \frac{1}{\sqrt2}\left(| \rho_{u,c} \rangle - | \rho_{d,c} \rangle\right)\ .
\label{udbasis}
\end{equation}
Since in this sector vector mesons do not mix with pseudoscalar or axial
vector mesons, the states $|\rho_{f,c}\rangle$ ($f=u,d$) with $c=1$ and
$c=3$ turn out to be the mass eigenstates that diagonalize the matrices
$\hat{\mathbf{G}}^{(1,+)}$ and $\hat{\mathbf{G}}^{(-1,+)}$, respectively.
This can be easily understood by noticing that the external magnetic field
distinguishes between quarks that carry different electric charges, and in
this case this represents the only source of breakdown of the $u$-$d$ flavor
degeneracy.

\section{The charged meson sector}

\label{chargedmes}

\subsection{Charged meson polarization functions}

\label{chargedpol}

We address now the analysis of the charged mesons, i.e.\ the states
$s^{\pm}=(s_{1}\mp is_{2})/\sqrt{2}$ and $v^{\pm\mu}=(v_{1}^{\mu}\mp iv_{2}^{\mu})/\sqrt{2}$,
with $s=\sigma,\pi$ and $v=\rho,a$. We concentrate on the positive
charge sector, noticing that the analysis of negatively charged mesons
is completely equivalent. The corresponding quadratic piece of the
bosonized action can be written as
\begin{equation}
S_{\mathrm{bos}}^{{\rm quad,+}}\ =\ -\frac{1}{2}\int d^{4}x\;d^{4}x'\sum_{M,M'}\,\delta M(x)^{\dagger}\ {\cal G}_{MM'}(x,x')\ \delta M'(x')\ ,\label{quadbosch}
\end{equation}
where, for notational convenience, we simply denote the positively
charged states by $M,M'=\sigma,\pi,\rho^{\mu},a^{\mu}$ (a proper
contraction of Lorentz indices of vector mesons is understood). The
functions ${\cal G}_{MM'}(x,x')$ can be separated in two terms, namely
\begin{equation}
{\cal G}_{MM'}(x,x')\ =\ \frac{1}{2g_{M}}\ \delta_{MM'}\,\delta^{(4)}(x-x')-{\cal J}_{MM'}(x,x')\ ,
\end{equation}
where
\begin{equation}
\frac{1}{g_{M}}\ \delta_{MM'}\ =\ \left\{ \begin{array}{cl}
1/g & \ \ \ \ \mbox{for}\ \ M=M'=\pi\\
1/[g(1-2\cc)] & \ \ \ \ \mbox{for}\ \ M=M'=\sigma\\
-\eta^{\mu\nu}/\gcoupr & \ \ \ \ \mbox{for}\ \ MM'=\rho^{\mu}\rho^{\nu},a^{\mu}a^{\nu}
\end{array}\right.\qquad,
\label{valgm}
\end{equation}
and $\delta_{MM'}=0$ otherwise. The polarization functions ${\cal J}_{MM'}(x,x')$
are given by
\begin{equation}
{\cal J}_{MM'}(x,x')\ =\ -2i\,N_{c}\,\mbox{tr}_{D}\left[iS_{x,x'}^{u}\,\Gamma^{M'}\,iS_{x',x}^{d}\,\Gamma^{M}\right]\ ,
\end{equation}
where, as in the case of neutral mesons, one has
$\Gamma^{\sigma}=\mathbbm{1}$, $\Gamma^{\pi}=i\gamma^{5}$,
$\Gamma^{\rho^{\mu}}=\gamma^{\mu}$ and
$\Gamma^{a^{\mu}}=\gamma^{\mu}\gamma^{5}$. Using Eq.~(\ref{uno}) we have
\begin{equation}
{\cal J}_{MM'}(x,x')\ =\
e^{i\Phi_{e}(x,x')}\int\frac{d^{4}t}{(2\pi)^{4}}\;e^{-it(x-x')}\,
{\cal J}_{MM'}(t)\ ,
\label{nueve}
\end{equation}
where
\begin{equation}
{\cal J}_{MM'}(t)\ =\ -2iN_{c}\int\frac{d^{4}p}{(2\pi)^{4}}\;
\mbox{tr}_{D}\left[i\bar{S}^{u}({p_{\parallel}^{+}},{p_{\perp}^{+}})\,
\Gamma^{M'}\,i\bar{S}^{d}({p_{\parallel}^{-}},{p_{\perp}^{-}})\,\Gamma^{M}\right]\ .
\label{jmmtt}
\end{equation}
Here we have defined $p_{a}^{\pm}=p_{a}\pm t_{a}/2$, where $a=\parallel,\perp$.
In addition, we have used $\Phi_{e}(x,x')=\Phi_{Q_{u}}(x,x')+\Phi_{Q_{d}}(x',x)$.
Thus, $\Phi_{e}$ is the Schwinger phase associated with positively
charged mesons.

Contrary to the neutral meson case discussed in the previous section, here
the Schwinger phases coming from quark propagators do not cancel, due to
their different flavors. As a consequence, the polarization functions in
Eq.~(\ref{nueve}) do not become diagonal when transformed to the momentum
basis. Instead of using the standard plane wave decomposition, to
diagonalize the polarization functions it is necessary to expand the meson
fields in terms of a set of functions associated to the solutions of the
corresponding equations of motion in the presence of a uniform magnetic
field. These functions can be specified by a set of four quantum numbers
that we denote by
\begin{eqnarray}
\bar{q}=(q^{0},\ell,\chi,q^{3})
\end{eqnarray}
(see e.g.~Ref.~\cite{GomezDumm:2023owj} for a detailed analysis). As in the
case of a free particle, $q^{0}$ and $q^{3}$ are the eigenvalues of the
components of the four-momentum operator along the time direction and the
magnetic field direction, respectively. The integer $\ell$ is related with
the so-called Landau level, while the fourth quantum number, $\chi$, can be
conveniently chosen (although this is not strictly necessary) according to
the gauge in which the eigenvalue problem is
analyzed~\cite{Wakamatsu:2022pqo,GomezDumm:2023owj}. In particular, since
for the standard gauges SG, LG1 and LG2 one has unbroken continuous
symmetries, in those cases it is natural to consider quantum numbers $\chi$
associated with the corresponding group generators. Usual choices are
\begin{align}
 & \mbox{SG:} &  & \chi=n\ , &  & \mbox{nonnegative integer, associated to \ensuremath{L_{3}}~\cite{GomezDumm:2023owj}};\\
 & \mbox{LG1:} &  & \chi=q^{1}\ , &  & \mbox{real number, eigenvalue of}\ -i\frac{\partial\ }{\partial x^{1}}\ ;\\
 & \mbox{LG2:} &  & \chi=q^{2}\ , &  & \mbox{real number, eigenvalue of}\ -i\frac{\partial\ }{\partial x^{2}}\ .
\end{align}
To sum or integrate over these quantum numbers, we introduce the shorthand
notation
\begin{equation}
\sumint_{\bar{q}}\ \equiv\ \dfrac{1}{2\pi}\sum_{\ell\,=\,\ell_{{\rm min}}}^{\infty}\int\frac{{dq^{0}\,dq^{3}}}{(2\pi)^{2}}\left\{ \begin{array}{cl}
{\displaystyle \frac{1}{2\pi}\sum_{n}} & \qquad\mbox{for SG}\\
[5mm]{\displaystyle \frac{1}{2\pi}\int dq^{i}} & \qquad\mbox{for LG\ensuremath{i}\,, \ensuremath{i=1,2}}
\end{array}\right.\label{notation2}
\end{equation}
where $\ell_{{\rm min}}=0$~$(-1)$ for spin 0 (spin 1) particles.

In this way, we can write
\begin{eqnarray}
 &  & \delta\sigma(x)\ =\sumint_{\bar{q}}\ \mathbb{F}(x,\bar{q})\,\delta\sigma(\bar{q})\ ,\qquad\qquad\delta\pi(x)\ =\sumint_{\bar{q}}\ \mathbb{F}(x,\bar{q})\,\delta\pi(\bar{q})\ ,\nonumber \\
 &  & \delta\rho^{\mu}(x)\ =\sumint_{\bar{q}}\ \mathbb{R}^{\mu\nu}(x,\bar{q})\,\delta\rho_{\nu}(\bar{q})\ ,\qquad\qquad\delta a^{\mu}(x)\ =\sumint_{\bar{q}}\ \mathbb{R}^{\mu\nu}(x,\bar{q})\,\delta a_{\nu}(\bar{q})\ ,\label{diez}
\end{eqnarray}
where
\begin{eqnarray}
\mathbb{F}(x,\bar{q})={\cal F}_{e}(x,\bar{q})\ ,\qquad\qquad\mathbb{R}^{\mu\nu}(x,\bar{q})=
\sum_{\lambda=-1,0,1}{\cal F}_{e}(x,\bar{q}_{\lambda})\ \Upsilon_{\lambda}^{\mu\nu}\ ,
\label{FandR}
\end{eqnarray}
with $\bar{q}_{\lambda}=(q^{0},\ell-s\lambda,\chi,q^{3})$,
$s=\mbox{sign}(B)$. The function ${\cal F}_{Q}(x,\bar{q})$ depends on the
gauge choice; the explicit forms that correspond to the standard gauges are
given in App.~\ref{functionF}. Regarding the tensors
$\Upsilon_{\lambda}^{\mu\nu}$, one has various possible choices; here we
take
\begin{equation}
\Upsilon_{0}^{\mu\nu}=\eta_{\parallel}^{\mu\nu}\ ,\qquad\qquad\Upsilon_{\pm1}^{\mu\nu}=\frac{1}{2}(\eta_{\perp}^{\mu\nu}\mp S_{3}^{\mu\nu})\ .\label{upsidef}
\end{equation}

Given Eqs.~(\ref{diez}) we introduce the polarization functions
in $\bar{q}$-space (or Ritus space). They read
\begin{eqnarray}
{\cal J}_{ss'}(\bar{q},\bar{q}') & = & \int d^{4}x\,d^{4}x'\ \mathbb{F}(x,\bar{q})^{\ast}\,
{\cal J}_{ss'}(x,x')\,\mathbb{F}(x',\bar{q}')\ ,\nonumber \\
{\cal J}_{sv^\mu}^{\mu}(\bar{q},\bar{q}') & = & \int d^{4}x\,d^{4}x'\ \mathbb{F}(x,\bar{q})^{\ast}\,
{\cal J}_{sv^\alpha}^{\,\alpha}(x,x')\,\mathbb{R}_\alpha^{\ \mu}(x',\bar{q}')\ ,\nonumber \\
{\cal J}_{v^\mu s}^{\mu}(\bar{q},\bar{q}') & = & \int d^{4}x\,d^{4}x'\ {\mathbb{R}_\alpha^{\ \mu}(x,\bar{q})}^{\ast}\,
{\cal J}_{v^\alpha s}^{\,\alpha}(x,x')\,\mathbb{F}(x',\bar{q}')\ ,\nonumber \\
{\cal J}_{v^\mu v^{\prime\nu}}^{\mu\nu}(\bar{q},\bar{q}') & = & \int d^{4}x\,d^{4}x'
\ {\mathbb{R}_\alpha^{\ \mu}(x,\bar{q})}^{\ast}\,{\cal J}_{v^\alpha v^{\prime\beta}}^{\alpha\beta}(x,x')\,
\mathbb{R}_\beta^{\ \nu}(x',\bar{q}')\ ,
\label{chargeJbarq}
\end{eqnarray}
where $s,s'$ stand for the states $\sigma$ or $\pi$, while $v,v'$
stand for $\rho$ or $a$. After a somewhat long calculation one can
show that all these $\bar{q}$-space polarization functions are diagonal,
i.e., one has
\begin{eqnarray}
{\cal J}_{MM'}(\bar{q},\bar{q}')\ =\ \hat{\delta}_{\bar{q}\bar{q}'}\,J_{MM'}(\ell,q_{\parallel})\ ,
\label{jdiag}
\end{eqnarray}
where
\begin{eqnarray}
\hat{\delta}_{\bar{q}\bar{q}'}=\left(2\pi\right)^{4}\delta(q^{0}-q^{\prime\,0})\,
\delta_{\ell\ell'}\,\delta_{\chi\chi'}\,\delta(q^{3}-q^{\prime\,3})\ .
\end{eqnarray}
Here, $\delta_{\chi\chi'}$ stands for $\delta_{nn'}$, $\delta(q^{1}-q^{\prime\,1})$
and $\delta(q^{2}-q^{\prime\,2})$ for SG, LG1 and LG2, respectively.
It is important to stress that Eq.~(\ref{jdiag}) holds for all three
gauges; moreover, the functions $J_{MM'}(\ell,q_{\parallel})$ are
independent of the gauge choice. The explicit form of these functions
for the various possible $MM'$ combinations, together with some details
of the calculations, are given in App.~\ref{polcharged}. The quadratic
piece of the bosonized action in Eq.~(\ref{quadbosch}) can now be
expressed as
\begin{equation}
S_{\mathrm{bos}}^{{\rm quad,+}}\ =\ - \dfrac{1}{2} \sumint_{\bar{q}}\sum_{M,M'}
\,\delta M(\bar{q})^{\dagger}\ {G}_{MM'}(\ell,q_{\parallel})\ \delta M^{\,\prime}(\bar{q})\ ,
\end{equation}
where
\begin{equation}
{G}_{MM'}(\ell,q_{\parallel})\ =\ \frac{1}{2g_{M}}\,\delta_{MM'}
-{J}_{MM'}(\ell,q_{\parallel})\ .
\label{eqn70}
\end{equation}

As in the case of neutral mesons, to determine the charged meson masses
it is convenient to write the vector and axial vectors states in a
polarization basis. A suitable set of polarization vectors
$\epsilon^{\mu}(\ell,q^{3},c)$, where $c=1,2,3,L$ is the polarization index,
is given in App.~\ref{pol_vec}. Here, $c=L$ corresponds to the
``longitudinally polarized'' charged mesons, which will be denoted by
$\rho_{L}$ and $a_{L}$; for these states the polarization vector
$\epsilon^{\mu}(\ell,q^{3},L)$ is defined only for $\ell\geq0$, and it is
proportional to the four-vector $\Pi^{\mu}$ defined by Eq.~(\ref{Pigrande}),
evaluated at $q^{0}=\sqrt{m^{2}+(2\ell+1)B_{e}+(q^{3})^{2}}$. Next, to get
rid of the Lorentz indices of vector and axial vector states, we consider
the mass matrix $\mathbf{G}$ and the polarization functions $\mathbf{J}$
obtained in the basis given by the corresponding
projections onto the polarization vector states. We have
\begin{eqnarray}
{\mathbf{G}}_{ss'}(\ell,\Pi^{2}) & = & \frac{1}{2g_{s}}\,\delta_{ss'}-\mathbf{J}_{ss'}(\ell,\Pi^{2})\ ,\nonumber \\
{\mathbf{G}}_{sv_{c}}(\ell,\Pi^{2}) & = & -\,{\mathbf{J}}_{sv_{c}}(\ell,\Pi^{2})\ ,\nonumber \\
{\mathbf{G}}_{v_{c}s}(\ell,\Pi^{2}) & = & -\,{\mathbf{J}}_{v_{c}s}(\ell,\Pi^{2})\ ,\nonumber \\
{\mathbf{G}}_{v_{c}{v'}_{c'}}(\ell,\Pi^{2}) & = & -\frac{1}{2g_{v}}\,\zeta_{c}\,
\delta_{v_{c}{v'}_{c'}}-\,{\mathbf{J}}_{v_{c}{v'}_{c'}}(\ell,\Pi^{2})\ ,
\label{gyj}
\end{eqnarray}
where
\begin{eqnarray}
\mathbf{J}_{ss'}(\ell,\Pi^{2}) & = & {J}_{ss'}(\ell,q_{\parallel})\ ,\nonumber \\
{\mathbf{J}}_{sv_{c}}(\ell,\Pi^{2}) & = & {J}_{sv^\mu}^{\,\mu}(\ell,q_{\parallel})\,\epsilon_{\mu}(\ell,q^{3},c)\ ,\nonumber \\
{\mathbf{J}}_{v_{c}s}(\ell,\Pi^{2}) & = & \epsilon_{\mu}(\ell,q^{3},c)^{\ast}\,{J}_{v^\mu s}^{\,\mu}(\ell,q_{\parallel})\ ,\nonumber \\
{\mathbf{J}}_{v_c {v'}_{c'}}(\ell,\Pi^{2}) & = & \epsilon_{\mu}(\ell,q^{3},c)^{\ast}
\,{J}_{v^\mu v^{\prime\nu}}^{\,\mu\nu}(\ell,q_{\parallel})\,\epsilon_{\nu}(\ell,q^{3},c')\ .\label{boldj}
\end{eqnarray}
In the above equations, $s$ and $s'$ stand for the scalar or pseudoscalar
states $\pi,\sigma$, while $v$ and $v'$ stand for the vector or axial vector
states $\rho,a$. We use once again the definitions $g_{\pi}=g$,
$g_{\sigma}=g(1-2\alpha)$, whereas $\zeta_c$ is defined as $\zeta_{c}=1$ for
$c=L$ and $\zeta_{c}=-1$ for $c=1,2,3$. Moreover, we have defined
$\Pi^2 = \Pi_\mu^\ast\, \Pi^\mu$. From Eq.~(\ref{Pigrande}), one has $\Pi^{2} =
q_{\parallel}^{2}-(2\ell+1)B_{e}$.

To determine the physical meson pole masses corresponding to a given Landau
level $\ell$, we need to evaluate the matrix elements of
${\mathbf{G}}(\ell,\Pi^{2})$ at $\Pi^{2}=m^{2}$. However, as in the case of
the neutral meson sector, it turns out that many of the corresponding
polarization functions are divergent. Once again, we consider the magnetic
field independent regularization scheme, according to which we have
\begin{equation}
\mathbf{J}^{{\rm reg}}(\ell,\Pi^{2})\ =\ \mathbf{J}^{0,\,{\rm reg}}(\Pi^{2})\,+\,\mathbf{J}^{{\rm mag}}(\ell,\Pi^{2})\ .
\label{jregjmagch}
\end{equation}
To obtain the regularized ``$B=0$'' matrix $\mathbf{J}^{0,\,{\rm reg}}(\Pi^{2})$
we calculate the projections over polarization states as in Eqs.~(\ref{boldj}),
replacing the functions $J_{MM'}(\ell,q_{\parallel})$ by their regularized
expressions. The latter are obtained by taking the corresponding regularized
functions $J_{MM'}^{{\rm reg}}(q)$ in App.~\ref{b0loops}, and performing
the replacement $q^{\mu}\to\Pi^{\mu}$. On the other hand, to determine
the ``magnetic'' contribution $\mathbf{J}^{{\rm mag}}(\ell,\Pi^{2})$
we calculate the matrix elements of $\mathbf{J}(\ell,\Pi^{2})$
according to Eqs.~(\ref{boldj}) (as stated, the functions $J_{MM'}(\ell,q_{\parallel})$
in that equation are quoted in App.~\ref{polcharged}), and then
we subtract the corresponding unregularized expressions in the
above defined $B\to 0$ limit. These can be obtained from the unregularized
functions $J_{MM'}^{{\rm unreg}}(q)$ in App.~\ref{b0loops}, following
the same procedure as for the regularized ones.

\subsection{Box structure of the charged meson mass matrix}

As in the case of neutral mesons, the symmetries of the system imply that
not all charged mesons states mix with each other. Firstly, it is clear that
the mass matrix can be separated into two equivalent sectors of positive and
negative charges. Next, restricting ourselves to positively charged mesons,
it is seen that one can exploit the symmetry of the system under the
reflection on the plane perpendicular to the magnetic field to classify the
meson states into two groups. This is discussed in detail in
App.~\ref{pbsymmetry}, where the action of the operator $\mathcal{P}_{3}$,
associated to this symmetry transformation, is studied. Considering the
polarization basis introduced in the previous subsection, it is found that
charged meson states $M$ transform under $\mathcal{P}_{3}$ by getting phases
$\eta_{\mathcal{P}_{3}}^{M}=\pm1$.
In a similar way as in the case of neutral meson states,
the $10\times10$ mass matrix $\mathbf{G}(\ell,\Pi^{2})$ can be written
as a direct sum of two $5\times5$ submatrices,
\begin{equation}
\mathbf{G}\ =\ \mathbf{G}^{(-)}\oplus\,\mathbf{G}^{(+)}\ ,
\end{equation}
where the corresponding meson subspaces are
\begin{eqnarray}
\mathbf{G}^{(-)} & ,\  & {\rm states}\ \ \pi,\rho_{2},a_{L},a_{1},a_{3}\ ;\\
\mathbf{G}^{(+)} & ,\  & {\rm states}\ \ \sigma,a_{2},\rho_{L},\rho_{1},\rho_{3}\ .
\end{eqnarray}

Now, it is worth noticing that while the above discussion holds for
Landau levels $\ell\geq1$, one should separately consider the particular
cases $\ell=-1$ and $\ell=0$. As mentioned above, one has $\ell_{{\rm min}}=0$
for pseudoscalar and scalar fields; moreover, as discussed in App.~\ref{pol_vec},
for $\ell=-1$ there is only one nontrivial polarization vector, $\epsilon_{\mu}(-1,q^{3},1)$.
Therefore, the charged mass matrix $\mathbf{G}(-1,\Pi^{2})$ is given
by a direct sum of two $1\times1$ matrices $\mathbf{G}^{(-)}$ and
$\mathbf{G}^{(+)}$ corresponding to the states $a_{1}$ and $\rho_{1}$,
respectively. These do not mix with any other state. The case $\ell=0$
is also a particular one, since, as stated in App.~\ref{pol_vec},
one cannot have a vector or axial vector meson field polarized in
the direction $\epsilon_{\mu}(0,q^{3},c)$ with $c=3$. In this way,
the charged mass matrix $\mathbf{G}(0,\Pi^{2})$ is given by a direct
sum of two $4\times4$ matrices.

\subsection{Charged meson masses and wave-functions}

Taking into account the results in the previous subsections, the pole
masses of charged mesons can be obtained, for each value of $\ell$,
by solving the equations
\begin{equation}
\mbox{det}\,\mathbf{G}^{(\pm)}(\ell,m^{2})\ =\ 0\ .
\end{equation}
Here we are interested in the determination of the energies of the lowest
lying meson states. As stated, for the Landau mode $\ell=-1$ the only
available states are the vector meson $\rho_{1}$ and the axial vector meson
$a_{1}$, which do not mix with each other. In turn, for $\ell=0$ one gets
the lowest energy charged pion, which gets coupled through
$\mathbf{G}^{(-)}$ to the $\ell=0$ vector and axial vector mesons. In what
follows we analyze these two modes in detail.

As mentioned above, for $\ell=-1$ the matrix $\mathbf{G}^{(+)}$
has dimension 1. Thus, according to Eqs.~(\ref{gyj}) and (\ref{jregjmagch}),
the pole mass of the $\rho$ state can be obtained from
\begin{equation}
\frac{1}{2g_{v}}\,-\,\mathbf{J}_{\rho_{1}\rho_{1}}^{{\rm reg}}(-1,m^{2})\ =\ 0\ ,
\label{rhopolemass}
\end{equation}
where
\begin{equation}
\mathbf{J}_{\rho_{1}\rho_{1}}^{{\rm reg}}(-1,m^{2})\ =\ \mathbf{J}_{\rho_{1}\rho_{1}}^{0,\,{\rm reg}}(m^{2})\,+
\,\mathbf{J}_{\rho_{1}\rho_{1}}^{{\rm mag}}(-1,m^{2})\ .
\end{equation}
The functions on the r.h.s.\ of this equation can be obtained from
the definitions in Sec.~\ref{chargedpol}; one has
\begin{eqnarray}
\mathbf{J}_{\rho_{1}\rho_{1}}^{0,\,{\rm reg}}(-1,m^{2}) & = & -2\,b_{\rho\rho,1}^{ud,\,{\rm reg}}(m^{2})\ ,\nonumber \\
\mathbf{J}_{\rho_{1}\rho_{1}}^{{\rm mag}}(-1,m^{2}) & = & -2\,\left[d_{\rho\rho,2}(-1,m^{2}-B_{e})-b_{\rho\rho,1}^{ud,\,{\rm unreg}}(m^{2})\right]\ ,
\end{eqnarray}
where $b_{\rho\rho,1}^{ud,\,{\rm reg}}$ and $b_{\rho\rho,1}^{ud,\,{\rm
unreg}}$ are given in App.~\ref{b0loops}, while the expression of
$d_{\rho\rho,2}$ can be found in App.~\ref{polcharged}. Once the solution
$m^{2}=m_{\rho^{+}}^{2}$ has been determined, we can obtain the energy
$E_{\rho^{+}}$ of the lowest charged $\rho$ state as
\begin{equation}
E_{\rho^{+}}\,=\,\sqrt{m_{\rho^{+}}^{2}+(2\ell+1)B_{e}+(q^{3})^{2}}\Big|_{\ell=-1,q^{3}=0}\ =\ \sqrt{m_{\rho^{+}}^{2}-B_{e}}\ .
\label{rhoenergy}
\end{equation}

In the case of the lowest charged pion state ($\ell=0$), we consider
the $4\times4$ mass matrix $\mathbf{G}^{(-)}(0,m^{2})$ that couples
the states $\pi$, $\rho_{2}$, $a_{1}$ and $a_{L}$. The pole mass
can be found from
\begin{eqnarray}
\mbox{det}\,\left[\mbox{diag}\left(\frac{1}{2g}\,,\,\frac{1}{2g_{v}}\,,\,\frac{1}{2g_{v}}\,,
\,-\frac{1}{2g_{v}}\right)-\,\mathbf{J}^{{\rm reg}}(0,m^{2})\right]\ =\ 0\ ,
\label{pimasseq}
\end{eqnarray}
where, according to Eq.~(\ref{jregjmagch}),
\begin{equation}
\mathbf{J}^{{\rm reg}}(0,m^{2})\ =\ \mathbf{J}^{0,\,{\rm reg}}(m^{2})\,+\,\mathbf{J}^{{\rm mag}}(0,m^{2})\ .
\label{jmagl0}
\end{equation}
The nonvanishing matrix elements of $\mathbf{J}^{0,\,{\rm reg}}(m^{2})$
read
\begin{eqnarray}
\mathbf{J}_{\pi\pi}^{0,\,{\rm reg}}(m^{2}) & = & 2\,b_{\pi\pi,1}^{ud,\,{\rm reg}}(m^{2})\ ,\nonumber \\
\mathbf{J}_{\rho_{2}\rho_{2}}^{0,\,{\rm reg}}(m^{2}) & = & -2\,b_{\rho\rho,1}^{ud,\,{\rm reg}}(m^{2})\ ,\nonumber \\
\mathbf{J}_{a_L a_L}^{0,\,{\rm reg}}(m^{2}) & = & 2\,b_{aa,2}^{ud,\,{\rm reg}}(m^{2})\ ,\nonumber \\
\mathbf{J}_{a_{1}a_{1}}^{0,\,{\rm reg}}(m^{2}) & = & -2\,b_{aa,1}^{ud,\,{\rm reg}}(m^{2})\ ,\nonumber \\
\mathbf{J}_{\pi a_L}^{0,\,{\rm reg}}(m^{2}) & = & \mathbf{J}_{a_L\pi}^{0,\,{\rm reg}}(m^{2})^{\ast}\,=\,2\,m\,b_{\pi a,1}^{ud,\,{\rm reg}}(m^{2})\ ,
\end{eqnarray}
where the functions on the right hand sides are given in App.~\ref{b0loops}.
The matrix elements of $\mathbf{J}^{{\rm mag}}(0,m^{2})$, obtained
from the general expressions quoted in App.~\ref{polcharged}, are
given in App.~\ref{matl0}. The lowest solution of Eq.~(\ref{pimasseq})
can be identified with the charged pion pole mass squared, $m_{\pi^+}^{2}$.
Then the energy of the lowest charged pion reads
\begin{equation}
E_{\pi^{+}}\,=\,\sqrt{m_{\pi^{+}}^{2}+(2\ell+1)B_{e}+(q^{3})^{2}}\Big|_{\ell=0,q^{3}=0}\ =\ \sqrt{m_{\pi^{+}}^{2}+B_{e}}\ .
\label{pimasenergy}
\end{equation}
In the same way, higher solutions of Eq.~(\ref{pimasseq})
are to be identified with vector meson pole masses; a similar analysis
can be done for the sector corresponding to the $4\times4$ matrix
$\mathbf{G}^{(-)}(0,m^{2})$ (which involves the $\sigma$ meson).
In addition, one can obtain pole masses of other higher charged meson
states by considering Landau levels $\ell\geq1$ (as stated, the mass
matrix separates in those cases into two boxes of dimension 5).

Together with the determination of meson pole masses, we can also obtain the
spin-isospin composition of the physical meson states as in the case of
neutral mesons. For $\ell=-1$ there are just two states, $\rho_{1}$ and
$a_{1}$, which do not get mixed due to the above described reflection
symmetry. On the other hand, for $\ell\geq0$, one gets in general a
decomposition similar to that obtained in the case of neutral states. Thus,
in the particular case of the lowest lying charged pion, the physical state
$\pi^+$ can be written as a combination of $\ell = 0$ states
\begin{equation}
|\pi^+\rangle\ =\ c_{\pi}^{\pi^+}\,|\pi\rangle\, +
\,i\,c_{\rho_{2}}^{\pi^+}\,|\rho_{2}\rangle\, + \,
c_{a_{1}}^{\pi^+}\,|a_{1}\rangle\, +\,
\,i\,c_{a_{L}}^{\pi^+}\,|a_{L}\rangle\ .
\label{physmesch}
\end{equation}

\section{Numerical Results}
\label{sec:Num}

\subsection{Model parametrization and magnetic catalysis}

To obtain numerical results for particle properties it is necessary to fix
the model parameters. In addition to the usual requirements for the
description of low energy phenomenology, we find it adequate to choose a
parameter set that also takes into account LQCD results for the behavior of
quark-antiquark condensates under an external magnetic field. As stated, in
our framework divergent quantities are regularized using the MFIR scheme,
with a proper time cutoff. Within this scenario, we take the parameter set
$m_c = 7.01$~MeV, $\Lambda = 842$~MeV, $g=5.94/\Lambda^2$ and
$\alpha=0.114$. For vanishing external field, this parametrization leads to
effective quark masses $M_f = 400$~MeV and quark-antiquark condensates
$\phi_{u,d}^0 = (-227$~MeV$)^3$. Moreover, it properly reproduces the
empirical values the pion mass, the eta mass and the pion decay constant in
vacuum, namely $m_\pi = 140$~MeV, $m_\eta=548$~MeV and $f_\pi=92.2$~MeV,
respectively. Regarding the vector couplings, we take $\gcoupr =
3.947/\Lambda^2$, which for $B=0$ leads to the empirical value
$m_\rho=775$~MeV and to a phenomenologically acceptable value of about
1020~MeV for the a$_1$ mass. Notice that, as usual in this type of model,
the a$_1$ mass is found to lie above the quark-antiquark production
threshold and can be determined only after some extrapolation. For the sake
of simplicity, the remaining coupling constants of the vector and axial
vector sector are taken to be $\gcoupw=\gcoupa=\gcoupr$, which leads to
$m_\omega=m_\rho$ and $m_{f_1}=m_{{\rm a}_1}$.

As mentioned in the Introduction, while most NJL-like models are able to
reproduce the effect of magnetic catalysis at vanishing temperature, they
fail to describe the inverse magnetic catalysis effect observed in lattice
QCD at finite temperature (an interesting exception is the case of models which include nonlocal
interactions~\cite{Pagura:2016pwr,GomezDumm:2017iex}). One of the simplest
approaches to partially cure this behavior consists of allowing the model
couplings to depend on the magnetic field, so as to incorporate the sea
effect produced by the backreaction of gluons to magnetized quarks loops.
Thus, we consider here both the situation in which the couplings are
constant and the one in which they vary with the magnetic field. For
definiteness, we adopt for $g(B)$ the form proposed in
Ref.~\cite{Avancini:2016fgq}, namely
\begin{equation}
g(B) \ =\ g\,\mathcal{F}(B) \ ,
\label{gdeb}
\end{equation}
where
\begin{equation}
\mathcal{F}(B) \ =\ \kappa_1 + (1-\kappa_1)\,e^{-\kappa_2(eB)^2} \ ,
\label{FeB}
\end{equation}
with $\kappa_1= 0.321$ and $\kappa_2= 1.31$~GeV$^{-2}$. Concerning the
vector couplings, given the common gluonic origin of $g$ and $\gcoupr$, we
assume that they get affected in a similar way by the magnetic field; hence,
we take $\gcoupr(B)=\gcoupr \mathcal{F}(B)$.

\begin{figure}[h]
\centering{}\includegraphics[width=0.6\textwidth]{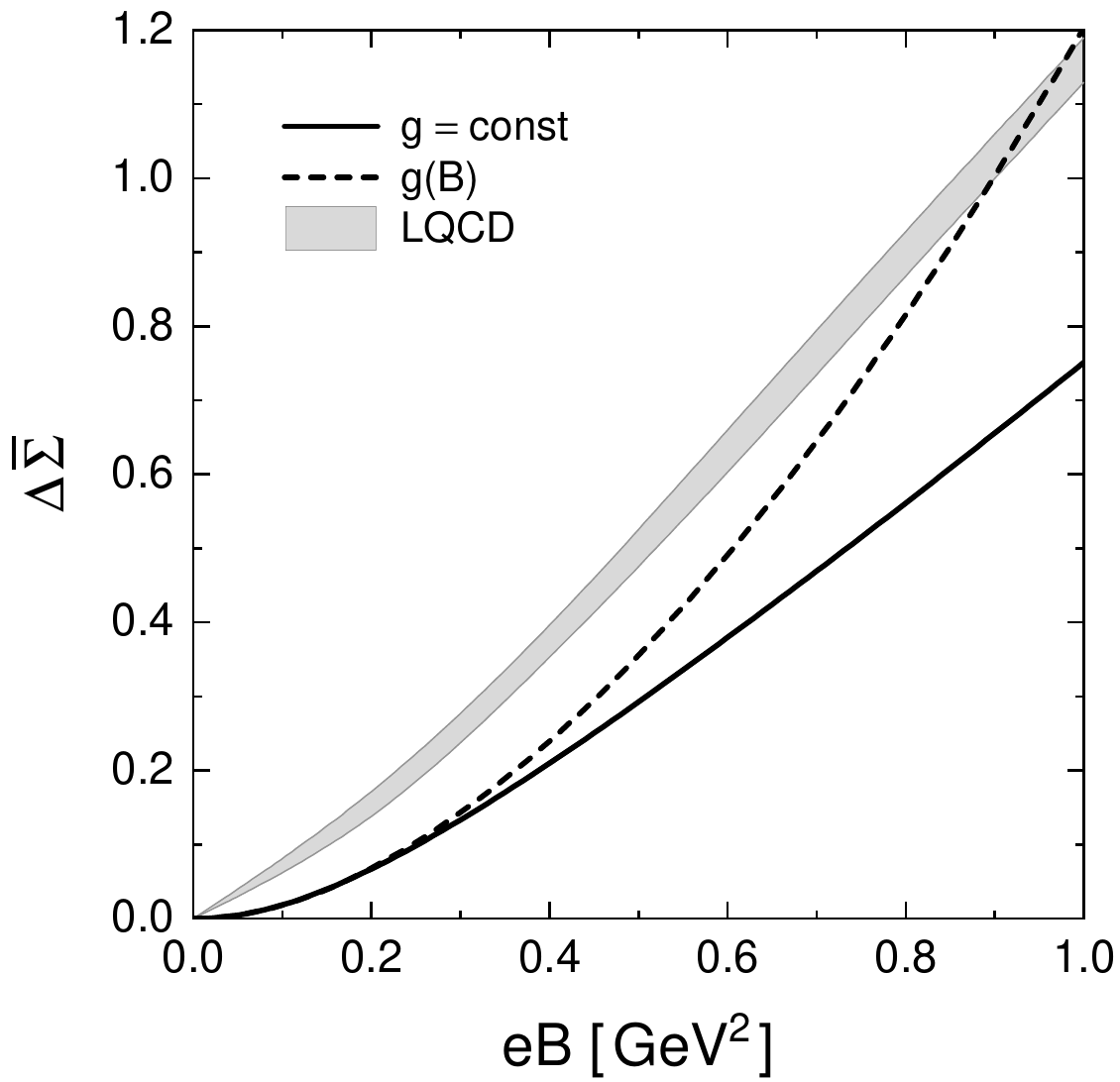}
\caption{Normalized average $q\bar q$ condensate as a function of
$eB$. Solid and dashed lines correspond to constant and
$B$-dependent couplings. LQCD results from Ref.~\cite{Bali:2012zg}
(gray band) are added for comparison.}
\label{fig:sigmas}
\end{figure}

The effect of magnetic catalysis can be observed from Fig.~\ref{fig:sigmas},
where we show the behavior of the normalized averaged light quark condensate
as a function of the magnetic field, for $eB$ up to 1~GeV$^2$. Following
Ref.~\cite{Bali:2012zg}, we use the definitions
\begin{equation}
\Delta \bar\Sigma(B) \ =\ \dfrac{\Delta \Sigma_u (B)+\Delta \Sigma_d (B)}{2} \ ,
\qquad\qquad
\Delta\Sigma_f(B) =  - \frac{2\,m_c \, [\phi_f(B) - \phi_f^0]}{D^4}\ ,
\label{Nor_Cond}
\end{equation}
where $D=(135\times 86)^{1/2}$~MeV is a phenomenological normalization
constant. Solid and dashed lines correspond to constant and $B$-dependent
couplings, respectively. Although the curves do not show an accurate fit to
lattice data (gray band, taken from Ref.~\cite{Bali:2012zg}), it is seen
that the model is able to reproduce qualitatively the effect of magnetic
catalysis. We have seen that a better agreement could be achieved using a
parameter set that leads to lower values of the quark masses; however, this
would hinder the analysis of the rho meson mass, since the latter would lie
below the quark-antiquark production threshold even for $B=0$.
Additionally, we have checked that the choice of a 3D-cutoff (within the
MFIR scheme) leads in general to even lower values of $\Delta \bar\Sigma$,
increasing the difference with LQCD results.

\subsection{Neutral mesons}

Let us analyze our results for the effect of the magnetic field on meson
masses. We start with the neutral sector. As well known, for vanishing
external field pseudoscalar mesons mix with ``longitudinal'' axial vector
mesons. Now, as discussed in Sec.~\ref{subs:neutralbox}, for nonzero $B$ the
mixing also involves neutral vector mesons with spin projection $S_z=0$
(corresponding to the polarization state $c=2$). The four lowest mass states
of this sector are to be identified with the physical states $\tilde\pi^0$,
$\tilde\eta$, $\tilde\rho^{\,0}$ and $\tilde\omega$, where the particle
names are chosen according to the spin-isospin composition of the states in
the limit of vanishing external field, see Eq.~(\ref{physmes}).

\begin{figure}[h]
\centering{}\includegraphics[width=0.7\textwidth]{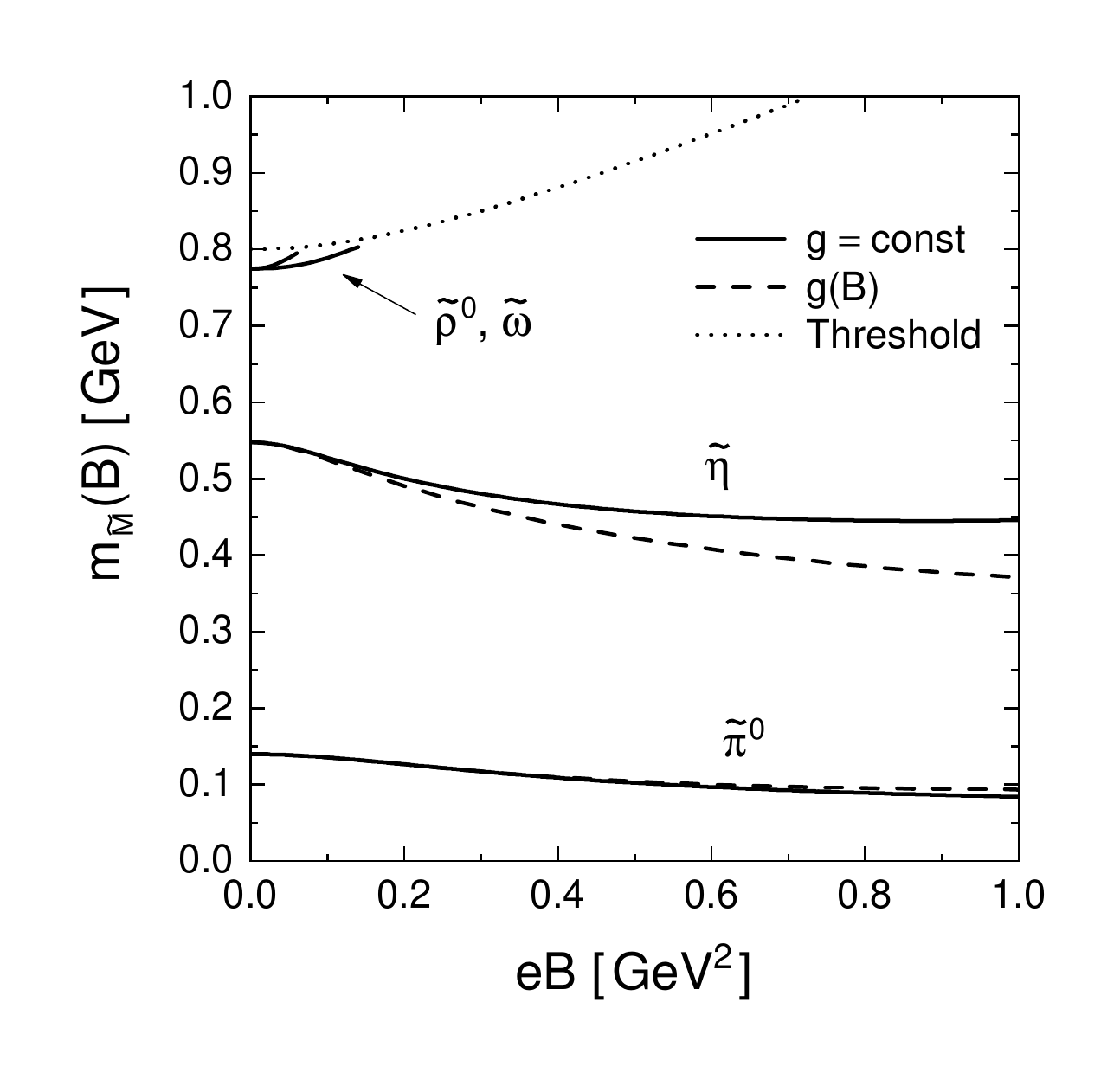}
\vspace*{-5mm}
\caption{Masses of neutral mesons with spin projection $S_z=0$ as
functions of $eB$. Solid (dashed) lines correspond to constant
($B$-dependent) couplings.} \label{fig:neutralps}
\end{figure}

The masses of these particles can be determined from Eq.~(\ref{detgpar}). In
Fig.~\ref{fig:neutralps} we show their behavior with the magnetic field, for
constant and $B$-dependent couplings (solid and dashed lines, respectively).
In the case of $\tilde\rho^{\,0}$ and $\tilde\omega$ mesons, for $B=0$ one
has $m_\rho = m_\omega = 775$~MeV, close to the quark-antiquark production
threshold ---which arises from the lack of confinement of the
model--- given by $2M_d(B=0)=800$~MeV. As can be seen from the figure,
since $m_{\tilde\rho^0}$ and $m_{\tilde\omega}$ increase with the magnetic
field, they overcome the threshold (shown by the dotted line) at relatively
low values of $eB$. Beyond this limit, although one could obtain some
results through analytic
continuation~\cite{Avancini:2021pmi,Carlomagno:2022inu}, pole masses would
include an unphysical absorptive part, becoming relatively less reliable.
For clarity, we display in Fig.~\ref{fig:neutralps} just the curves for
$m_{\tilde\rho^0}$ and $m_{\tilde\omega}$ that correspond to the case of a
constant value of the coupling $g$; in the case of the $B$-dependent
coupling $g(B)$, the situation is entirely similar. It is also worth
mentioning that the results for the $\tilde{\eta}$ and $\tilde\omega$ masses
should be taken only as indicative, since a more realistic calculation would
require a three-flavor version of the model in which flavor-mixing effects
could be fully taken into account.

\begin{figure}[h]
\centering{}\includegraphics[width=1.0\textwidth]{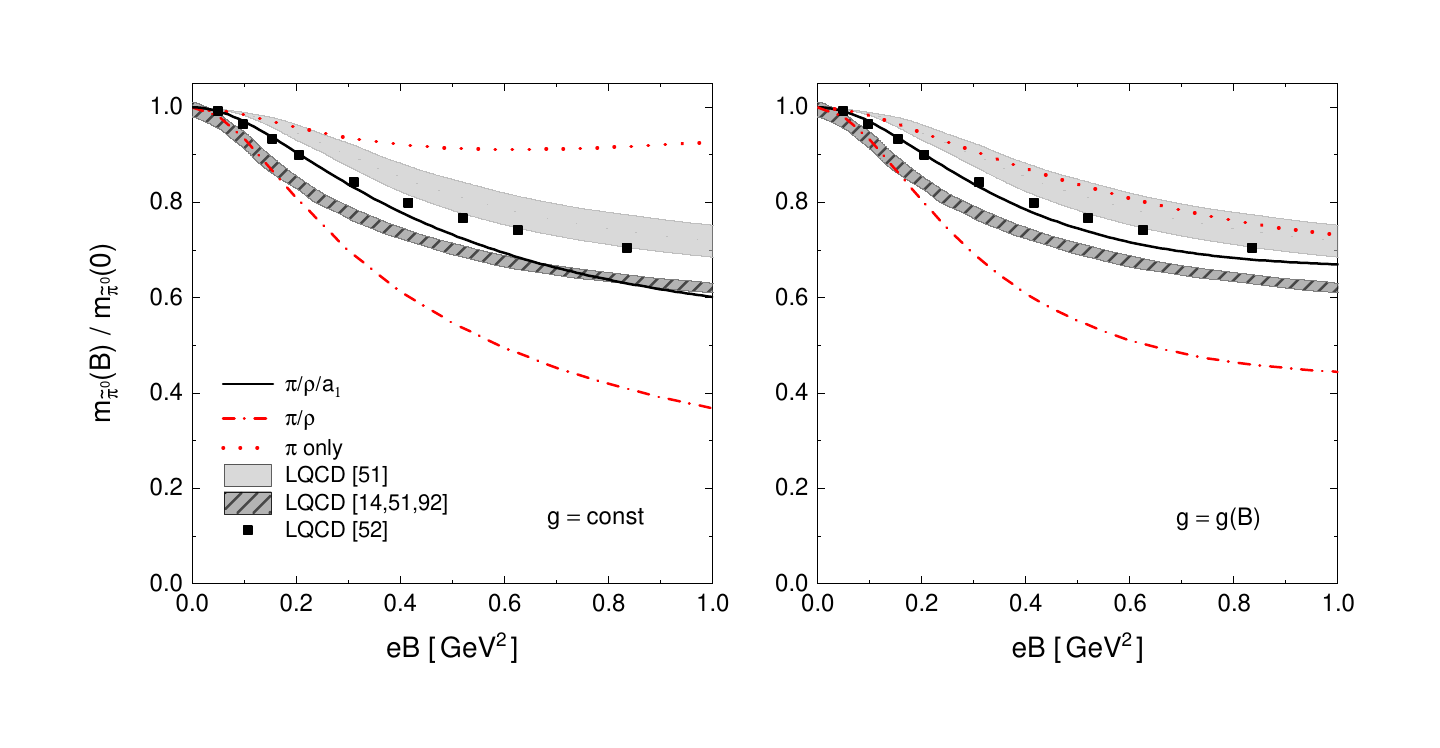}
\vspace*{-1.3cm}
\caption{Normalized mass of the $\tilde{\pi}^0$ meson as a function of $eB$,
for the case of constant (left panel) and $B$-dependent couplings (right
panel). Red dotted and dash-dotted lines show the results from models that do
not include the axial vector meson sector. The bands and the fat squares
correspond to LQCD results quoted in
Refs.~\cite{Bali:2011qj,Bali:2017ian,Borsanyi:2010cj,Ding:2020hxw}.}
\label{fig:neutralpion}
\end{figure}

Regarding the neutral pion mass, in Fig.~\ref{fig:neutralpion} we compare
our results with those obtained in previous
works~\cite{Avancini:2016fgq,Carlomagno:2022inu} and those corresponding to
LQCD calculations, in which quenched Wilson fermions~\cite{Bali:2017ian},
dynamical staggered quarks~\cite{Bali:2011qj,Bali:2017ian,Borsanyi:2010cj}
and improved staggered quarks~\cite{Ding:2020hxw} are considered. Although
LQCD studies do not take into account flavor mixing (they deal with
individual flavor states), according to the analysis in
Ref.~\cite{Carlomagno:2022inu} the lightest meson mass is expected to be
approximately independent of the value of the mixing parameter $\alpha$. It
is also worth noticing that LQCD results have been obtained using different
methods and values of the pion mass at $B=0$. In the figure we show the
results obtained for NJL-like models in which different meson sectors have
been taken into account. Left and right panels correspond to $g={\rm
constant}$ and $g=g(B)$ [given by Eqs.~(\ref{gdeb}-\ref{FeB})],
respectively. If one considers just the pseudoscalar sector (red dotted
lines), when $g$ is kept constant the behavior of $m_{\tilde\pi^0}$ with the
magnetic field is found to be non monotonic, deviating just slightly from
its value at $B=0$. In contrast, as seen from the right panel of
Fig.~\ref{fig:neutralpion}, if one lets $g$ to depend on the magnetic field
the mass shows a monotonic decrease, reaching a reduction of about 30\% at
$eB=1$ GeV$^2$. This suppression is shown to be in good agreement with LQCD
results. When the mixing with the vector sector is considered, the results
for both constant and $B$-dependent couplings (red dash-dotted lines in left
and right panels) are similar to each other and monotonically decreasing,
lying however quite below LQCD predictions. Finally, if the mixing
with axial vector mesons is also included (solid lines) we obtain, for both
constant and $B$-dependent couplings, a monotonic decrease which is in good
qualitative agreement with LQCD calculations for the studied range of $eB$.
One may infer that the incorporation of axial vector mesons, being the
chiral partners of vector mesons, leads to cancellations that help to
alleviate the magnitude of the neutral pion mass suppression. Their
inclusion into the full picture leads to relatively more robust results, in
good agreement with LQCD calculations, and is in fact one of the main
takeaways of this work.

Let us discuss the composition of the $\tilde\pi^0$ state. The values of the
coefficients associated with the spin-isospin decomposition given in
Eq.~(\ref{physmes}) are quoted in the upper part of Table~\ref{taba} for
$eB=0$, 0.5~GeV$^2$ and 1~GeV$^2$. Those associated with the spin-flavor
decomposition, defined in the same way as in Eq.~(\ref{udbasis}), are given
in the lower part of the table. We quote the values corresponding to the
model in which the couplings constants do not depend on the magnetic field;
the results are qualitatively similar for the case of $B$-dependent
couplings. One finds that while the mass eigenvalues do not depend on
whether $B$ is positive or negative, the corresponding eingenvectors do; the
relative signs in Table~\ref{taba} correspond to the choice $B>0$. We
consider first the results for vanishing magnetic field. It is seen that,
due to the wellknown $\pi$-a$_1$ mixing, the $\tilde\pi^0$ state has already
some axial vector component. We also note that even though $\alpha$ is
relatively small (in our parametrization we have taken $\alpha=0.114$, to be
compared with its maximum possible value 1/2), the effect of flavor mixing
is already very strong; the spin-isospin composition is clearly dominated by
the $\pi_3$ component, which is given by an antisymmetric equal-weight
combination of $u$ and $d$ quark flavors. This can be understood by noticing
that, as soon as $\alpha$ is different from zero, the U(1)$_A$ symmetry gets
broken. The state $\pi_3$ is then the only one that remains being a
pseudo-Goldstone boson, which forces the lowest-mass state $\tilde\pi^0$ to
be dominated by the $\pi_3$ component. In the presence of the magnetic
field, the mixing is expected to be modified, since the external field
distinguishes between flavor components $\pi_u$ and $\pi_d$ instead of
isospin states. From the upper part of Table~\ref{taba} it is seen that,
even for the relatively small value of $\alpha$ considered here, the mass
state $\tilde\pi^0$ is dominated by the $\pi_3$ component
($|c^{\tilde\pi^0}_{\pi_3}|^2 \gtrsim 0.97$) for the full range of
values of $eB$ up to 1~GeV$^2$. This means that the dominance of the
flavor composition over the isospin composition will occur only for
extremely large values of $eB$. In any case, from the values in
Table~\ref{taba} one can still observe some effect of the magnetic field on
the composition of the $\tilde\pi^0$ state: when $eB$ increases, it is
found that there is a slight decrease of the $\pi_3$ component in favor of
the others. In addition, a larger weight is gained by the $u$-flavor
components, as one can see by looking at the entries corresponding to the
spin-flavor states (lower part of Table~\ref{taba}): one has
$|c^{\tilde\pi^0}_{\pi_u}|^2+|c^{\tilde \pi^0}_{\rho_{u,2}}|^2+ |c^{\tilde
\pi^0}_{a_{u,L}}|^2 = 0.50 (0.64)$ for $eB=0 (1.0)$~GeV$^2$. This can be
understood by noticing that the magnetic field is known to reduce the mass
of the lowest neutral meson
state~\cite{Luschevskaya:2014lga,Bali:2017ian,Ding:2020hxw}; for large $eB$
one expects the lowest mass state ($\tilde\pi^0$) to have a larger component
of the quark flavor that couples more strongly to the magnetic field (i.e.,
the $u$ quark). Concerning the vector meson components of the $\tilde\pi^0$
state, it is seen that they are completely negligible at low values of $eB$,
reaching a contribution similar to the one of the axial vector meson
($\simeq 0.5 \ \%$) at $eB= 1$~GeV$^2$.

\begin{table}[h]
\begin{center}
\begin{tabular}{cc cc cc cc}
\hline \hline
 $eB\ [{\rm GeV}^2]$ & &\multicolumn{6}{c}{Spin-isospin composition}
\\
 & &  $c^{\tilde \pi^0}_{\pi_3}$ &
      $c^{\tilde \pi^0}_{\pi_0}$ &
      $c^{\tilde \pi^0}_{\rho_{3,2}}$ &
      $c^{\tilde \pi^0}_{\rho_{0,2}}$ &
      $c^{\tilde \pi^0}_{a_{3,L}}$ &
      $c^{\tilde \pi^0}_{a_{0,L}}$
\\
\hline
 0   & &   0.998   &    0        &    0             &  0          &    -0.067     &   0
 \\
 0.5 & &   0.993   &    0.084    &    0.016         &  0.060      &    -0.063     &   -0.011
\\
 1.0 & &   0.987   &    0.141    &    0.010         &  0.057      &    -0.058     &   -0.012
\\
\hline \hline
&& && && &\\ \hline \hline
  $eB\ [{\rm GeV}^2]$   & &\multicolumn{6}{c}{Spin-flavor composition}
\\
 && \hspace*{5mm}  $c^{\tilde \pi^0}_{\pi_u}$\hspace*{5mm}&
    \hspace*{5mm}  $c^{\tilde \pi^0}_{\pi_d}$\hspace*{5mm} &
    \hspace*{5mm}  $c^{\tilde \pi^0}_{\rho_{u,2}}$\hspace*{5mm} &
    \hspace*{5mm}  $c^{\tilde \pi^0}_{\rho_{d,2}}$\hspace*{5mm} &
    \hspace*{5mm}  $c^{\tilde \pi^0}_{a_{u,L}}$\hspace*{5mm} &
    \hspace*{5mm}  $c^{\tilde \pi^0}_{a_{d,L}}$\hspace*{5mm}
\\
\hline
 0   & & 0.706 &   -0.706         & 0           &    0     & -0.047  & 0.047 \\
 0.5   & & 0.707 &  -0.704         & 0.011           &    0.006     & -0.047  & 0.047 \\
 1.0     & & 0.798  &   -0.598     & 0.048      & 0.033  &  -0.050 & 0.032 \\
\hline \hline
\end{tabular}
\caption{Composition of the $\tilde\pi^0$ meson mass eigenstate for selected values
of $eB$. Relative signs hold for the choice $B>0$. The upper
table corresponds to the spin-isospin decomposition, as given in
Eq.~(\ref{physmes}), while the lower one corresponds to a spin-flavor
decomposition.}
\label{taba}
\end{center}
\end{table}

In addition, as discussed in Sec.~\ref{subs:neutralbox}, the neutral sector
includes states with spin projections $S_z = \pm 1$, i.e., spin parallel to
the direction of the magnetic field. We consider here the effect of the
magnetic field on vector meson states, whose masses can be obtained from the
submatrices $\hat{\mathbf{G}}^{(\pm 1,+)}$ in Eq.~(\ref{gboxes}). Since in
this sector vector meson and axial vector meson states do not mix, the
analysis is entirely equivalent to the one carried out in
Ref.~\cite{Carlomagno:2022inu}, where the axial vector sector was not taken
into account. As stated in Sec.~\ref{neumass}, it is easy to see that
the mass matrices involving the states $\rho_{0,c}$ and $\rho_{3,c}$, with
$c=1,3$ are diagonalized by rotating from the isospin basis to a flavor
basis $(\rho_{u,c},\rho_{d,c})$ given by Eq.~(\ref{udbasis}); moreover, the
masses of these mesons turn out to be equal for polarization states $c=1$
($S_z = +1$) and $c=3$ ($S_z = -1$).

\begin{figure}[t]
\centering{}\includegraphics[width=0.7\textwidth]{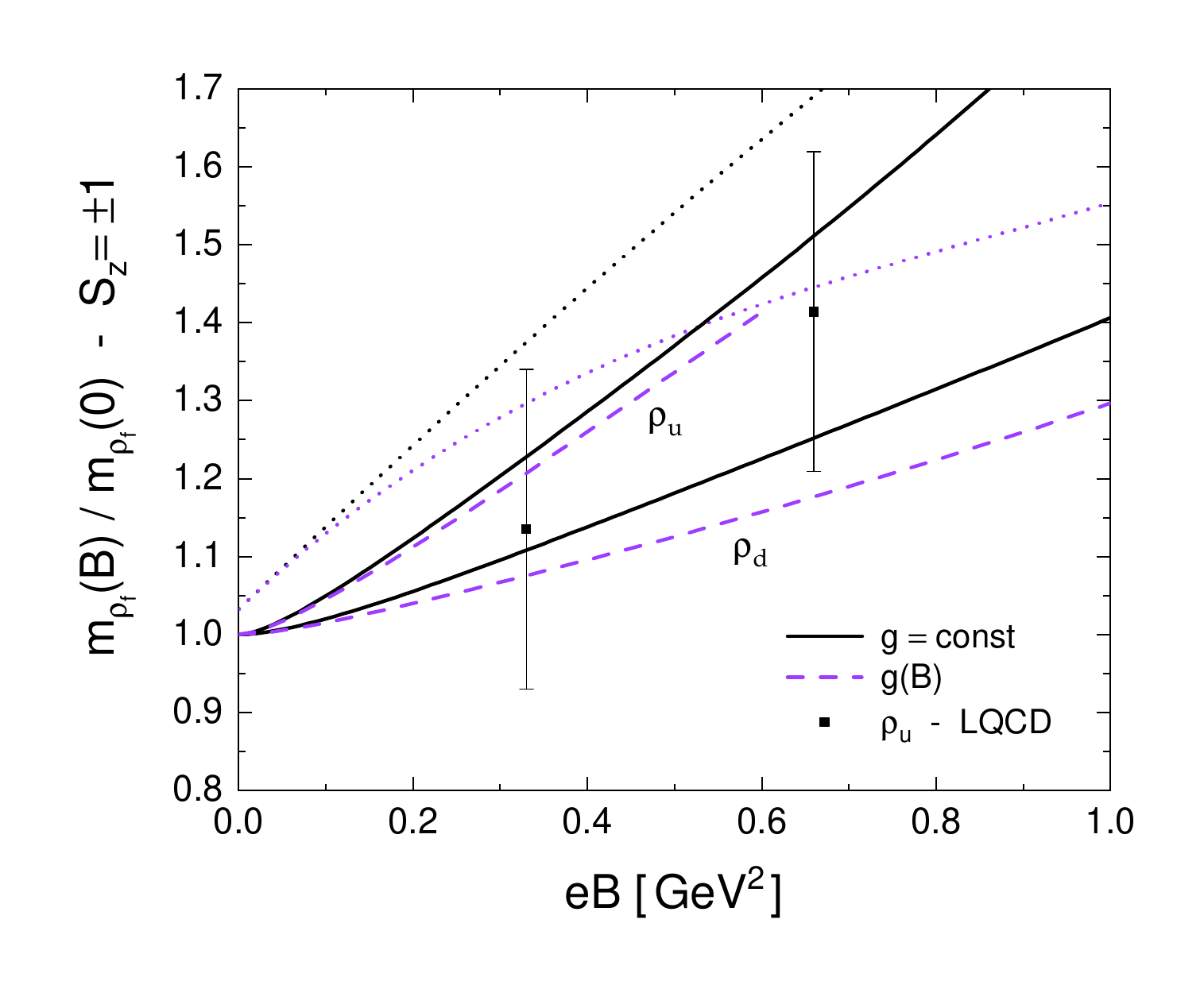}
\vspace*{-5mm}
\caption{Masses of $\rho$ mesons with spin projection $S_z=\pm 1$ as
functions of $eB$. Solid and dashed lines correspond to constant and
$B$-dependent couplings; dotted lines indicate $q\bar q$ production
thresholds. LQCD data for $\rho_u$ from Ref.~\cite{Bali:2017ian} are
included for comparison.}
\label{fig:neutralrho}
\end{figure}

The numerical results for $\rho_u$ and $\rho_d$ meson masses as functions of
the magnetic field are shown in Fig.~\ref{fig:neutralrho}. It is seen that
both masses increase with $B$, the enhancement being larger in the case of
the $\rho_{u}$ mass; this can be understood from the larger (absolute) value
of the $u$-quark charge, which measures the coupling with the magnetic
field. The results are similar for the case of constant and $B$-dependent
couplings, corresponding to solid and dashed lines in the figure,
respectively. The dotted lines indicate the mass thresholds for $q\bar q$
pair production, given by $m_{\rho_{f}}^{({\rm th})} =
M_d+\sqrt{M_d^2+2B_d}$. As discussed in Ref.~\cite{Carlomagno:2022inu}, this
threshold is given by a situation in which the spins of both the quark and
antiquark components of the $\rho_{f}$ meson are aligned (or anti-aligned)
with the magnetic field; thus, one of the fermions lies in its lowest Landau
level, while the other one lies in its first excited Landau level. In
comparison with the $S_z=0$ threshold $2M_d$, for $S_z=\pm 1$ the threshold
$m_{\rho_{f}}^{({\rm th})}$ grows faster with $B$. For a constant
coupling $g$, this allows the values of $m_{\rho_{u}}$ and $m_{\rho_{d}}$
to remain below the threshold for the studied range of magnetic fields. On
the other hand, in the case of a $B$-dependent coupling $g(B)$ the $\rho_u$
meson is found to become unstable for $eB$ somewhat larger than
$0.6$~GeV$^2$. Our results for the $\rho_u$ mass are found to be in
agreement, within errors, with values obtained through LQCD calculations,
also shown in Fig.~\ref{fig:neutralrho}~\cite{Bali:2017ian}.

\subsection{Charged mesons}
\label{reschmes}

As discussed in Sec.~\ref{chargedmes}, to study the lowest lying charged
meson states in the presence of the magnetic field one has to consider the
Landau modes $\ell=-1$ and $\ell = 0$. For $\ell = -1$,
the lowest mass state is the one that we have denoted as $\rho_1$, which
does not get mixed with any other state. The corresponding pole mass
$m_{\rho^+}$ can be obtained from Eq.~(\ref{rhopolemass}), while the lowest
energy for this state is given by $E_{\rho^+} = \sqrt{m_{\rho^+}^2-B_e}$,
see Eq.~(\ref{rhoenergy}).

\begin{figure}[t]
\centering{}\includegraphics[width=0.7\textwidth]{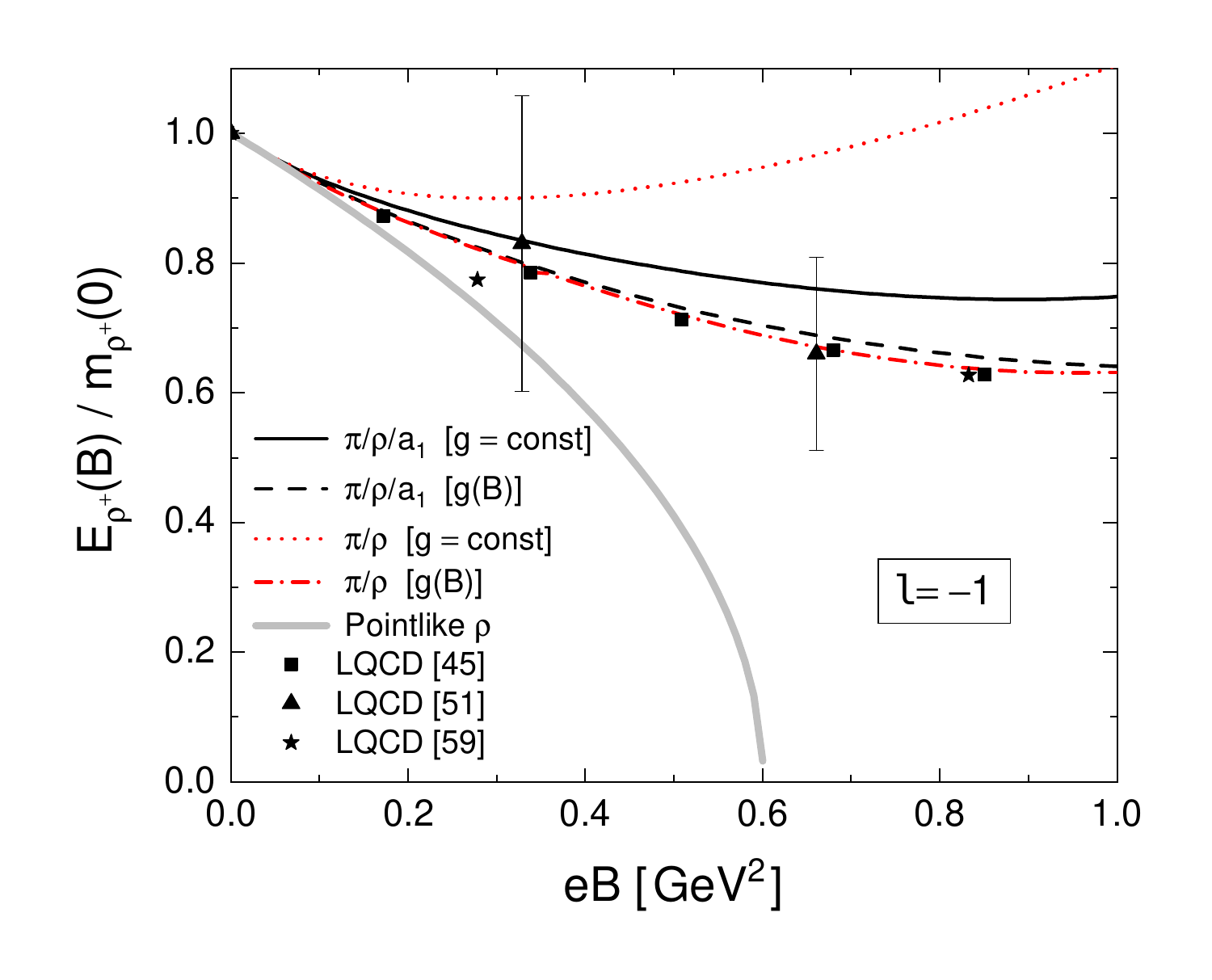}
\vspace*{-5mm}
\caption{Energy of the $\rho^+$ meson as a function of $eB$ for the lowest
Landau mode $\ell=-1$ and vanishing component of the momentum in the
direction of $\vec B$. Values are normalized to the $\rho^+$ mass at zero
external field. Black solid and dashed lines correspond to constant and
$B$-dependent couplings, respectively. Red dotted and dash-dotted lines show
results from models that do not include axial vector mesons, while the light
gray line corresponds to a pointlike $\rho^+$. For comparison, lattice QCD
data quoted in Refs.~\cite{Hidaka:2012mz,Andreichikov:2016ayj,Bali:2017ian}
are also included.}
\label{fig:chargedrho}
\end{figure}

In Fig.~\ref{fig:chargedrho} we show our numerical results for $E_{\rho^+}$
as a function of $eB$, normalized by the value of the $\rho$ mass at $B=0$.
Black solid and dashed lines correspond to the cases of constant and
$B$-dependent couplings, respectively, where $g(B)$ is given by
Eqs.~(\ref{gdeb}-\ref{FeB}). It can be seen that for $g=\mbox{constant}$ the
results differ considerably from those obtained in a similar
model~\cite{Carlomagno:2022arc} which instead does not take into account the presence
of axial vector mesons (red dotted line in the figure). On the contrary, for
$g=g(B)$ (red dash-dotted line) they remain basically unchanged. In fact,
here the differences between models that include or not axial vector mesons
do not arise from direct mixing effects (the $\rho_1$ state does not mix
with axial vectors) but from the fact that axial vector states mix with
pions already for $B=0$; this leads to some change in the model parameters
so as to get consistency with the phenomenological inputs. In any case, it
is found that ---as in the case of neutral mesons--- the results from the
full model (black solid and dashed lines) appear to be rather robust: they
show a similar behavior either for constant or $B$-dependent couplings, and
this behavior is shown to be in good agreement with LQCD
calculations~\cite{Hidaka:2012mz,Andreichikov:2016ayj,Bali:2017ian}, also
shown in the figure. Notice that our results, as those from LQCD, are not
consistent with $\rho^+$ condensation for the considered range of values of
$eB$. The curve corresponding to the lowest energy state of a pointlike
$\rho^+$ meson as a function of $eB$ is shown for comparison.

It is worth mentioning that our results are qualitatively different from
those obtained in other works in the framework of two-flavor NJL-like
models~\cite{Liu:2014uwa,Cao:2019res}, which do find $\rho^+$ meson
condensation for $eB \sim 0.2$ to $0.6$~GeV$^2$. As discussed in
Refs.~\cite{Carlomagno:2022arc,GomezDumm:2023owj}, in those works Schwinger
phases are neglected and it is assumed that charged $\pi$ and $\rho$ mesons
lie in zero three-momentum states (i.e., meson wavefunctions are
approximated by plane waves). Here we use, instead, an expansion of meson
fields in terms of the solutions of the corresponding equations of motion
for nonzero $B$, taking properly into account the presence of Schwinger
phases in quark propagators. In fact, as shown in
Ref.~\cite{GomezDumm:2023owj}, the plane wave approximation may have a
dramatic incidence on these numerical results, implying a substantial change
in the behavior of the $\rho^+$ mass for the $\ell = -1$ Landau mode.

\begin{figure}[t]
\centering{}\includegraphics[width=0.7\textwidth]{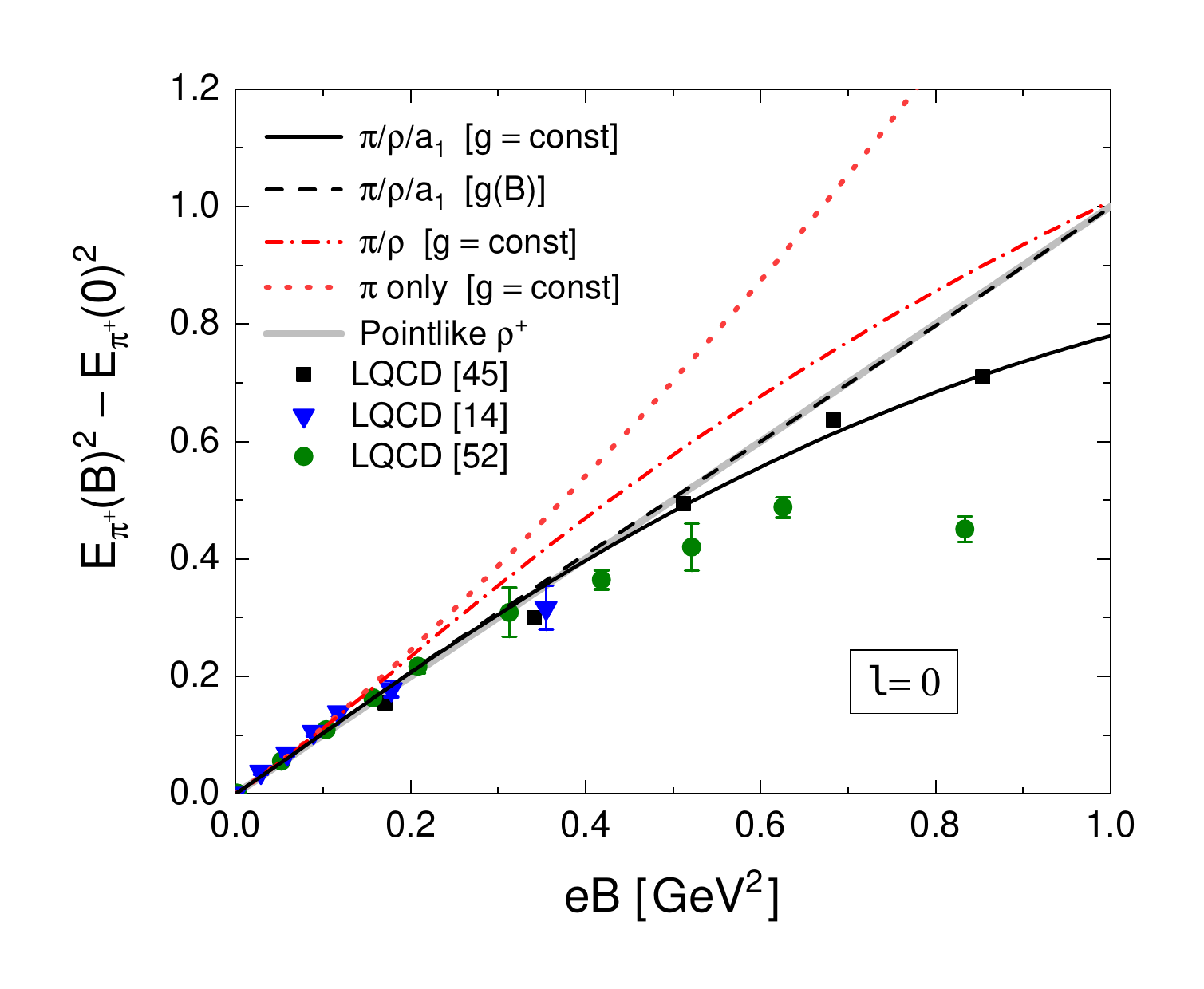}
\vspace*{-5mm}
\caption{Squared energy of the $\pi^+$ mass eigenstate for the Landau mode
$\ell =0$ and vanishing component of the momentum in the direction of $\vec
B$. Values are given with respect to the squared $\pi^+$ mass for vanishing
external field. Black solid and dashed lines correspond to constant and
$B$-dependent couplings, respectively. Red dotted and dash-dotted lines show
results from models that do not include axial vector mesons, while the light
gray line corresponds to a pointlike $\pi^+$. For comparison, lattice QCD
data quoted in Refs.~\cite{Bali:2011qj,Andreichikov:2016ayj,Ding:2020hxw}
are also included.}
\label{fig:chargedpi}
\end{figure}

In the case of the mode $\ell=0$, as discussed in Sec.~\ref{chargedmes}, the
lowest mass state $\pi^+$ is given in general by a mixing between the states
that we have denoted as $\pi$, $\rho_2$, $a_L$ and $a_1$. The corresponding
pole mass $m_{\pi^+}$ can be obtained from Eq.~(\ref{pimasseq}), while the
lowest energy for this state is given by $E_{\pi^+} =
\sqrt{m_{\pi^+}^2+B_e}$, see Eq.~(\ref{pimasenergy}). Our numerical results
are presented in Fig.~\ref{fig:chargedpi}, where, for the sake of comparison
with LQCD values, we plot the values of the difference
$E_{\pi^+}(B)^2-E_{\pi^+}(0)^2$ as a function of $eB$. Once again, black
solid and dashed lines correspond to the cases of constant and $B$-dependent
couplings, respectively. We also include for comparison the results obtained
from similar NJL-like models that just include the pseudoscalar meson sector
(red dotted line), or just include the mixing between the pseudoscalar and
vector meson sectors (red dash-dotted line), neglecting the effect of the
presence of axial vector mesons. It can be seen that the inclusion of the
axial vector meson sector leads to an improvement of the agreement with LQCD
data quoted in Refs.~\cite{Bali:2011qj,Andreichikov:2016ayj,Ding:2020hxw},
also shown in the figure.

It is interesting to point out that, for large external magnetic fields, the
values from LQCD shown in Fig.~\ref{fig:chargedpi} lie well below the curve
that corresponds to a pointlike pion. From Eq.~(\ref{pimasenergy}), it is
easy to see that to reproduce these results one should get a negative value
of the pole mass squared, $m_{\pi^+}^2<0$. In fact, this is what we obtain
from our NJL-like model if we assume that the coupling constants do not
depend on $B$ (solid line in the figure). The appearance of an imaginary
pole mass does not signal the existence of a meson condensation, since meson
energies are still positive quantities; indeed, the presence of the magnetic
field generates a zero-point motion in the plane perpendicular to $\vec B$
that induces an ``effective magnetic mass'' $\sqrt{m_{\pi^+}^2+B_e}$. Notice
that in this case some analytical expressions have to be revised. The
corresponding changes, basically related with the normalization of
polarization vectors, are indicated in App.~\ref{pol_vec}. In contrast, for
$B$-dependent couplings one does not observe a large variation of the
$\pi^+$ pole mass for the studied range of $eB$; the energy is essentially
dominated by the magnetic field. Thus, the curve shown in
Fig.~\ref{fig:chargedpi} (black dashed line) turns out to be approximately
coincident with the one corresponding to a pointlike charged pion.  We
remark that our numerical results indicate a monotonic enhancement of the
charged pion energy with the magnetic field, in contrast with the nonmonotic
behavior found in some recent LQCD simulations (green circles in the
figure)~\cite{Ding:2020hxw}. It would be interesting to get more insight on
this open issue from other effective models and further LQCD calculations.

To conclude this section, let us discuss the state composition of the
charged pion mass state. In Table~\ref{tabb} we quote our results for the
coefficients of the linear combination in Eq.~(\ref{physmesch}) for some
values of $eB$, considering both the cases $g=\mbox{constant}$ and $g(B)$
(upper and lower parts of the table, respectively). We also include the
values of the normalized squared $\pi^+$ pole masses. For $B=0$, as well
known, in these type of model the pion mass eigenstate is obtained from a
mixing between the pseudoscalar state $\pi$ and the longitudinal part of the
axial-vector state ($a_L$, in our notation). Then, for nonzero $B$, the
mixing between the states $\rho_2$ and $a_1$ is also turned on. As stated,
for $g=\mbox{constant}$ the value of $m_{\pi^+}^2$ becomes negative if the
magnetic field is increased; this occurs at $eB\simeq 0.5$~GeV$^2$. As shown
in the upper part of Table~\ref{tabb}, when approaching this point the mass
eigenstate turns out to be strongly dominated by the axial vector states
$a_1$ and $a_L$, which have similar weights. For larger values of $eB$ the
absolute value of $m_{\pi^+}^2$ gets increased, and once again the $\pi^+$
state becomes dominated by the pseudoscalar $\pi$ contribution. Notice,
however, that for $eB=1$~GeV the contributions of other states are
nonnegligible; moreover, it is seen that the coefficients $c^{\pi^+}_{a_1}$
and $c^{\pi^+}_{a_L}$ become imaginary. On the contrary, as shown in the
lower part of the table, for $g=g(B)$ these effects are not observed in the
studied range of values of the external field. As mentioned above, in this
case the $\pi^+$ pole mass does not show qualitative changes with $eB$; the
main effect of the magnetic field is the enhancement of axial vector
components, each of them reaching about $1/4$ of the state composition at
$eB=1$~GeV$^2$, while  the remaining 1/2 fraction is almost saturated by the
$\pi$ component. As stated, recent LQCD data support a negative value
of $m_{\pi^+}^2$ for large magnetic fields. It would be also interesting to
get information from lattice calculations on the state composition, in
particular, in the region $eB\sim 1$~GeV$^2$.

\begin{table}[h]
\begin{center}
\begin{tabular}{cc ccccc}
\hline \hline
 $eB$~[GeV$^2$] & \hspace*{5mm} $m_{\pi^+}(B)^2/m_{\pi^+}(0)^2$ \hspace*{5mm} & \multicolumn{4}{c}{State composition ($g=\mbox{constant}$)}
\\
 & & & \hspace*{5mm} $c^{\pi^+}_{\pi}\ $ \hspace*{5mm}&
      \hspace*{5mm} $c^{\pi^+}_{\rho_2}\ $\hspace*{5mm} &
     \hspace*{5mm} $c^{\pi^+}_{a_1}\ $ \hspace*{5mm} &
     \hspace*{5mm} $c^{\pi^+}_{a_L}\ $ \hspace*{5mm}
\\
\hline
 0    & 1  & &    0.998        &  0          &    0      & $-0.067$
\\
 0.5  & 0.006 & &    0.174        &  $-0.025$     &  0.697    & 0.697
\\
 1.0  & -10.29 & &    0.879        &  $-0.210$  & $-0.201\,i$   & $-0.378\,i$
\\
\hline \hline
&& && && \\ \hline \hline
 $eB$~[GeV$^2$] & \hspace*{5mm} $m_{\pi^+}(B)^2/m_{\pi^+}(0)^2$ \hspace*{5mm} & \multicolumn{4}{c}{State composition ($g=g(B)$)}
\\
 & & & \hspace*{5mm} $c^{\pi^+}_{\pi}\ $ \hspace*{5mm}&
      \hspace*{5mm} $c^{\pi^+}_{\rho_2}\ $\hspace*{5mm} &
     \hspace*{5mm} $c^{\pi^+}_{a_1}\ $ \hspace*{5mm} &
     \hspace*{5mm} $c^{\pi^+}_{a_L}\ $ \hspace*{5mm}
\\
\hline
 0    & 1  & &    0.998        &  0          &    0      & $-0.067$
\\
 0.5  & 1.18  & &    0.924        &  $-0.137$   &  0.287    & 0.214
\\
 1.0  & 0.95  & &    0.651        &  $-0.168$   &  0.545    & 0.501
\\
\hline \hline
\end{tabular}
\caption{Normalized squared pole mass and composition of the $\pi^+$ meson mass
eigenstate for selected values of $eB$.}
\label{tabb}
\end{center}
\end{table}

\section{Summary \& Conclusions}
\label{sec:conclu}

In this work we have studied the mass spectrum of light pseudoscalar and
vector mesons in the presence of an external uniform and static magnetic
field $\vec B$, introducing the effects of the mixing with the axial vector
meson sector. The study has been performed in the framework of a two-flavor
NJL-like model that includes isoscalar and isovector couplings in the
scalar-pseudoscalar and vector-axial vector sector, as well as a flavor
mixing term in the scalar-pseudoscalar sector. For simplicity, the coupling
constants of the vector and axial vector sector have been taken to be equal.
The ultraviolet divergences associated to the nonrenormalizability of the
model have been regularized using the magnetic field independent
regularization method, which has been shown to be free from unphysical
oscillations and to reduce the dependence of the results on the model
parameters~\cite{Avancini:2019wed}. Additionally, we have explored the
possibility of using magnetic field dependent coupling constants $g(B)$ to
account for the effect of the magnetic field on sea quarks.

As well known, for vanishing external field pseudoscalar mesons mix with
``longitudinal'' axial vector mesons. Now, the presence of an external
uniform magnetic field breaks isospin (due to the different quark electric
charges) and full rotational symmetry, allowing for a more complex meson
mixing pattern than in vacuum. The mixing structure is constrained by the
remaining unbroken symmetries, in such a way that the mass matrices
---written in a basis of polarization states--- can be separated into
several ``boxes''.

In the case of neutral mesons, Schwinger phases cancel and the polarization
functions become diagonal in the usual momentum basis. Since mesons can be
taken at rest, rotational invariance around $\hat{B}$ implies that $S_z$
(the spin in the field direction) is a good quantum number to characterize
these states. The aforementioned symmetries restrict the allowed mixing in
the original $20 \times 20$ mass matrix, which can be decomposed as a direct
sum of subspaces of states with $s_z=-1$, 0, and 1. For $s_z=\pm 1$ (spin
parallel to $\vec B$), it is seen that vector mesons do not mix with other
sectors, and the mass eigenstates are those of the flavor basis
($\rho_u,\rho_d$). We have shown that the corresponding masses increase with
$B$ in qualitative agreement with LQCD, within uncertainties. For $s_z=0$
(spin perpendicular to $\vec B$), scalar mesons turn out to get decoupled
from other states and therefore have been disregarded in our analysis.
Meanwhile, pseudoscalar mesons mix with vector and axial vector mesons whose
polarization states are parallel to $\vec B$. The four lowest mass states of
this sector are to be identified with the physical states $\tilde{\pi}^0$,
$\tilde{\eta}$, $\tilde\rho^{\,0}$ and $\tilde{\omega}$. Regarding
$m_{\tilde\rho^{0}}$ and $m_{\tilde{\omega}}$, we have found that they get
increased with the magnetic field, in such a way that they overcome a
$q\bar{q}$ decay threshold ---which arises from the lack of confinement of
the model--- at relatively low values of $eB$. Concerning
$m_{\tilde{\eta}}$, a slight decrease with $B$ is observed.

The impact of the inclusion of the axial vector meson sector on the mass of
the lightest state $\tilde\pi^0$, identified with the neutral pion, is
actually one of the main focus of our work. We have found that when axial
vector mesons are taken into account, $m_{\tilde\pi^0}$ displays a monotonic
decreasing behavior with $B$ in the studied range $eB<1$~GeV$^2$, which is
in good qualitative agreement with LQCD calculations for both $g={\rm
constant}$ and $g(B)$. Thus, our current results represent an improvement
over previous analyses that take into account just the mixing with the
vector meson sector, or no mixing at all. When no mixing is considered, the
behavior of $m_{\tilde\pi^0}$ with $B$ is non monotonic when $g$ is kept
constant, deviating just slightly from its value at $B=0$.  Only when $g$ is
allow to depend on the magnetic field one obtains a decreasing behavior
which resembles LQCD results. Even though the inclusion of the vector sector
leads to a reduction in $m_{\tilde\pi^0}$ together with a consistent
decreasing trend, the values lie quite below LQCD predictions, for both $g$
and $g(B)$. We therefore conclude that the inclusion of axial mesons is
important since it leads to more robust results for the neutral pion mass,
even independently of the assumption of a magnetic field dependent coupling
constant. Regarding the composition of the $\tilde{\pi}^0$ state, we have
found that it is largely dominated by the isovector component $\pi_3$
($|c^{\tilde\pi^0}_{\pi_3}|^2 \gtrsim 0.97$) for the studied range of values
of $eB$. In terms of flavor composition, a larger weight is gained by
$u$-flavor components for large values of $B$, which can be understood from
the fact that the $u$ quark couples more strongly to the magnetic field.

In the case of charged mesons, the corresponding polarization functions are
diagonalized by expanding meson fields in appropriate Ritus-like bases, so
as to account for the effect of nonvanishing Schwinger phases. Once again,
the symmetries of the system constrain the allowed mixing matrices, which
also depend on the value of the meson Landau level $\ell$. For $\ell=-1$ one
has only one vector and one axial vector polarization states. Moreover, they
do not mix with any other particle state. Thus, for $\ell=-1$ the effect of
the inclusion of axial vector mesons on the $\rho^+$ mass comes solely from
the model parametrization, which is affected by the presence of $\pi$-a$_1$
mixing at $B=0$. Our results show that when the axial vector sector is
included, the energy $E_{\rho^+}=\sqrt{m_{\rho^+}^2-B_e}$ of this state
undergoes a considerable reduction, leading to a decreasing behavior which
is in qualitative agreement with LQCD predictions, independently of the
assumption of a $B$-dependent coupling constant. However ---in
accordance to LQCD calculations and with our previous results within
NJL-like models that do not include axial
vectors~\cite{Carlomagno:2022arc,GomezDumm:2023owj}--- we find that
$E_{\rho^+}$ does not vanish for any considered value of the magnetic field,
a fact that can be relevant in connection with the occurrence of $\rho^+$
meson condensation for strong magnetic fields.

For $\ell=0$ only three polarization vectors are linearly independent, and
the pion mixing subspace is given by $\pi^+\mhyphen\rho^+\mhyphen {\rm
a}_1^+$ for only certain polarizations states. The lowest mass state in this
sector can be identified with the $\pi^+$, whose lowest energy is given by
$E_{\pi^+}=\sqrt{m_{\pi^+}^2+B_e}$. Our results show that, even though
vector mixing already induces a softening in the enhancement of the pion
energy with $B$, the inclusion of the axial vector meson sector reinforces
this softening, leading to an improved agreement with LQCD predictions.
Remarkably, for a constant coupling $g$ and magnetic fields stronger than
$eB=0.4$~GeV$^2$, we obtain values of the pion energy which lie well below
the ones correspoding to a pointlike pion, in concordance with LQCD results
in Refs.~\cite{Andreichikov:2016ayj,Ding:2020hxw}. On the other hand, in the case of
a $B$-dependent coupling we find that the pole mass becomes approximately 
constant; as a result, the energy is basically coincident
with the one corresponding to a pointlike charged pion. As for the $\pi^+$
state composition, we have seen that in general the magnetic field induces a
mixing between all states by increasing the contribution from vector and
axial vector components.

In view of the above results, one can conclude that the inclusion of axial
vector mesons leads to more robust results and improves the agreement between
NJL-like models and LQCD calculations. Still, issues about meson masses and
mass eigenstate compositions at large magnetic fields are still open, and
further results from LQCD and effective models of strong interactions would
be welcome.

\begin{acknowledgments}
NNS would like to thank the Department of Theoretical Physics of the
University of Valencia, where part of this work has been carried out, for
their hospitality within the Visiting Professor program of the University of
Valencia. This work has been partially funded by CONICET (Argentina) under
Grant No. PIP 2022-2024 GI-11220210100150CO, by ANPCyT (Argentina) under
Grant No.~PICT20-01847, by the National University of La Plata (Argentina),
Project No.~X824, by Ministerio de Ciencia e Innovaci\'on and Agencia
Estatal de Investigaci\'on (Spain), and European Regional Development Fund
Grant No.~PID2019-105439GB-C21, by EU Horizon 2020 Grant No.~824093
(STRONG-2020), and by Conselleria de Innovaci\'on, Universidades, Ciencia y
Sociedad Digital, Generalitat Valenciana, GVA PROMETEO/2021/083.
\end{acknowledgments}

\section*{Appendices}

\setcounter{section}{0}
\global\long\def\thesection{\Alph{section}}%
\global\long\def\theequation{\thesection.\arabic{equation}}%
\setcounter{equation}{0}
\global\long\def\thesubsection{\thesection.\arabic{subsection}}%

\section{Conventions and notation}

\label{conventions} \setcounter{equation}{0}

Throughout this section we use the Minkowski metric
$\eta^{\mu\nu}=\mbox{diag}(1,-1,-1,-1)$, while for a space-time coordinate
four-vector $x^{\mu}$ we adopt the notation $x^{\mu}=(t,\vec{x})$, with
$\vec{x}=(x^{1},x^{2},x^{3})$.

We study interactions between charged particles and an external
electromagnetic field ${\cal A}^{\mu}(x)$. The electromagnetic field
strength $F^{\mu\nu}$ and its dual ${{\tilde{F}}^{\mu\nu}}$ are given by
\begin{eqnarray}
F^{\mu\nu}=\partial^{\mu}\mathcal{A}^{\nu}-\partial^{\nu}\mathcal{A}^{\mu}\ ,
\qquad\qquad
{{\tilde{F}}^{\mu\nu}}=\frac{1}{2}\,\epsilon^{\mu\nu\alpha\beta}F_{\alpha\beta}\ ,
\end{eqnarray}
where the convention $\epsilon^{0123}=+1$ is used. We consider in particular
the situation in which one has a static and uniform magnetic field
$\vec{B}$; without losing generality, we choose the axis 3 to be parallel to
$\vec B$, i.e., we take $\vec B = (0,0,B)$ (note that $B$ can be either
positive or negative). Moreover, defining
\begin{eqnarray}
\hat{F}^{\mu\nu}=\frac{1}{B}\,F^{\mu\nu}\ ,\qquad\qquad
{\hat{\tilde{F}}^{\mu\nu}}=\frac{1}{B}\,{\tilde F}^{\mu\nu}
\end{eqnarray}
for $i,j=1,2,3$ one has
\begin{eqnarray}
\hat{F}^{0\nu}=0\ , & & \qquad\hat{F}^{ij}=-\epsilon_{ij3}\ , \nonumber \\
{\hat{\tilde{F}}^{ij}}=0\ , & & \qquad{\hat{\tilde{F}}^{0k}}=-\,\epsilon^{0k ij}\epsilon_{ij3}/2\ ,
\end{eqnarray}
i.e.\ the relevant components of the tensors are
$\hat{F}^{12}=-\hat{F}^{21}=-1$,
${\hat{\tilde{F}}^{03}}=-{\hat{\tilde{F}}^{30}}=-1$.

Since isotropy is broken by the particular direction of the external field
$\vec B$, it is convenient to separate the metric tensor into ``parallel''
and ``perpendicular'' pieces,
\begin{equation}
\eta_{\parallel}^{\mu\nu}=\mbox{diag}(1,0,0,-1)\ ,\qquad\qquad
\eta_{\perp}^{\mu\nu}=\mbox{diag}(0,-1,-1,0)\ .
\end{equation}
In addition, given a four-vector $v^\mu$, it is useful to define
``parallel'' and ``perpendicular'' vectors
\begin{equation}
v_\parallel^\mu = (v^0,0,0,v^3)\ ,\qquad\qquad
v_\perp^\mu = (0,v^1,v^2,0)\ .
\end{equation}

\section{Neutral meson polarization functions}
\label{polneutral}

\setcounter{equation}{0}

According to Eq.~(\ref{jotas2}), the polarization functions for neutral
mesons can be written as a sum of flavor-dependent functions
$\Sigma_{MM'}^{f}(q)$. The latter, in turn, can be written in terms of a set
of Lorentz covariant tensors as
\begin{eqnarray}
\Sigma_{MM'}^{f}(q)=\sum_{i=1,n_{mm'}}c_{mm',i}^{f}(q_{\perp}^{2},q_{\parallel}^{2})
\ \mathbb{O}_{MM'}^{(i)}(q)\ .
\label{suma-o}
\end{eqnarray}
Here, $M=\pi_b,\rho_b^{\mu},a_b^{\mu}$ correspond to $m=\pi,\rho,a$, and the
same is understood for $M'$ and $m'$. The coefficients $c_{mm',i}^{f}$ are
scalar functions, while the tensors $\mathbb{O}_{MM'}^{(i)}$ carry the
corresponding Lorentz structures. Notice that the number of terms in the
sum, $n_{mm'}$, depends on the combination $mm'$ considered. The scalar
coefficients can be expressed as
\begin{eqnarray}
c_{mm',i}^{f}(q_{\perp}^{2},q_{\parallel}^{2}) \ =
\ \frac{N_{c}}{8\pi^{2}}\int_{0}^{\infty}dz\,\int_{-1}^{1}d\xi\
e^{-z\,\phi_{0}^f(q_{\perp},q_{\parallel},\xi,z)}\;
\gamma_{mm',i}^{f}(q_{\perp}^{2},q_{\parallel}^{2},\xi,z)\ ,
\end{eqnarray}
where
\begin{equation}
\phi_{0}^f(q_{\perp},q_{\parallel},\xi,z) \ = \ M_{f}^{2}-\frac{1-\xi^{2}}{4}\;q_{\parallel}^{2}
+\frac{\cosh(z\,B_{f})-
\cosh(\xi\,z\,B_{f})}{2\,z\,B_{f}\,\sinh(z\,B_{f})}\;\vec q_{\perp}^{\;2}\ ,
\end{equation}
with $B_f = |BQ_f|$. In the following we list the sums associated to each
polarization function, together with the explicit expressions of the
functions $\gamma_{mm',i}^{f}(q_{\perp}^{2},q_{\parallel}^{2},\xi,z)$
corresponding to the coefficients
$c_{mm',i}^{f}(q_{\perp}^{2},q_{\parallel}^{2})$. For brevity, the arguments
of $c_{mm',i}^{f}$ and $\gamma_{mm',i}^{f}$ are not explicitly written.

The $\pi\pi$ polarization function is a scalar, therefore there is only one
coefficient $c_{\pi\pi,1}^{f}$, and $\mathbb{O}_{\pi_b \pi_{b'}}^{(1)}(q)=1$.
One has
\begin{eqnarray}
\Sigma_{\pi_{b}\pi_{b'}}^{f}(q)\ = \ c_{\pi\pi,1}^{f}\ ,
\end{eqnarray}
while the associated function $\gamma_{\pi\pi,1}^{f}$ is given by
\begin{eqnarray}
\gamma_{\pi\pi,1}^{f} & = &
\left(M_{f}^{2}+\frac{1}{z}+\frac{1-\xi^{2}}{4}\;q_{\parallel}^{2}\right)
\,\frac{B_{f}}{\tanh(z\,B_{f})} \nonumber \\
& & +\,\frac{B_{f}^{2}}{\sinh^{2}(z\,B_{f})}
\left[1-\frac{\cosh(z\,B_{f})-\cosh(\xi\,z\,B_{f})}{2B_{f}\sinh(z\,B_{f})}\;
{\vec q}_{\perp}^{\;2}\right]\ .
\end{eqnarray}
Analogously, for the $\sigma\sigma$ polarization function we have
\begin{eqnarray}
\Sigma_{\sigma_{b}\sigma_{b'}}^{f}(q) \ = \ c_{\sigma\sigma,1}^{f}\ ,
\end{eqnarray}
while
\begin{eqnarray}
\gamma_{\sigma\sigma,1}^{f} & = &
\left(-M_{f}^{2}+\frac{1}{z}+\frac{1-\xi^{2}}{4}\;q_{\parallel}^{2}\right)\,
\frac{B_{f}}{\tanh(z\,B_{f})}\nonumber \\
& & +\,\frac{B_{f}^{2}}{\sinh^{2}(z\,B_{f})}\left[1-\frac{\cosh(z\,B_{f})-\cosh(\xi\,z\,B_{f})}{2B_{f}\sinh[z\,B_{f}]}
\;{\vec q}_{\perp}^{\;2}\right]\ .
\end{eqnarray}

For the $\rho\rho$ polarization the sum in Eq.~(\ref{suma-o}) includes five
terms. We find
\begin{eqnarray}
\Sigma_{\rho_{b}^{\mu}\rho_{b'}^{\nu}}^{f\;\mu\nu}(q) \ = \
c_{\rho\rho,1}^{f}\,\eta_{\parallel}^{\mu\nu}+c_{\rho\rho,2}^{f}\,\eta_{\perp}^{\mu\nu}+
c_{\rho\rho,3}^{f}\,q_{\parallel}^{\mu}\,q_{\parallel}^{\nu}+
c_{\rho\rho,4}^{f}\,q_{\perp}^{\mu}\,q_{\perp}^{\nu}+
c_{\rho\rho,5}^{f}\,\big(q_{\perp}^{\mu}\,q_{\parallel}^{\nu}+q_{\parallel}^{\mu}\,q_{\perp}^{\nu}\big)\ ,
\end{eqnarray}
while the functions $\gamma_{\rho\rho,i}^{f}$ read
\begin{eqnarray}
\gamma_{\rho\rho,1}^{f} & = & -\left(M_{f}^{2}+\frac{1-\xi^{2}}{4}\;q_{\parallel}^{2}\right)
 \frac{B_{f}}{\tanh(z\,B_{f})}-\frac{B_{f}^{2}}{\sinh^{2}(z\,B_{f})}\nonumber \\
& & + \,\frac{B_f\big[\cosh(z\,B_{f})-\cosh(\xi\,z\,B_{f})\big]}
 {2\,\sinh^{3}[z\,B_{f}]}\;{\vec q}_{\perp}^{\;2}\ ,\nonumber \\
\gamma_{\rho\rho,2}^{f} & = & -\left(M_{f}^{2}+\frac{1}{z}+\frac{1-\xi^{2}}{4}\;q_{\parallel}^{2}\right)
 \frac{B_{f}\cosh(\xi\,z\,B_{f})}{\sinh(z\,B_{f})}\nonumber \\
 & & + \,\frac{B_f\big[\cosh(z\,B_{f})-
 \cosh(\xi\,z\,B_{f})\big]}{2\,\sinh^{3}[z\,B_{f}]}\;{\vec q}_{\perp}^{\;2}\ ,\nonumber \\
\gamma_{\rho\rho,3}^{f} & = & (1-\xi^{2})\,\frac{B_{f}}{2\,\tanh[z\,B_{f}]}\ ,\nonumber \\
\gamma_{\rho\rho,4}^{f} & = & B_{f}\,\frac{\cosh(z\,B_{f})-\cosh(\xi\,z\,B_{f})}{\sinh^{3}(z\,B_{f})}\ ,\nonumber \\
\gamma_{\rho\rho,5}^{f} & = & B_{f}\,
\frac{\cosh(\xi\,z\,B_{f})-\xi\coth(z\,B_{f})\sinh(\xi\,z\,B_{f})}{2\,\sinh(z\,B_{f})}\ .
\end{eqnarray}

For the $aa$ polarization function we get
\begin{equation}
\Sigma_{a_{b}^{\mu}a_{b'}^{\nu}}^{f\;\mu\nu}(q)\ = \
c_{aa,1}^{f}\,\eta_{\parallel}^{\mu\nu}+c_{aa,2}^{f}\,\eta_{\perp}^{\mu\nu}+
c_{aa,3}^{f}\,q_{\parallel}^{\mu}\,q_{\parallel}^{\nu}+c_{aa,4}^{f}\,q_{\perp}^{\mu}\,q_{\perp}^{\nu}+
c_{aa,5}^{f}\,\left(q_{\perp}^{\mu}\,q_{\parallel}^{\nu}+q_{\parallel}^{\mu}\,q_{\perp}^{\nu}\right)\ ,
\end{equation}
while the functions $\gamma_{aa,i}^{f}$ are given by
\begin{eqnarray}
\gamma_{aa,1}^{f} & = & -\left(- M_{f}^{2}+\frac{1-\xi^{2}}{4}\;q_{\parallel}^{2}\right)
 \frac{B_{f}}{\tanh(z\,B_{f})}-\frac{B_{f}^{2}}{\sinh^{2}(z\,B_{f})}\nonumber \\
& & + \,\frac{B_f\big[\cosh(z\,B_{f})-\cosh(\xi\,z\,B_{f})\big]}
 {2\,\sinh^{3}[z\,B_{f}]}\;{\vec q}_{\perp}^{\;2}\ ,\nonumber \\
\gamma_{aa,2}^{f} & = & -\left(- M_{f}^{2}+\frac{1}{z}+\frac{1-\xi^{2}}{4}\;q_{\parallel}^{2}\right)
 \frac{B_{f}\cosh(\xi\,z\,B_{f})}{\sinh(z\,B_{f})}\nonumber\\
& & + \,\frac{B_f\big[\cosh(z\,B_{f})-
 \cosh(\xi\,z\,B_{f})\big]}{2\,\sinh^{3}[z\,B_{f}]}\;{\vec q}_{\perp}^{\;2}\ ,\nonumber \\
\gamma_{aa,i}^{f} & = & \gamma_{\rho\rho,i}^{f}\quad\mbox{for}\quad i=3,4,5\ .
\end{eqnarray}

For the $\pi\rho$ and $\rho\pi$ polarization functions we get
\begin{equation}
\Sigma_{\pi_{b}\rho_{b'}^{\mu}}^{f\;\mu}(q) \ = \
\Sigma_{\rho_{b}^{\mu}\pi_{b'}}^{f\;\mu}(q)^{\,\ast} \ = \
c_{\pi\rho,1}^{f}\; \hat{\tilde{F}}^{\mu\alpha}\,q_{\parallel\alpha}
\end{equation}
and
\begin{equation}
\gamma_{\pi\rho,1}^{f} \ = \ -is_{f}\,B_{f}M_{f}\ ,
\end{equation}
with $s_f = {\rm sign}(B Q_f)$.

For the $\pi a$ and $a\pi$ polarization functions we get
\begin{equation}
\Sigma_{\pi_{b}a_{b'}^{\mu}}^{f\;\mu}(q) \ = \
\Sigma_{a_{b}^{\mu}\pi_{b'}}^{f\;\mu}(q)^{\,\ast} \ = \
c_{\pi a,1}^{f}\, q_{\parallel}^{\mu}+c_{\pi a,2}^{f}\, q_{\perp}^{\mu}\ ,
\end{equation}
and
\begin{eqnarray}
\gamma_{\pi a,1}^{f} & = & -iM_{f}\, \frac{B_{f}}{\tanh(z\,B_{f})} \ ,\nonumber \\
\gamma_{\pi a,2}^{f} & = & -iM_{f}\,
\frac{B_{f}\,\cosh(\xi\,z\,B_{f})}{\sinh(z\,B_{f})}\ .
\end{eqnarray}

For the $\sigma\rho$ and $\rho\sigma$ polarization functions we get
\begin{equation}
\Sigma_{\sigma_{b}\rho_{b'}^{\mu}}^{f\;\mu}(q) \ = \
\Sigma_{\rho_{b}^{\mu}\sigma_{b'}}^{f\;\mu}(q)^{\,\ast} \ = \
c_{\sigma\rho,1}^{f}\, \hat{F}^{\mu\alpha}\,q_{\perp\alpha}\ ,
\end{equation}
and
\begin{equation}
\gamma_{\sigma\rho,1}^{f} \ = \
is_{f}\,B_{f}M_{f}\,\frac{\cosh(z\,B_{f})\,\cosh(v\,z\,B_{f})-1}{\sinh^{2}(z\,B_{f})}\ .
\end{equation}

Finally, for the $a\rho$ and $\rho a$ polarization functions we have
\begin{eqnarray}
 \Sigma_{\rho_b^\mu a_{b'}^\nu}^{f\;\mu\nu}(q) & = &
 \Sigma_{a_{b}^{\nu}\rho_{b'}^{\mu}}^{f\;\nu\mu}(q) \,
= \, c_{a\rho,1}^{f}\left(\hat{\tilde{F}}^{\mu\alpha}\,q_{\alpha\parallel}\,q_{\parallel}^{\nu}
- q_{\parallel}^{\mu}\,q_{\alpha\parallel}\,\hat{\tilde{F}}^{\alpha\nu}\right)\nonumber \\
 & & \qquad\qquad\quad +\, c_{a\rho,2}^{f}\left(\hat{\tilde{F}}^{\mu\alpha}\,q_{\alpha\parallel}\,q_{\perp}^{\nu}
 -q_{\perp}^{\mu}\,q_{\alpha\parallel}\,\hat{\tilde{F}}^{\alpha\nu}\right)
 +\, c_{a\rho,3}^{f}\,\hat{\tilde{F}}^{\mu\nu}\ ,
\end{eqnarray}
and
\begin{eqnarray}
\gamma_{a\rho,1}^{f} & = & \frac{s_{f}}{4}\ B_{f}\,(1-\xi^{2})\ ,\nonumber \\
\gamma_{a\rho,2}^{f} & = & -\,\frac{s_{f}}{2}\,B_{f}
\left[\frac{\xi\,\sinh(\xi\,z\,B_{f})}{\sinh(z\,B_{f})}+
\frac{1-\cosh(z\,B_{f})\cosh(\xi\,z\,B_{f})}{\sinh^{2}(z\,B_{f})}\right]\ ,\nonumber \\
\gamma_{a\rho,3}^{f} & = & -\,s_{f}\,B_{f}M_{f}^{2}\ .
\end{eqnarray}

\section{The ``B=0'' polarization functions}

\label{b0loops} \setcounter{equation}{0}

To perform the MFIR we need to obtain the meson ``$B=0$'' polarization
functions $J_{MM'}^0(q)$ in both their unregularized (unreg) and regularized
(reg) forms. As stated in Sec.~\ref{sect2a}, although these polarization
functions are calculated from the propagators in the $B\to 0$ limit, they
still depend implicitly on $B$ through the values of the magnetized dressed
quark masses $M_{f}$. Hence, they should not be confused with the
polarization functions that one would obtain in the case of vanishing
external field. Moreover, they will be in general different for neutral and
charged mesons. In the case of neutral mesons (i.e.,
$M,M'=\sigma_b,\pi_b,\rho_b^\mu,a_b^\mu$, with $b=0,3$) one can write
\begin{equation}
J_{MM'}^{0,\lambda}(q) \ = \
F_{MM'}^{uu,\lambda}(q)\,+\,\varepsilon_{M}\,\varepsilon_{M'}\,F_{MM'}^{dd,\lambda}(q)\ ,
\end{equation}
where $\lambda$ stands for ``reg'' or ``unreg'', and $\varepsilon_{M}$ is
equal to either 1 or $-1$ [see text below Eq.~(\ref{jotas})]. On the other
hand, for charged mesons ($M,M'=\sigma,\pi,\rho^{\mu},a^{\mu}$) one has
\begin{equation}
J_{MM'}^{0,\lambda}(q) = 2\,F_{MM'}^{ud,\lambda}(q)\ .
\end{equation}
The functions $F_{MM'}^{ff',\lambda}(q)$ can be written in terms of scalar
functions $b_{mm',i}^{ff',\lambda}(q^2)$, with $m,m'=\pi,\rho,a$, as
follows:
\begin{eqnarray}
F_{\pi\pi}^{ff',\lambda}(q) & = & b_{\pi\pi,1}^{ff',\lambda}(q^{2})\ ,\\[5mm]
F_{\rho^{\mu}\rho^{\nu}}^{ff',\lambda\;\mu\nu}(q) & = &
b_{\rho\rho,1}^{ff',\lambda}(q^{2})
\left(\eta^{\mu\nu}-\frac{q^{\mu}q^{\nu}}{q^{2}}\right)+\,
b_{\rho\rho,2}^{ff',\lambda}(q^{2})\; \frac{q^{\mu}q^{\nu}}{q^{2}}\ ,\\[5mm]
F_{a^{\mu}a^{\nu}}^{ff',\lambda\;\mu\nu}(q) & = & b_{aa,1}^{ff',\lambda}(q^{2})
\left(\eta^{\mu\nu}-\frac{q^{\mu}q^{\nu}}{q^{2}}\right)+\,
b_{aa,2}^{ff',\lambda}(q^{2})\; \frac{q^{\mu}q^{\nu}}{q^{2}}\ ,\\[5mm]
F_{\pi a^{\mu}}^{ff',\lambda\;\mu}(q) & = &
F_{a^{\mu}\pi}^{ff',\lambda\;\mu}(q)^{\,\ast}\,
= \, b_{\pi a,1}^{ff',\lambda}(q^{2})\,q^{\mu}\ .
\end{eqnarray}
For the unregularized functions we find
\begin{equation}
b_{mm',i}^{ff',\,{\rm unreg}}(q^{2}) \ = \
\frac{N_{c}}{8\pi^{2}}\int_{-1}^{1}dv\int_{0}^{\infty}\frac{dz}{z}\ e^{-z\,\phi^{ff'}\!\!(v,q^{2})}
\ \omega_{mm',i}^{ff'}(q^{2},v,z)\ ,
\end{equation}
where
\begin{eqnarray}
\phi^{ff'}\!\!(v,q^{2}) \ = \ \frac{1}{2}\,(M_{f}^{2}+M_{f'}^{2})\,-\,
\frac{v}{2}\,(M_{f}^{2}-M_{f'}^{2})\,-\,\frac{(1-v^{2})}{4}\,q^{2}
\label{phiffp}
\end{eqnarray}
and
\begin{eqnarray}
\omega_{\pi\pi,1}^{ff'} & = & M_{f}M_{f'}+\,\frac{2}{z}+\frac{1-v^{2}}{4}\;q^{2}\ , \nonumber \\
\omega_{\rho\rho,1}^{ff'} & = & - M_{f}M_{f'}-\,\frac{1}{z}-\frac{1-v^{2}}{4}\;q^{2}\ , \nonumber \\
\omega_{\rho\rho,2}^{ff'} & = & - M_{f}M_{f'}-\,\frac{1}{z}+\frac{1-v^{2}}{4}\;q^{2}\ , \nonumber \\ 
\omega_{aa,1}^{ff'} & = & M_{f}M_{f'}-\,\frac{1}{z}-\frac{1-v^{2}}{4}\;q^{2}\ , \nonumber \\
\omega_{aa,2}^{ff'} & = & M_{f}M_{f'}-\,\frac{1}{z}+\frac{1-v^{2}}{4}\;q^{2}\ , \nonumber \\ 
\omega_{\pi a,1}^{ff'} & = & -\,\frac{i}{2}\big[(1-v)M_{f}+(1+v)M_{f'}\big]\ .
\label{d0paureg}
\end{eqnarray}

To express the regularized functions $b_{mm',i}^{ff',\,{\rm reg}}(q^{2})$ it
is convenient to introduce the ultraviolet divergent integrals
\begin{eqnarray}
I_{1f} & = & 4i\int\frac{d^{4}p}{(2\pi)^{4}}\;\frac{1}{p^{2}-M_{f}^{2}\,+\,i\epsilon}\ ,\\
I_{2ff'}(q^{2}) & = &
2i\int\frac{d^{4}p}{(2\pi)^{4}}\;
\frac{1}{(p_{+}^{2}-M_{f}^{2}+i\epsilon)(p_{-}^{2}-M_{f'}^2+i\epsilon)}\ ,
\end{eqnarray}
where $p_{\pm}=p\pm q/2$. Now we can consider some regularization scheme to
obtain regularized integrals $I_{1f}^{\rm reg}$ and $I_{2ff'}^{\rm
reg}(q^{2})$. Using the definitions
\begin{equation}
\bar{M}=\frac{M_{f}+M_{f'}}{2}\ ,\qquad\qquad\Delta=M_{f}-M_{f}'\ ,
\end{equation}
and introducing the shorthand notation
\begin{eqnarray}
& & \bar{I}_{1}=\frac{I_{1f}^{\rm reg}+I_{1f'}^{\rm reg}}{2}\ ,\qquad I_{2}=I_{2ff'}^{\rm reg}(q^{2})\ ,\nonumber \\
& & I_{2}^{0}=I_{2ff'}^{\rm reg}(0)\ ,\qquad
 I_{2}^{'0}=\frac{dI_{2ff'}^{\rm reg}(q^{2})}{dq^{2}}\Big|_{q^{2}=0}\ ,
\end{eqnarray}
we obtain
\begin{eqnarray}
b_{\pi\pi,1}^{ff',\,{\rm reg}} & = & N_{c}\left[\bar{I}_{1}-(q^{2}-\Delta^{2})I_{2}\right]\ ,\nonumber\\ [2mm] 
b_{\rho\rho,1}^{ff',\,{\rm reg}} & = &
\frac{N_{c}}{3}\left[\left(4\bar{M}^{2}+\Delta^{2}-\frac{4\bar{M}^{2}\Delta^{2}}{q^{2}}\right)(I_{2}-I_{2}^{0})-
(3\Delta^{2}-2q^{2})I_{2}+16\bar{M}^{2}\Delta^{2}I_{2}^{'0}\right]\ ,\nonumber\\ [2mm] 
b_{\rho\rho,2}^{ff',\,{\rm reg}} & = & -N_{c}\ \Delta^{2}\left[I_{2}-\frac{4\bar{M}^{2}}{q^{2}}\left(I_{2}-I_{2}^{0}\right)\right]\ ,\nonumber\\ [2mm] 
b_{aa,1}^{ff',\,{\rm reg}} & = & \frac{N_{c}}{3}
\left[\left(4\bar{M}^{2}+\Delta^{2}-\frac{4\bar{M}^{2}\Delta^{2}}{q^{2}}\right)(I_{2}-I_{2}^{0})
-(12\bar{M}^{2}-2q^{2})I_{2}+16\bar{M}^{2}\Delta^{2}I_{2}^{'0}\right]\ ,\nonumber\\ [2mm] 
b_{aa,2}^{ff',\,{\rm reg}} & = & -4N_{c}\ \bar{M}^{2}
\left[I_{2}-\frac{\Delta^{2}}{q^{2}}\left(I_{2}-I_{2}^{0}\right)\right]\ ,\nonumber\\ [2mm] 
b_{\pi a,1}^{ff',\,{\rm reg}} & = & 2i\ N_{c}
\ \bar{M}\left[I_{2}-\frac{\Delta^{2}}{q^{2}}\left(I_{2}-I_{2}^{0}\right)\right]\ . 
\end{eqnarray}

To regularize the vacuum loop integrals $I_{1f}$ and $I_{2ff'}(q^2)$ we use
the proper time scheme. We get in this way
\begin{eqnarray}
I_{1f}^{\rm reg} & = & \frac{\Lambda^2}{4\pi^2}\,E_2(M_f^2/\Lambda^2)\ , \label{I1freg} \\
I_{2ff'}^{\rm reg}(q^{2}) & = & -\frac{1}{16\pi^2}\int_{-1}^1 dv\
E_1(\phi^{ff'}\!\!(v,q^2)/\Lambda^2)\ ,
\end{eqnarray}
where $E_n(x) = \int_1^\infty dt\,t^{-n}\,\exp(-tx)$ is the exponential
integral function. The regularization requires the introduction of a
dimensionful parameter $\Lambda$, which plays the role of an ultraviolet
cutoff.

\section{Useful relations}

\label{useful} \setcounter{equation}{0}

We quote here a few relations that are found to be useful in order to obtain
the neutral meson polarization functions, see Sec.~\ref{neutralmesons}.
These are~\cite{gradshteyn}
\begin{eqnarray}
\int_{-1}^{1}\ d\xi\ (1-\xi^{2})\ e^{z(1-\xi^{2})/4} & = &
\frac{4}{z}+\left(1-\frac{2}{z}\right)\int_{-1}^{1}\ d\xi\ e^{z(1-\xi^{2})/4}\ ,\\[2mm]
\!\!\!\int_{0}^{\infty}\ \frac{dz}{z}\ e^{-\beta z}\left[\frac{\cosh[\xi z]}{\sinh[z]}-\frac{1}{z}\right]
 & = & \beta\left(1-\ln\frac{\beta}{2}\right)-\ln2\pi+\sum_{s=\pm1}\ln\Gamma\left(\frac{\beta+s\,\xi+1}{2}\right)\ ,\\[2mm]
\!\!\!\int_{0}^{\infty}\ dz\ e^{-\beta z}\left[\frac{\cosh[\xi z]}{\sinh[z]}-\frac{1}{z}\right] & = &
\ln\frac{\beta}{2}-\frac{1}{2}\sum_{s=\pm1}\psi\left(\frac{\beta+s\,\xi+1}{2}\right)\ ,
\label{muuno}
\end{eqnarray}
with Re$\,\beta> 0$. For $\xi=1$, the last relation leads to
\begin{equation}
\int_{0}^{\infty}\ dz\ e^{-\beta z}\left[\coth z-\frac{1}{z}\right] \ = \
\ln\frac{\beta}{2}-\frac{1}{\beta}-\psi\Big(\frac{\beta}{2}\Big)\ .
\label{fco}
\end{equation}

\section{Polarization vectors}

\label{pol_vec} \setcounter{equation}{0}

\subsection{Neutral mesons}

For arbitrary three-momentum $\vec{q}$, a convenient choice for the
polarization vectors of neutral mesons is
\begin{eqnarray}
\epsilon^{\mu}(\vec{q},1) & = & \frac{1}{\sqrt{2}\,m^{(0)}_{\perp}\,m^{(0)}_{2\perp}}
\left[\,q_{+}(E,0,0,q^{3})+{m^{(0)}_{\perp}}^{2}\left(0,1,i,0\right)\right]\nonumber \\
\epsilon^{\mu}(\vec{q},2) & = & \frac{1}{m^{(0)}_{\perp}}\left(q^{3},0,0,E\right)\nonumber \\
\epsilon^{\mu}(\vec{q},3) & = & \frac{1}{\sqrt{2}\,m\,m^{(0)}_{2\perp}}
\left[\,q_{-}(E,0,0,q^{3})+\frac{q_{+}^\ast q_{-}}{2}\left(0,1,i,0\right)+
{m^{(0)}_{2\perp}}^{2}\left(0,1,-i,0\right)\right]\ ,
\label{pol_vec_neut}
\end{eqnarray}
where $q_{\pm}=q^{1}\pm i\,q^{2}$, and we have used the definitions
\begin{eqnarray}
m^{(0)}_{\perp} = \sqrt{m^{2}+{\vec q}_\perp^{\;2}}\ ,\qquad\qquad
m^{(0)}_{2\perp} = \sqrt{m^{2}+{\vec q}_\perp^{\;2}/2}\ ,
\end{eqnarray}
with ${\vec q}_\perp^{\;2} = (q^{1})^{2}+(q^{2})^{2}$. One has in this case
\begin{equation}
E^2 \ = \ (q^3)^2 + {\vec q}_\perp^{\;2} + m^2\ .
\end{equation}
In addition, as stated in the main text, one can introduce a fourth
``longitudinal'' polarization vector
\begin{eqnarray}
\epsilon^{\mu}(\vec{q},L) & = & \frac{1}{m}\,(E,q^{1},q^{2},q^{3})\ .
\label{pol_vec_neutL}
\end{eqnarray}
These four polarization vectors satisfy
\begin{equation}
\epsilon^{\mu}(\vec{q},c)^\ast\,\epsilon_{\mu}(\vec{q},c')\ = \ \left\{
\begin{array}{cl}
\zeta_c & \qquad {\rm for} \ c=c' \\
0 & \qquad {\rm for}\ c\neq c'
\end{array}
\right.\ ,
\label{epsmodn}
\end{equation}
where $\zeta_c = -1$ for $c=1,2,3$ and $\zeta_c = 1$ for $c=L$. Note that
for $\vec{q}=0$ they reduce to those given by Eqs.~(\ref{polneu_a}) and
(\ref{polneu_b}).

\subsection{Charged mesons}

In the case of charged mesons, for $\ell\geq 1$ one finds three linearly
independent polarization vectors. A convenient choice is
\begin{eqnarray}
\epsilon^{\mu}(\ell,q^{3},1) & = & \frac{1}{\sqrt{2}\;m_{\perp}\,m_{2\perp}}\,
\bigg[\Pi_{+}\,(E,\,0,\,0,\,\,q^{3})+m_{\perp}^{2}\,(0,\,1,\,is,\,0)\bigg]\ ,\nonumber \\[2mm]
\epsilon^{\mu}(\ell,q^{3},2) & = & \frac{1}{m_{\perp}}\,(q^{3},\,0,\,0,\,E)\ ,\nonumber \\[2mm]
\epsilon^{\mu}(\ell,q^{3},3) & = &
\frac{1}{\sqrt{2}\,m\,m_{2\perp}}\left[\Pi_{-}(E,0,0,q^{3})+
\frac{\Pi_{+}^{*}\Pi_{-}}{2}\left(0,1,is,0\right)+m_{2\perp}^{2}(0,1,-is,0)\right]\ ,\
\label{chvecpol}
\end{eqnarray}
where we have used the definitions
\begin{eqnarray}
m_{\perp} & = & \sqrt{m^{2}+(2\ell+1)B_{e}}\ ,\nonumber \\
m_{2\perp} & = & \sqrt{m^{2}+\ell B_{e}}\ ,\nonumber \\
\Pi_{+} & = & -\Pi^{1}(\ell,q_{\parallel})+is\,\Pi^{2}(\ell,q_{\parallel})=-i\sqrt{2(\ell+1)B_{e}}\ ,\nonumber \\
\Pi_{-} & = & -\Pi^{1}(\ell,q_{\parallel})-is\,\Pi^{2}(\ell,q_{\parallel})=i\sqrt{2\ell B_{e}}\ ,
\end{eqnarray}
with $B_{Q}=|QB|$. One has
\begin{equation}
E^2 \ = \ (q^3)^2 + (2\ell +1) B_Q + m^2\ ,
\end{equation}
while the four-vector $\Pi^\mu$ is given in Eq.~(\ref{Pigrande}).

For $\ell=0$, two independent nontrivial transverse polarization vectors
can be constructed. A suitable choice is
\begin{eqnarray}
\epsilon^{\mu}(0,q^{3},1) & = & \frac{1}{\sqrt{2}\;m_{\perp}\,m_{2\perp}}\,
\bigg[\Pi_{+}\,(E,\,0,\,0,\,\,q^{3})+m_{\perp}^{2}\,(0,\,1,\,is,\,0)\bigg]\ ,\nonumber \\[2mm]
\epsilon^{\mu}(0,q^{3},2) & = & \frac{1}{m_{\perp}}\,(q^{3},\,0,\,0,\,E)\ ,
\end{eqnarray}
where $m_{\perp}$, $m_{2\perp}$, $\Pi_{+}$ and $E$ are understood
to be evaluated at $\ell=0$.

For $\ell=-1$, there is only one nontrivial polarization vector, which can
be conveniently written as
\begin{equation}
\epsilon^{\mu}(-1,q^{3},1)=\frac{1}{\sqrt{2}}\Big(0,\,1,\,is,\,0\Big)\ .\label{polm1}
\end{equation}

Finally, for $\ell\ge0$ one can also define a ``longitudinal'' polarization
vector that we denote as $\epsilon^{\mu}(\ell,q^{3},L)$; it is given by
\begin{equation}
\epsilon^{\mu}(\ell,q^{3},L)\ = \ \frac{1}{m}\,\Pi^{\mu}(\ell,q_{\parallel})\big|_{q^{0}=E}\ .
\label{long}
\end{equation}
For $\ell=-1$ no longitudinal vector is introduced (notice that $\Pi^\mu$
has been defined only for $\ell\geq 0$).

In a similar way as in the neutral case, the above polarization vectors
satisfy
\begin{equation}
\epsilon^{\mu}(\ell,q^{3},c)^\ast\,\epsilon_{\mu}(\ell,q^{3},c')\ = \ \left\{
\begin{array}{cl}
\zeta_c & \qquad {\rm for} \ c=c' \\
0 & \qquad {\rm for}\ c\neq c'
\end{array}
\right.\ ,
\label{vecnorm}
\end{equation}
where the indices $c$ and $c'$ run only over the allowed polarizations for
the corresponding value of $\ell$, while $\zeta_c$ is defined below Eq.~(\ref{epsmodn}).

As stated in Sec.~\ref{reschmes}, for some range of values of the magnetic
field one can get $m^2<0$. In that case, in Eq.~(\ref{vecnorm}) one
should replace $\zeta_c\to \tilde \zeta_c$, where $\tilde \zeta_c$ depends
on the value of $\ell$. For $\ell=0$, $\tilde \zeta_c = -1~(+1)$ for
$c=L,2~(1)$.

\section{Reflection symmetry and box structure of $\mathbf{G}$ matrices}

\label{pbsymmetry}
\setcounter{equation}{0}

\subsection{Reflection at the plane perpendicular to the magnetic field}

As well known, the electromagnetic interaction is invariant under a
parity transformation
\begin{equation}
x^{\mu}\overset{\mathcal{P}}{\longrightarrow}x_{\mu}\ .
\end{equation}
However, it is not easy to deal with this transformation in the presence of
an external uniform magnetic field. This is due to the fact that the
description of spatial reflections at a plane parallel to the magnetic field
requires to choose a gauge. Instead, we can focus on the spatial reflection
at the plane perpendicular to the magnetic field, say
$\mathcal{P}_{\hat{B}}$. Since, as customary (and without losing
generality), we choose the axis 3 to be in the direction of the magnetic
field, in what follows we denote this transformation by
$\mathcal{P}_{\hat{3}}$.

The transformation $\mathcal{P}_{\hat{3}}$ distinguishes between the
parallel and perpendicular components of $x^{\mu}$. Namely,
\begin{equation}
x_{\parallel}^{\mu}\ \overset{\mathcal{P}_{\hat{3}}}{\longrightarrow}\ x_{\parallel\mu}\ ,
\qquad\qquad x_{\perp}^{\mu}\ \overset{\mathcal{P}_{\hat{3}}}{\longrightarrow}\
x_{\perp}^{\mu}\ .
\end{equation}
For a plane wave associated to a neutral particle we have
\begin{equation}
e^{\pm i\,q\, (\mathcal{P}_{\hat{3}}x)}\ = \ e^{\pm i\,(\mathcal{P}_{\hat{3}}q)\, x}\ ,
\label{PWParity}
\end{equation}
with
\begin{equation}
q_{\parallel}^{\mu}\ \overset{\mathcal{P}_{\hat{3}}}{\longrightarrow}\ q_{\parallel\mu}\ ,\qquad\qquad
q_{\perp}^{\mu}\ \overset{\mathcal{P}_{\hat{3}}}{\longrightarrow}\ q_{\perp}^{\mu}\ .
\end{equation}
The wavefunctions of charged particles can be written in terms of the
functions $\mathcal{F}_{Q}(x,\bar{q})$ discussed in App.~\ref{functionF}. In
this case we have
\begin{equation}
\mathcal{F}_{Q}(\mathcal{P}_{\hat{3}}x,\bar{q}) \ = \
\mathcal{F}_{Q}(x,\mathcal{P}_{\hat{3}}\bar{q}) \ ,
\label{FunctionFParity}
\end{equation}
with $\mathcal{P}_{\hat{3}}\bar{q}=\left(q^{0},\,\ell,\,\chi,-q^{3}\right)$,
independently of the chosen gauge.

It is easy to see that the transformation $\mathcal{P}_{\hat{3}}$ is
equivalent to a parity transformation (denoted by $\mathcal{P}$) followed by
a rotation of angle $\pi$ around the axis 3, i.e.,
$\mathcal{P}_{\hat{3}}=R_{\hat{3}}(\pi)\,\mathcal{P}$. Therefore, the action
of the transformation $\mathcal{P}_{\hat{3}}$ on meson fields can be obtain
as a combination of these two operations. For sigma and pion mesons we have
\begin{eqnarray}
\mathcal{P}_{\hat{3}}\,\sigmaa(x)\,\mathcal{P}_{\hat{3}}^{-1} &
= & \sigmaa\left(\mathcal{P}_{\hat{3}}x\right)\ ,\qquad b=0,1,2,3\ ,
\label{SigmaFieldParity}
\\
\mathcal{P}_{\hat{3}}\,\pib(x)\,\mathcal{P}_{\hat{3}}^{-1} &
= & -\pib(\mathcal{P}_{\hat{3}}x)\ ,\qquad b=0,1,2,3\ ,
\label{PionFiledParity}
\end{eqnarray}
while for vector and axial vector fields we get
\begin{eqnarray}
\mathcal{P}_{\hat{3}}\,\rho_{b\parallel}^{\mu}(x)\,\mathcal{P}_{\hat{3}}^{-1} &
= & \rho_{b\parallel\mu}(\mathcal{P}_{\hat{3}}x)\ ,\quad
\mathcal{P}_{\hat{3}}\,\rho_{b\perp}^{\mu}(x)\,\mathcal{P}_{\hat{3}}^{-1}\,
=\,\rho_{b\perp}^{\mu}(\mathcal{P}_{\hat{3}}x)\ ,\qquad b=0,1,2,3\ ,
\label{RhoFieldParity}
\\
\mathcal{P}_{\hat{3}}\,a_{b\parallel}^{\mu}(x)\,\mathcal{P}_{\hat{3}}^{-1} &
= & -a_{b\parallel\mu}(\mathcal{P}_{\hat{3}}x)\ ,\quad
\mathcal{P}_{\hat{3}}\,a_{b\perp}^{\mu}(x)\,\mathcal{P}_{\hat{3}}^{-1} \,
=\,-a_{b\perp}^{\mu}(\mathcal{P}_{\hat{3}}x)\ ,\qquad b=0,1,2,3\ .
\label{AxialFieldParity}
\end{eqnarray}
We emphasize that Eqs.~(\ref{SigmaFieldParity}-\ref{AxialFieldParity}) are
valid for both neutral and charged mesons.

In the case of fermionic fields there is an ambiguity, since one can take a
rotation of angle $\pi$ or $-\pi$. One has
\begin{equation}
\mathcal{P}_{\hat{3}}\,\psi_{f}(x)\,\mathcal{P}_{\hat{3}}^{-1}\,
=\,\pm\, i\,\eta_{f}\,\mathscr{P}\,\psi_{f}(\mathcal{P}_{\hat{3}}x)\ ,
\label{QuarkFieldParity}
\end{equation}
where $\mathscr{P}=\Sigma^{3}\,\gamma^{0}$, with
$\Sigma^3=i\gamma^{1}\gamma^{2}$. Anyway, since in our calculations quark
fields always appear in bilinear operators, we can choose all fermionic
phases in such a way that $\pm\, i\,\eta_{f}=1.$ It is important to notice
that the fermion propagator $S_{f}(x,x')$ satisfies
\begin{equation}
S_{f}(\mathcal{P}_{\hat{3}}x,\mathcal{P}_{\hat{3}}x')\,
=\,\mathscr{P}\,S_{f}(x,x')\,\mathscr{P}^{\dagger}\ .
\label{proptransp3}
\end{equation}

\subsection{Particle states under reflection at the plane perpendicular to the magnetic
field}

In terms of creation and annihilation operators, the fields describing
neutral scalar and vector mesons can be written as
\begin{eqnarray}
s_{b}(x) & = &
\int\frac{d^{3}q}{(2\pi)^{3}2E_{s}}\,
\Big[\,a_{s_{b}}(\vec{q})\,e^{-iq x}+a_{s_{b}}^{\dagger}(\vec{q})\,e^{iq x}\,\Big]\ ,
\\
v_{b}^{\mu}(x) & = & \int\frac{d^{3}q}{\left(2\pi\right)^{3}2E_{v}}\,\sum_{c=1}^{3}\,
\Big[\,a_{v_{b}}(\vec{q},c)\,e^{-iq x}\,\epsilon^{\mu}(\vec{q},c)+
a_{v_{b}}^{\dagger}(\vec{q},c)\,e^{iq x}\,\epsilon^{\mu}(\vec{q},c)^\ast\,\Big]\ ,
\end{eqnarray}
where $q^0=E=\sqrt{{\vec q}^{\;2}+m^2}$, $s=\sigma,\pi$ and
$v=\rho,a$, while $b=0$ ($b=3$) for isoscalar (isovector) states. The
polarization vectors $\epsilon^{\mu}(\vec{q},c)$ are given in
Eqs.~(\ref{pol_vec_neut}); as stated, we can also define a ``longitudinal''
polarization $\epsilon^{\mu}(\vec{q},L)$ given by Eq.~(\ref{pol_vec_neutL}),
which can be obtained from a derivative of the scalar field.

In the case of the scalar and pseudoscalar fields, the action of
$\mathcal{P}_{\hat{3}}$ yields
\begin{eqnarray}
s_{b}(\mathcal{P}_{\hat{3}}x) & = & \int\frac{d^{3}q}{(2\pi)^{3}2E_{s}}\
\Big[ a_{s_{b}}(\vec{q})\,e^{-iq\,\mathcal{P}_{\hat{3}}x} \, + \,
a_{s_{b}}^{\dagger}(\vec{q})\,e^{iq\,\mathcal{P}_{\hat{3}}x}\Big]\nonumber \\
 & = & \int\frac{d^{3}q}{(2\pi)^{3}2E_{s}}\ \Big[ a_{s_{b}}(\mathcal{P}_{\hat{3}}\vec{q})
 \,e^{-iq\,x}\, + \, a_{s_{b}}^{\dagger}(\mathcal{P}_{\hat{3}}\vec{q})\,e^{iq\,x}\Big]\ ,
\end{eqnarray}
where we have used Eq.~(\ref{PWParity}) followed by a change $q^{3}\rightarrow-q^{3}$
in the integral. Then, from Eqs.~(\ref{SigmaFieldParity}) and (\ref{PionFiledParity})
we conclude
\begin{equation}
\mathcal{P}_{\hat{3}}\, a_{\sigma_{b}}^{\dagger}(\vec{q})\,\mathcal{P}_{\hat{3}}^{-1}
\, = \, a_{\sigma_{b}}^{\dagger}(\mathcal{P}_{\hat{3}}\vec{q})\ , \qquad\qquad
\mathcal{P}_{\hat{3}}\, a_{\pi_{b}}^{\dagger}(\vec{q})\, \mathcal{P}_{\hat{3}}^{-1}
\, = \, -a_{\pi_{b}}^{\dagger}(\mathcal{P}_{\hat{3}}\vec{q})\ .
\end{equation}

In the case of vector and axial vector fields, we have to consider the
behavior of the polarization vectors under the $\mathcal{P}_{\hat{3}}$
transformation. From Eqs.~(\ref{pol_vec_neut}) and (\ref{pol_vec_neutL}) we
have
\begin{equation}
\epsilon^{\mu}(\mathcal{P}_{\hat{3}}\vec{q},c)=\left\{ \begin{array}{lcl}
\epsilon_{\parallel\mu}(\vec{q},c)+\epsilon_{\perp}^{\mu}(\vec{q},c) & \qquad & {\rm for}\ \ c=1,3,L\ ,\\
-\epsilon_{\parallel\mu}(\vec{q},c) &  & {\rm for}\ \ c=2\ .
\end{array}\right.\label{PolTransP3}
\end{equation}
Using these relations together with Eq.~(\ref{PWParity}) one has
\begin{eqnarray}
\!\!\!\!\!\!\! v_{b\parallel}^{\mu}(\mathcal{P}_{\hat{3}}x) & = &
\int\!\frac{d^{3}q}{(2\pi)^{3}2E_{v}}\sum_{c=1,3,L}\Big[
a_{v_{b}}(\mathcal{P}_{\hat{3}}\vec{q},c)\,e^{-iq\, x}\,
\epsilon_{\parallel\mu}(\vec{q},c)\, + \,
a_{v_{b}}^{\dagger}(\mathcal{P}_{\hat{3}}\vec{q},c)\,e^{iq\, x}\,
\epsilon_{\parallel\mu}(\vec{q},c)^\ast\Big]\nonumber \\
 & & +\,
\int\!\frac{d^{3}q}{(2\pi)^{3}2E_{v}}\,\Big[
-a_{v_{b}}(\mathcal{P}_{\hat{3}}\vec{q},2)\,e^{-iq\, x}\,
\epsilon_{\parallel\mu}(\vec{q},2)\,-\,
a_{v_{b}}^{\dagger}(\mathcal{P}_{\hat{3}}\vec{q},2)\,
e^{iq\, x}\,\epsilon_{\parallel\mu}(\vec{q},2)^\ast\Big]\ ,\nonumber \\
\!\!\!\!\!\!\! v_{b\perp}^{\mu}(\mathcal{P}_{\hat{3}}x) & = &
\int\!\frac{d^{3}q}{(2\pi)^{3}2E_{v}}\sum_{c=1,3,L}\Big[
a_{v_{b}}(\mathcal{P}_{\hat{3}}\vec{q},c)\,e^{-iq\, x}\,
\epsilon_{\perp}^{\mu}(\vec{q},c)+
a_{v_{b}}^{\dagger}(\mathcal{P}_{\hat{3}}\vec{q},c)\,
e^{iq\, x}\,\epsilon_{\perp}^{\mu}(\vec{q},c)^\ast\Big]\ .
\end{eqnarray}
This leads a to a different behavior of creation operators depending
on the polarization state, namely
\begin{eqnarray}
& \mathcal{P}_{\hat{3}}\,a_{\rho_{b}}^{\dagger}(\vec{q},c)\,\mathcal{P}_{\hat{3}}^{-1} =
a_{\rho_{b}}^{\dagger}(\mathcal{P}_{\hat{3}}\vec{q},c)\ ,\quad
\mathcal{P}_{\hat{3}}\,a_{a_{b}}^{\dagger}(\vec{q},c)\,\mathcal{P}_{\hat{3}}^{-1} =
-a_{a_{b}}^{\dagger}(\mathcal{P}_{\hat{3}}\vec{q},c) & \quad {\rm for}\ c=1,3,L\ ; \nonumber \\
& \mathcal{P}_{\hat{3}}\,a_{\rho_{b}}^{\dagger}(\vec{q},c)\,\mathcal{P}_{\hat{3}}^{-1} =
-a_{\rho_{b}}^{\dagger}(\mathcal{P}_{\hat{3}}\vec{q},c)\ ,\quad
\mathcal{P}_{\hat{3}}\,a_{a_{b}}^{\dagger}(\vec{q},c)\,\mathcal{P}_{\hat{3}}^{-1} =
a_{a_{b}}^{\dagger}(\mathcal{P}_{\hat{3}}\vec{q},c) & \quad {\rm for}\ \ c=2\ .
\end{eqnarray}

This analysis can be extended to charged scalar and vector mesons. A
detailed description of charged meson fields can be found in
Ref.~\cite{GomezDumm:2023owj}. Briefly, for $s=\sigma,\pi$ and $v=\rho,a$
one has (as in the main text, we consider positively charged mesons)
\begin{eqnarray}
s(x) & = & \sumint_{\left\{ \bar{q}_{E}\right\}}\,\frac{1}{2E_{s}}\,
\Big[a_{s}^+(\breve{q})\,\mathcal{F}_e(x,\bar{q})\,+\,
a_{s}^{-}(\breve{q})^{\dagger}\,\mathcal{F}_{-e}(x,\bar{q})^\ast\Big]\ , \\
v^{\mu}(x) & = & \sumint_{\left\{ \bar{q}_{E}\right\} }\,\frac{1}{2E_{v}}\,
\sum_{c=1}^{3}\,\Big[a_{v}^+(\breve{q},c)\,W_e^{\mu}(x,\bar{q},c)\, + \,
a_{v}^{-}(\breve{q},c)^{\dagger}\,W_{-e}^{\mu}(x,\bar{q},c)^\ast\Big]\ ,
\end{eqnarray}
where $\breve{q}=(\ell,\chi,q^3)$ and
$W_{e}^{\mu}(x,\bar{q},c)=\mathbb{R}^{\mu\nu}(x,\bar{q})
\,\epsilon_\nu(\ell,q^{3},c)$, with $\mathbb{R}^{\mu\nu}(x,\bar{q})$ and
$\epsilon^\mu(\ell,q^{3},c)$ given by Eqs.~(\ref{FandR}) and
(\ref{chvecpol}), respectively. We have also used the notation
\begin{equation}
\sumint_{\left\{ \bar{q}_{E}\right\} }\ \equiv\ \sumint_{\bar{q}}\
2\,\pi\,\delta(q^{0}-E)\ ,
\end{equation}
where $E=\sqrt{m^{2}+B_{e}(2\ell+1)+(q^{3})^{2}}$. As stated, for
$\ell\geq 0$ one can also define a ``longitudinal'' polarization vector,
given by Eq.~(\ref{long}). Taking into account Eq.~(\ref{FunctionFParity})
and the explicit forms of the polarization vectors, one can show the
relations
\begin{equation}
W^{\mu}(\mathcal{P}_{\hat{3}}x,\bar{q},c)\, =\, \left\{
\begin{array}{lcl}
W_{\parallel\mu}(x,\mathcal{P}_{\hat{3}}\bar{q},c)+W_{\perp}^{\mu}(x,\mathcal{P}_{\hat{3}}\bar{q},c) &
\qquad & {\rm for}\ \ c=1,3,L\ , \\
-W_{\parallel\mu}(x,\mathcal{P}_{\hat{3}}\bar{q},c) &  & {\rm for}\ \ c=2\ .
\end{array}\right.
\label{ChPolTransP3}
\end{equation}
Taking into account these equations together with
Eq.~(\ref{FunctionFParity}) for the case of scalar and pseudoscalar
particles, we obtain
\begin{align}
& \mathcal{P}_{\hat{3}} a_{\sigma}^{Q\,\dagger (\breve{q})}\mathcal{P}_{\hat{3}}^{-1} =
a_{\sigma}^{Q\,\dagger}\left(\mathcal{P}_{\hat{3}}\breve{q}\right)\ ,\quad
\mathcal{P}_{\hat{3}}a_{\pi}^{Q\,\dagger}(\breve{q})\mathcal{P}_{\hat{3}}^{-1} =
-a_{\pi}^{Q\,\dagger}\left(\mathcal{P}_{\hat{3}}\breve{q}\right)\ ; & & \nonumber \\
& \mathcal{P}_{\hat{3}}a_{\rho}^{Q\,\dagger}(\breve{q},c)\mathcal{P}_{\hat{3}}^{-1} =
a_{\rho}^{Q\,\dagger}\left(\mathcal{P}_{\hat{3}}\breve{q},c\right)\ , \quad
\mathcal{P}_{\hat{3}}a_{a}^{Q\,\dagger}(\breve{q},c)\mathcal{P}_{\hat{3}}^{-1} =
-a_{a}^{Q\,\dagger}\left(\mathcal{P}_{\hat{3}}\breve{q},c\right) & & {\rm for}\ \ c=1,3,L\ ; \nonumber \\
& \mathcal{P}_{\hat{3}}a_{\rho}^{Q\,\dagger}(\breve{q},c)\mathcal{P}_{\hat{3}}^{-1} =
-a_{\rho}^{Q\,\dagger}\left(\mathcal{P}_{\hat{3}}\breve{q},c\right)\ , \quad
\mathcal{P}_{\hat{3}}a_{a}^{Q\,\dagger}(\breve{q},c)\mathcal{P}_{\hat{3}}^{-1} =
a_{a}^{Q\,\dagger}\left(\mathcal{P}_{\hat{3}}\breve{q},c\right) & & {\rm for}\ \ c=2\ .
\end{align}

These transformation laws indicate how meson states transform under
$\mathcal{P}_{\hat{3}}$, namely
\begin{align}
\mathcal{P}_{\hat{3}} |M(q)\rangle &
=\,\mathcal{P}_{\hat{3}}\,a_{M}^{\dagger}(\vec{q})|0\rangle\, =\,
\eta_{\mathcal{P}_{3}}^{M}\,|M(\mathcal{P}_{\hat{3}}q)\rangle &  & \text{for neutral mesons}\ ,\\
\mathcal{P}_{\hat{3}}|M(\bar{q})\rangle &
=\,\mathcal{P}_{\hat{3}}\,a_{M}^{\dagger}(\breve{q})|0\rangle\, =\,
\eta_{\mathcal{P}_{3}}^{M}\,|M(\mathcal{P}_{\hat{3}}\bar{q})\rangle &  & \text{for charged mesons}\ ,
\end{align}
with
\begin{equation}
\eta_{\mathcal{P}_{3}}^{M}=\left\{ \begin{array}{rcl}
1 & \; {\rm for} \; & M=\sigmaa,\,\rho_{b,1},\,\rho_{b,3},\,\rho_{b,L},\,a_{b,2}\ ,\\
-1 & \; {\rm for} \; & M=\pib,\,a_{b,1},\,a_{b,3},\,a_{b,L},\,\rho_{b,2}\ .
\end{array}\right.
\end{equation}
Here the index $b$ runs from 0 to 3, covering both charged and neutral
mesons.

The fact that our system is invariant under the reflection in the plane
perpendicular to the magnetic field implies that particles with different
parity phase $\eta_{\mathcal{P}_{3}}^{M}$ cannot mix.

\subsection{Box structure of meson mass matrices}

We outline here how the previous assertion is realized in our model.
The masses of charged and neutral mesons are obtained by equations of the
form $\mbox{det}\,G=0$, where $G_{MM'}=
(2g_{M})^{-1}\,\delta_{MM'}-J_{MM'}.$ From Eqs.~(\ref{jotas2}),
(\ref{jdiag}) and (\ref{eq:h1}-\ref{eq:h4}), it is seen that the matrices
$J$ can be written in terms of the functions
\begin{equation}
\Sigma_{MM'}^{ff'}(q)\ = \
-i\,N_{c}\,\int\frac{d^{4}p}{(2\pi)^{4}}\ \trmin_{D}\left[i\,\tilde{S}^{f}(p_{\parallel}^{+},p_{\perp}^{+})\,
\Gamma^{M'}i\,\tilde{S}^{f'}(p_{\parallel}^{-},p_{\perp}^{-})\,\Gamma^{M}\right]\ ,
\end{equation}
[notice that for charged particles $\mathcal{J}_{MM'}(q) = 2\,\Sigma_{MM'}^{ud}(q)$,
see Eq.~(\ref{jmmtt})]. Now, if the system is invariant under a reflection at
the plane perpendicular to the axis 3, the solutions of $\mbox{det}\,G=0$
should be invariant under the change $q\to \mathcal{P}_{\hat{3}}q$ and $\bar
q\to \mathcal{P}_{\hat{3}}\bar q$ for neutral and charged mesons,
respectively. Performing such a transformation on the functions
$\Sigma_{MM'}^{ff'}(q_\parallel,q_\perp)$ one has
\begin{equation}
\Sigma_{MM'}^{ff'}(\mathcal{P}_{\hat{3}}q)\ =\
-i\,N_{c}\,\int\frac{d^{4}p}{(2\pi)^{4}}\
\trmin_{D}\left[i\,\tilde{S}^{f}(\mathcal{P}_{\hat{3}}p_{\parallel}^{+},p_{\perp}^{+})\,
\Gamma^{M'}i\,\tilde{S}^{f'}(\mathcal{P}_{\hat{3}}p_{\parallel}^{-},p_{\perp}^{-})\,\Gamma^{M}\right]\ ,
\end{equation}
where a change $p^{3}\rightarrow -p^{3}$ has been performed in the integral.
Taking into account the result in Eq.~(\ref{proptransp3}) we get
\begin{equation}
\Sigma_{MM'}^{ff'}(\mathcal{P}_{\hat{3}}q)\ =\
-i\,N_{c}\,\int\frac{d^{4}p}{(2\pi)^{4}}\
\trmin_{D}\left[i\,\tilde{S}^{f}(p_{\parallel}^{+},p_{\perp}^{+})\,
\bar{\Gamma}^{M'}i\,\tilde{S}^{f'}(p_{\parallel}^{-},p_{\perp}^{-})\,\bar{\Gamma}^{M}\right]\ ,
\end{equation}
where we have defined
\begin{equation}
\bar{\Gamma}^{M} \ =\ \mathscr{P}^{\dagger}\,\Gamma^{M}\,\mathscr{P}\ .
\end{equation}
For the cases of our interest we have
\begin{align}
& \mathscr{P}^{\dagger}\,\mathbbm{1}\,\mathscr{P} = \mathbbm{1}\ , & &
\mathscr{P}^{\dagger}\,\gamma^{\mu}\,\mathscr{P} = \gamma_{\parallel\mu}+\gamma_{\perp}^{\mu}\ ,
\nonumber \\
& \mathscr{P}^{\dagger}\,i\gamma^{5}\,\mathscr{P} = -i\gamma^{5}\ , \qquad & &
\mathscr{P}^{\dagger}\,\gamma^{\mu}\gamma^{5}\,\mathscr{P} =
-\left(\gamma_{\parallel\mu}+\gamma_{\perp}^{\mu}\right)\gamma^{5}\ .
\nonumber
\end{align}
For neutral and charged mesons, the above changes are complemented by the
transformations of the polarization vectors and the functions
$W_{Q}^{\mu}\left(x,\bar{q},c\right)$, respectively [see
Eqs.~(\ref{PolTransP3}) and (\ref{ChPolTransP3})]. In this way it is easy to
see that for $\eta_{\mathcal{P}_{3}}^{M}\neq\eta_{\mathcal{P}_{3}}^{M'}$ one
has $\Sigma_{MM'}^{ff'}(\mathcal{P}_{\hat{3}}q) = -\Sigma_{MM'}^{ff'}(q)$,
and consequently $\Sigma_{MM'}^{ff'}(q)=0$ and $J_{MM'}=0$.

\section{Functions ${\cal F}_{Q}(x,\bar{q})$ in standard gauges}

\label{functionF} 
\setcounter{equation}{0}

In this appendix we quote the expressions for the functions
${\calF}(x,\bar{q})$ in the standard gauges SG, LG1 and LG2. As in the main
text, we choose the axis 3 in the direction of the magnetic field, and use
the notation $B_{Q}=|Q\,B|$, $s=\mbox{sign}(QB)$.

It is worth pointing out that the functions ${\calF}(x,\bar{q})$ can be
determined up to a global phase, which in general can depend on $\ell$. In
the following expressions for SG, LG1 and LG2 the corresponding phases have
been fixed by requiring ${\calF}(x,\bar{q})$ to satisfy Eqs.~(\ref{hPiInt})
and (\ref{hPiGauge}), with $f_{Q,\ell\ell\,'}(t_{\perp})$ given by
Eq.~(\ref{fkkp}).

\subsection{Symmetric gauge}

In the SG we take $\chi=n$, where $n$ is a nonnegative integer. Thus, the
set of quantum numbers used to characterize a given particle state is
$\bar{q}=(q^{0},\ell,n,q^{3})$. In addition, we introduce polar
coordinates $r,\varphi$ to denote the vector $\vec{x}_{\perp}=(x^{1},x^{2})$
that lies in the plane perpendicular to the magnetic field. The functions
${\cal F}_{Q}(x,\bar{q})$ in this gauge are given by
\begin{equation}
\calF(x,\bar{q})^{{\rm (SG)}}\ =\ \sqrt{2\pi}\ e^{-i(q^{0}\,x^{0}-q^{3}x^{3})}\,
e^{-is(\ell-n)\varphi}\,R_{\ell,n}(r)\ ,\label{efes}
\end{equation}
where
\begin{equation}
R_{\ell,n}(r)\ =\ N_{\ell,n}\,\xi^{(\ell-n)/2}\,e^{-\xi/2}\,L_{n}^{\ell-n}(\xi)\ ,
\end{equation}
with $\xi=B_{Q}\,r^{2}/2\,$. Here we have used the definition $N_{\ell,n}=(B_{Q}\ n!/\ell!)^{1/2}$,
while $L_{j}^{m}(x)$ are generalized Laguerre polynomials.

\subsection{Landau gauges LG1 and LG2}

For the gauges LG1 and LG2 we take $\chi=q^{j}$ with $j=1$ and $j=2$,
respectively. Thus, we have $\bar{q}=(q^{0},\ell,q^{j},q^{3})$.
The corresponding functions $\calF(x,\bar{q})$ are given by
\begin{eqnarray}
\calF(x,\bar{q})^{{\rm (LG1)}} & = & (-is)^{\ell}\,N_{\ell}\,e^{-i(q^{0}\,x^{0}-q^{1}x^{1}-q^{3}x^{3})}\,D_{\ell}(\rho_{s}^{(1)})\ ,\\
\calF(x,\bar{q})^{{\rm (LG2)}} & = & N_{\ell}\,e^{-i(q^{0}\,x^{0}-q^{2}x^{2}-q^{3}x^{3})}\,D_{\ell}(\rho_{s}^{(2)})\ ,
\end{eqnarray}
where $\rho_{s}^{(1)}=\sqrt{2B_{Q}}\,(x^{2}+s\,q^{1}/B_{Q})$, $\rho_{s}^{(2)}=\sqrt{2B_{Q}}\,(x^{1}-s\,q^{2}/B_{Q})$
and $N_{\ell}=\left(4\pi B_{Q}\right)^{1/4}/\sqrt{\ell!}\,$. The
cylindrical parabolic functions $D_{\ell}(x)$ in the above equations
are defined as
\begin{equation}
D_{\ell}\left(x\right)=2^{-\ell/2}\,e^{-x^{2}/4}\,H_{\ell}\big(x/\sqrt{2}\big)\ ,
\end{equation}
where $H_{\ell}(x)$ are Hermite polynomials, with the standard
convention $H_{-1}(x)=0$.

\section{Charged meson polarization functions}

\label{polcharged} \setcounter{equation}{0}

We quote here our results for the polarization functions of charged mesons.
Starting from Eq.~(\ref{chargeJbarq}) and using Eqs.~(\ref{FandR}) we get
\begin{eqnarray}
{\cal J}_{ss'}(\bar{q},\bar{q}') & = & \int\frac{d^{4}t}{(2\pi)^{4}}\ {\cal J}_{ss'}(t)\,
h_{e}(\bar{q},\bar{q}',t)\ ,
\label{eq:h1}\\
{\cal J}_{sv^{\mu}}^\mu(\bar{q},\bar{q}') & = &
\int\frac{d^{4}t}{(2\pi)^{4}}\,\sum_{\lambda}{\cal J}_{sv^{\alpha}}^\alpha(t)
\,(\Upsilon_{\lambda})_\alpha^{\ \mu}\,h_{e}(\bar{q},\bar{q}'_{\lambda},t)\ ,\\
{\cal J}_{v^{\mu}s}^\mu(\bar{q},\bar{q}') & = & \int\frac{d^{4}t}{(2\pi)^{4}}\,
\sum_{\lambda}\, (\Upsilon_{\lambda})^\mu_{\ \alpha}\,{\cal J}_{v^{\alpha}s}^\alpha(t)\,\,
h_{e}(\bar{q}_{\lambda},\bar{q}',t)\ ,\\
{\cal J}_{v^{\mu}{v'}^{\nu}}^{\mu\nu}(\bar{q},\bar{q}') & = &
\int\frac{d^{4}t}{(2\pi)^{4}}\, \sum_{\lambda,\lambda'}\,(\Upsilon_{\lambda})^\mu_{\ \alpha}\,
{\cal J}_{v^{\alpha}{v'}^{\beta}}^{\alpha\beta}(t)\
(\Upsilon_{\lambda'})_\beta^{\,\ \nu}\,
h_{e}(\bar{q}_{\lambda},\bar{q}'_{\lambda'},t)\ ,
\label{eq:h4}
\end{eqnarray}
where $s,s'=\sigma,\pi$ and $v,v'=\rho,a$. Here we have defined
\begin{equation}
h_{Q}(\bar{q},\bar{q}',t)\ =
\ \int d^{4}x\,d^{4}x'\,{\cal F}_{Q}(x,\bar{q})^{\,\ast}\,{\cal F}_{Q}(x',\bar{q}')\,e^{i\Phi_{Q}(x,x')}\,e^{-it(x-x')}\ .
\label{hPiInt}
\end{equation}
As shown in Ref.~\cite{GomezDumm:2023owj}, explicit calculations in
any of the standard gauges lead to
\begin{equation}
h_{Q}(\bar{q},\bar{q}',t)\ =\
\delta_{\chi\chi'}\left(2\pi\right)^{4}\,\delta^{\left(2\right)}(q_{\parallel}-q_{\parallel}^{\prime})
\,\left(2\pi\right)^{2}\,\delta^{\left(2\right)}(q_{\parallel}-t_{\parallel})\,f_{Q,\ell\ell\,'}(t_{\perp})\ ,
\label{hPiGauge}
\end{equation}
where $\delta_{\chi\chi'}$ stands for $\delta_{nn'}$,
$\delta(q^{1}-q^{\prime1})$ and $\delta(q^{2}-q^{\prime2})$ for SG, LG1 and
LG2, respectively, while
\begin{equation}
f_{Q,\ell\ell\,'}(t_{\perp}) \ = \
\frac{4\pi(-i)^{\ell+\ell'}}{B_{Q}}\,\sqrt{\frac{\ell!}{\ell^{\prime}!}}\,
\left(\frac{2\,\vec{t}_{\perp}^{\;2}}{B_{Q}}\right)^{\frac{\ell'-\ell}{2}}
L_{\ell}^{\ell^{\prime}-\ell}\Big(\frac{2\,\vec{t}_{\perp}^{\;2}}{B_{Q}}\Big)\;
e^{-\vec{t}_{\perp}^{\;2}/B_{Q}}\;e^{is(\ell-\ell')\varphi_{\perp}}\ .
\label{fkkp}
\end{equation}
We recall that here $B_{Q}=|QB|$ and $s=\mbox{sign}(QB)$. Since we are
considering positively charged mesons, we have $Q=e$, $e$ being the proton
charge. This implies that $B_{Q}=e|B|$ and $s=\mbox{sign}(B)$ for all
considered mesons.

As mentioned in the main text, using Eqs.~(\ref{hPiGauge}) and (\ref{fkkp}),
and after a somewhat long but straightforward calculation, one can show that
\begin{equation}
{\cal J}_{MM'}(\bar{q},\bar{q}') \ = \
\left(2\pi\right)^{4}\delta(q^{0}-q^{\prime\,0})\,\delta_{\ell\ell'}\,
\delta_{\chi\chi'}\,\delta(q^{3}-q^{\prime\,3})\,
J_{MM'}(\ell,q_{\parallel})\ ,
\end{equation}
where the function $J_{MM'}(\ell,q_{\parallel})$ can be written in general
as
\begin{equation}
J_{MM'}(\ell,q_{\parallel})=\sum_{i=1}^{n_{mm'}}d_{mm',i}(\ell,q_{\parallel}^{2})
\ \mathbb{P}_{MM'}^{(i)}(\Pi)\ .
\label{sumjmmp}
\end{equation}
Here, $m(m')=\pi,\rho,a$ correspond to $M(M')=\pi,\rho^{\mu},a^{\mu}$. The
Lorentz structure is carried out by the set of functions
$\mathbb{P}_{MM'}^{(i)}(\Pi)$, where the four-vector $\Pi^{\mu}$ is given by
\begin{equation}
\Pi^{\mu} \ = \ \left(q^{0},i\sqrt{\frac{B_{M}}{2}}\left(\sqrt{\ell+1}-\sqrt{\ell}\right),
-s\sqrt{\frac{B_{M}}{2}}\left(\sqrt{\ell+1}+\sqrt{\ell}\right),q^{3}\right)\ .
\label{Pigrande}
\end{equation}
In turn, the coefficients $d_{mm',i}(\ell,q_{\parallel}^2)$ can be
expressed as
\begin{eqnarray}
d_{mm',i}(\ell,q_{\parallel}^2)=\frac{N_{c}}{4\pi^{2}}\int_{0}^{\infty}dz\;
\int_{-1}^{1}d\xi\,\frac{e^{-z\phi^{ud}(\xi,q_{\parallel}^{2})}}{\alpha_{+}}
\left(\frac{\alpha_{-}}{\alpha_{+}}\right)^{\ell}\
\beta_{mm',i}(\ell,q_{\parallel}^2,\xi,z)\ ,
\end{eqnarray}
where $\phi^{ff'}\!(\xi,q^2)$ is given by Eq.~(\ref{phiffp}), and we have
introduced the definitions
\begin{equation}
t_{u}=\tanh\left[(1-\xi)zB_{u}/2\right]\ ,\qquad
t_{d}=\tanh\left[(1+\xi)zB_{d}/2\right]\ ,
\label{tf}
\end{equation}
together with
\begin{equation}
\alpha_{\pm}\ = \ \frac{t_{u}}{B_{u}}+\frac{t_{d}}{B_{d}}\pm B_{e}\frac{t_{u}}{B_{u}}\frac{t_{d}}{B_{d}}\ .
\label{alpha}
\end{equation}

The terms of the sum in Eq.~(\ref{sumjmmp}) for each $MM'$, as well as the
explicit form of the corresponding functions
$\beta_{mm',i}(\ell,q_{\parallel}^2,\xi,z)$ are listed in what follows.
Notice that the number of terms, $n_{mm'}$, depends on the $mm'$
combination. In addition, for $\ell=0$ and $\ell=-1$ some of the
coefficients $d_{mm',i}(\ell,q_{\parallel}^2)$ are zero; therefore, for each
function $\beta_{mm',i}(\ell,q_{\parallel}^2,\xi,z)$ we explicitly indicate
the range of values of $\ell$ to be taken into account. For brevity, the
arguments of $d_{mm',i}$ and $\beta_{mm',i}$ are omitted.

For the $\pi\pi$ polarization function one has only a scalar contribution,
i.e., $n_{\pi\pi}=1$. Thus,
\begin{equation}
J_{\pi\pi}(\ell,q_{\parallel}) = d_{\pi\pi,1}\ ;
\end{equation}
the corresponding function $\beta_{mm',1}$ is given by
\begin{eqnarray}
\beta_{\pi\pi,1} & = & (1-t_{u}t_{d})\left[M_{u}M_{d}+\frac{1}{z}+(1-\xi^{2})\frac{q_{\parallel}^{2}}{4}\right]\nonumber \\
 &  & +\, (1-t_{u}^{2})(1-t_{d}^{2})\,\frac{\alpha_{-}+\ell(\alpha_{-}-\alpha_{+})}{\alpha_{+}\alpha_{-}} \ ,
 \quad [\ \ell\ge 0\ ]\ .
\end{eqnarray}

For the $\rho^{\mu}\rho^{\nu}$ polarization function we find 7 terms, namely
\begin{eqnarray}
J_{\rho^{\mu}\rho^{\nu}}^{\mu\nu}(\ell,q_{\parallel}) & = & d_{\rho\rho,1}\,\eta_{\parallel}^{\mu\nu}+d_{\rho\rho,2}\,
\eta_{\perp}^{\mu\nu}+d_{\rho\rho,3}\,\Pi_{\parallel}^{\mu}\,\Pi_{\parallel}^{\nu\,*}+
d_{\rho\rho,4}\,\Pi_{\perp}^{\mu}\,\Pi_{\perp}^{\nu\,*}\nonumber \\
 &  & +\, d_{\rho\rho,5}\left(\Pi_{\parallel}^{\mu}\,\Pi_{\perp}^{\nu\,*}+\Pi_{\perp}^{\mu}\,\Pi_{\parallel}^{\nu\,*}\right)
 -d_{\rho\rho,6}\,is\,\hat{F}^{\mu\nu}\nonumber \\
 &  & +\, d_{\rho\rho,7}\,is\left(\hat{F}^{\mu}_{\ \alpha}\,\Pi_\perp^\alpha\,\Pi_{\parallel}^{\nu\,*}+\Pi_{\parallel}^{\mu}\,
 \Pi_{\perp}^{\alpha\,*}\hat{F}_\alpha^{\ \nu}\right)\ ;
\label{exprr}
\end{eqnarray}
the corresponding functions $\beta_{\rho\rho,i}$ are
\begin{eqnarray}
& & \beta_{\rho\rho,1} = \psi_{1}^{+}\ ,\qquad \beta_{\rho\rho,2} = \psi_{2}^{+}+\psi_{3}^{+}+\left(2\ell+1\right)\,\psi_{4}\ ,
\qquad \beta_{\rho\rho,3} = \psi_{5}\ ,\qquad \beta_{\rho\rho,4} = 2\,\psi_{4}/B_e\ ,\nonumber \\
& & \beta_{\rho\rho,5} = \psi_{6}^{+}+\psi_{7}^{+}\ ,\qquad
\beta_{\rho\rho,6} = \psi_{2}^{+}-\psi_{3}^{+}+\,\psi_{4}\ ,\qquad
\beta_{\rho\rho,7} = -\,\psi_{6}^{+}+\psi_{7}^{+}\ ,
\label{efes}
\end{eqnarray}
where
\begin{eqnarray}
\psi_{1}^{\pm} & = & -(1-t_{u}t_{d})\bigg[\pm M_{u}M_{d}+(1-\xi^{2})\frac{q_{\parallel}^{2}}{4}\bigg]\nonumber \\
& & -\,
\frac{\alpha_{-}+\ell(\alpha_{-}-\alpha_{+})}{\alpha_{+}\alpha_{-}}\,(1-t_{u}^{2})(1-t_{d}^{2})\ , \quad [\ \ell\ge 0\ ]\ ,\nonumber \\[1.5mm]
\psi_{2}^{\pm} & = & -\frac{1}{2}\,\frac{\alpha_{-}}{\alpha_{+}}\,
(1+t_{u})\,(1+t_{d})\,\bigg[\pm M_{u}M_{d}+\frac{1}{z}+(1-\xi^{2})\frac{q_{\parallel}^{2}}{4}\bigg]\ ,\quad [\ \ell\ge-1\ ]\ ,\nonumber \\[1.5mm]
\psi_{3}^{\pm} & = & -\frac{1}{2}\,\frac{\alpha_{+}}{\alpha_{-}}\,(1-t_{u})\,(1-t_{d})\,
\bigg[\pm M_{u}M_{d}+\frac{1}{z}+(1-\xi^{2})\frac{q_{\parallel}^{2}}{4}\bigg]\ ,\quad [\ \ell\ge 1\ ]\ ,\nonumber \\[1.5mm]
\psi_{4} & = & \frac{\alpha_{+}-\alpha_{-}}{2\alpha_{+}\alpha_{-}}\,(1-t_{u}^{2})(1-t_{d}^{2})\ ,\quad [\ \ell\ge 1\ ]\ ,\nonumber \\[1.5mm]
\psi_{5} & = & \frac{1-\xi^{2}}{2}\,(1-t_{u}t_{d})\ ,\quad [\ \ell\ge 0\ ]\ ,\nonumber \\[1.5mm]
\psi_{6}^{\pm} & = & \frac{1}{2\alpha_{+}}\bigg[\frac{1+\xi}{2}\,
\frac{t_{u}\,(1+t_{u})\,(1-t_{d}^{2})}{B_{u}}\pm\frac{1-\xi}{2}\,\frac{t_{d}\,
(1+t_{d})\,(1-t_{u}^{2})}{B_{d}}\bigg]\ ,\quad [\ \ell\ge 0 \ ]\ ,\nonumber \\[1.5mm]
\psi_{7}^{\pm} & = & \frac{1}{2\alpha_{-}}\bigg[\frac{1+\xi}{2}\,
\frac{t_{u}\,(1-t_{u})\,(1-t_{d}^{2})}{B_{u}}\pm\frac{1-\xi}{2}\,
\frac{t_{d}\,(1-t_{d})\,(1-t_{u}^{2})}{B_{d}}\bigg]\ ,\quad[\ \ell\ge 1\ ]\ .
\label{g14}
\end{eqnarray}

For the $a^{\mu}a^{\nu}$ polarization function we have
\begin{eqnarray}
J_{a^{\mu}a^{\nu}}^{\mu\nu}(\ell,q_{\parallel}) & = & d_{aa,1}\,\eta_{\parallel}^{\mu\nu}+
d_{aa,2}\,\eta_{\perp}^{\mu\nu}+d_{aa,3}\,\Pi_{\parallel}^{\mu}\,\Pi_{\parallel}^{\nu\,*}+
d_{aa,4}\,\Pi_{\perp}^{\mu}\,\Pi_{\perp}^{\nu\,*}\nonumber \\
 &  & +\, d_{aa,5}\left(\Pi_{\parallel}^{\mu}\,\Pi_{\perp}^{\nu\,*}+\Pi_{\perp}^{\mu}\,\Pi_{\parallel}^{\nu*}\right)-
d_{aa,6}\,is\,\hat{F}^{\mu\nu}\nonumber \\
 &  & +\, d_{aa,7}\,is\left(\hat{F}^\mu_{\ \alpha}\,\Pi^\alpha_\perp\,\Pi_{\parallel}^{\nu\,*}+
 \Pi_{\parallel}^{\mu}\,\Pi_\perp^{\alpha\,*}\,\hat{F}_\alpha^{\ \nu}\right)\ ;
\label{exprr}
\end{eqnarray}
the functions $\beta_{aa,i}$ are in this case given by
\begin{eqnarray}
 &  & \beta_{aa,1}=\psi_{1}^{-}\ ,\qquad
 \beta_{aa,2}=\psi_{2}^{-}+\psi_{3}^{-}+\left(2\ell+1\right)\,\psi_{4}\ , \qquad
 \beta_{aa,3}=\psi_{5}\quad,\quad\beta_{aa,4}=2\,\psi_{4}/B_e\ ,\nonumber \\
 &  & \beta_{aa,5}=\psi_{6}^{+}+\psi_{7}^{+}\ ,\qquad
 \beta_{aa,6}=\psi_{2}^{-}-\psi_{3}^{-}+\,\psi_{4}\ ,\qquad
 \beta_{aa,7}=-\,\psi_{6}^{+}+\psi_{7}^{+}\ .
 \label{betaaa}
\end{eqnarray}

For the $\pi\rho^{\mu}$ and $\rho^{\mu}\pi$ polarization functions
we obtain
\begin{equation}
J_{\pi\rho^{\mu}}^\mu(\ell,q_{\parallel}) \, = \,
J_{\rho^{\mu}\pi}^\mu(\ell,q_{\parallel})^{\,\ast}
\, =\, d_{\pi\rho,1}\,
s\,\hat{\tilde{F}}^\mu_{\ \alpha}\,\Pi_\parallel^{\alpha\,*}\ ;
\end{equation}
the function $\beta_{\pi\rho,1}$ reads
\begin{equation}
\beta_{\pi\rho,1} \ = \ -\frac{i}{2}(t_{u}-t_{d})\Big[(M_{u}+M_{d})-\xi(M_{u}-M_{d})\Big]\ ,\quad [\ \ell\ge 0\ ]\ .
\end{equation}

For the $\pi a^{\mu}$ and $a^{\mu}\pi$ polarization functions we have
\begin{equation}
J_{\pi a^{\mu}}^\mu(\ell,q_{\parallel}) \, = \, J_{a^{\mu}\pi}^\mu(\ell,q_{\parallel})^{\,\ast}\,
=\, d_{\pi a,1}\, \Pi_{\parallel}^{\mu\,\ast}+d_{\pi a,2}\, \Pi_{\perp}^{\mu\,\ast}-
d_{\pi a,3}\, is\, {\hat{F}}^\mu_{\ \alpha}\,\Pi_\perp^{\alpha\,\ast}\ ;
\end{equation}
the functions $\beta_{\pi a,i}$ are given by
\begin{eqnarray}
& & \beta_{\pi a,1} = -\,\frac{i}{2}\,(1-t_{u}t_{d})\Big[(M_{u}+M_{d})-\xi(M_{u}-M_{d})\Big]\ ,\quad [\ \ell\ge 0\ ]\ ,
\nonumber\\
& & \beta_{\pi a,2} = \psi_{8}+\psi_{9}\ , \qquad \beta_{\pi a,3} =
-\,\psi_{8}+\psi_{9}\ ,
\end{eqnarray}
where
\begin{eqnarray}
\psi_{8} & = & -\frac{i}{2\alpha_{+}}\,\bigg[M_{u}\,\frac{t_{u}\,(1+t_{u})\,
(1-t_{d}^{2})}{B_{u}}+M_{d}\,\frac{t_{d}\,(1+t_{d})\,(1-t_{u}^{2})}{B_{d}}\bigg]\ ,\quad [\ \ell\ge 0\ ]\ ,\nonumber \\[1.5mm]
\psi_{9} & = & -\frac{i}{2\alpha_{-}}\,\bigg[M_{u}\,\frac{t_{u}\,(1-t_{u})\,
(1-t_{d}^{2})}{B_{u}}+M_{d}\,\frac{t_{d}\,(1-t_{d})\,(1-t_{u}^{2})}{B_{d}}\bigg]\ ,\quad [\ \ell\ge 1\ ]\ .
\end{eqnarray}

Finally, for the $a^{\mu}\rho^{\nu}$ and $\rho^{\mu}a^{\nu}$ polarization
functions we get
\begin{eqnarray}
J_{a^{\mu}\rho^{\nu}}^{\mu\nu}(\ell,q_{\parallel}) & = &
 J_{\rho^{\nu}a^{\mu}}^{\nu\mu}(\ell,q_{\parallel})^{\,*}
\,=\,d_{a\rho,1}\,s\,{\hat{\tilde{F}}^{\mu\nu}}+
d_{a\rho,2}\,s\left({\hat{\tilde{F}}^\mu_{\ \alpha}}\,\Pi_\parallel^\alpha\,\Pi_{\parallel}^{\nu\,*}-
\Pi_{\parallel}^{\mu}\,\Pi_\parallel^{\ \alpha\,*}\hat{\tilde{F}}_\alpha^{\ \nu}\right)\nonumber \\
 &  & \ \qquad\qquad\qquad+\,d_{a\rho,3}\,s\left(\hat{\tilde{F}}^\mu_{\ \alpha}\,\Pi_\parallel^\alpha\,\Pi_{\perp}^{\nu\,*}-
 \Pi_{\perp}^{\mu}\,\Pi_\parallel^{\alpha\,*}\hat{\tilde{F}}_\alpha^{\ \nu}\right)\nonumber \\
 &  & \ \qquad\qquad\qquad+\,d_{a\rho,4}\,i\left(\hat{\tilde{F}}^\mu_{\ \alpha}\,\Pi_\parallel^\alpha\,
 \Pi_\perp^{\beta\,*}\,\hat{F}_\beta^{\ \nu}-\hat{F}^\mu_{\ \alpha}\,\Pi_\perp^\alpha\,\Pi_\parallel^{\beta\,*}
 \,\hat{\tilde{F}}_\beta^{\ \nu}\right)\ ;
\end{eqnarray}
the corresponding coefficients $\beta_{a\rho,i}$ are
\begin{eqnarray}
& & \beta_{a\rho,1} = -M_{u}M_{d}\, (t_{u}-t_{d})\ ,\quad [\ \ell\ge 0\ ]\ ,\nonumber \\
& & \beta_{a\rho,2} = \frac{1-\xi^{2}}{4}\, (t_{u}-t_{d})\ ,\quad [\ \ell\ge 0\ ]\ ,\nonumber \\
& & \beta_{a\rho,3} = \psi_{6}^{-}-\psi_{7}^{-}\ ,\qquad \beta_{a\rho,4} =
-\,\psi_{6}^{-}-\psi_{7}^{-}\ .
\end{eqnarray}

\section{Matrix elements of $\mathbf{J}^{{\rm mag}}(\ell,m^2)$ for $\ell=0$}

\label{matl0}\setcounter{equation}{0}

In this appendix we list the elements of the matrix $\mathbf{J}^{{\rm
mag}}(\ell,m^2)$ for $\ell=0$, i.e., the case considered in
Eq.~(\ref{jmagl0}). The expressions are given in terms of the coefficients
$b_{mm',i}^{ud,\,{\rm unreg}}(q^{2})$ given in App.~\ref{b0loops} and the
coefficients $d_{mm',i}(\ell,q_{\parallel}^{2})$ quoted in
App.~\ref{polcharged}. In the expressions below, it is understood that they
are evaluated at $q^{2}=m^{2}$ and
$(\ell,q_{\parallel}^{2})=(0,m^{2}+B_{e})$, respectively.

We obtain (for $m^2>0$)
\begin{eqnarray}
\mathbf{J}_{\pi\pi}^{\rm mag} & = & d_{\pi\pi,1}-2\,b_{\pi\pi,1}^{ud,\,{\rm unreg}}\ ,\\[3mm]
\mathbf{J}_{\pi\rho_{2}}^{\rm mag} & = & {\mathbf{J}_{\rho_{2}\pi}^{\rm mag}}^{\,*}=-s\ m_{\perp}\, d_{\pi\rho,1}\ ,\\[5mm]
\mathbf{J}_{\pi a_{L}}^{\rm mag} & = & {\mathbf{J}_{a_{L}\pi}^{\rm mag}}^{\,*}=
\frac{1}{m}\left(m_{\perp}^{2}\, d_{\pi a,1}-2B_{e}\, d_{\pi a,2}-2\,m^{2}\,b_{\pi a,1}^{ud,\,{\rm unreg}}\right)
\label{j0palmag}\ ,\\[3mm]
\mathbf{J}_{\pi a_{1}}^{\rm mag} & = & {\mathbf{J}_{a_{1}\pi}^{\rm mag}}^{\,*}=-i\,\frac{\sqrt{B_{e}}\ m_{\perp}}{m}
\big(d_{\pi a,1}-2\,d_{\pi a,2}\big)\ ,\\[3mm]
\mathbf{J}_{\rho_{2}\rho_{2}}^{\rm mag} & = & -\,d_{\rho\rho,1}+2\,b_{\rho\rho,1}^{ud,\,{\rm unreg}}\ ,\\[5mm]
\mathbf{J}_{a_{L}\rho_{2}}^{\rm mag} & = & {\mathbf{J}_{\rho_{2}a_{L}}^{\rm mag}}^{*}=
-\frac{s\,m_{\perp}}{m}\left(-\,d_{a\rho,1}+m_{\perp}^{2}\, d_{a\rho,2}-2\, B_{e}\, d_{a\rho,3}\right)\ ,\\[5mm]
\mathbf{J}_{a_{1}\rho_{2}}^{\rm mag} & = & {\mathbf{J}_{\rho_{2}a_{1}}^{\rm mag}}^{*}=
-i\,\frac{s\, \sqrt{B_{e}}}{m}\left(-\,d_{a\rho,1}+m_{\perp}^{2}\, d_{a\rho,2}-2\, m_{\perp}^{2}\, d_{a\rho,3}\right)\ ,\\[3mm]
\mathbf{J}_{a_{L}a_{L}}^{\rm mag} & = & \frac{1}{m^2}
\left(m_{\perp}^{2}\,d_{aa,1}-2\,B_{e}\,d_{aa,2}+m_{\perp}^{4}\,d_{aa,3}-
4\,m_{\perp}^{2}\,B_{e}\,d_{aa,5}-2\,m^{2}\,b_{aa,2}^{ud,\,{\rm unreg}}\,\right)\ ,\\[3mm]
\mathbf{J}_{a_{L}a_{1}}^{\rm mag} & = & {\mathbf{J}_{a_{1}a_{L}}^{\rm mag}}^{*}=
-i\frac{\sqrt{B_{e}}\,m_{\perp}}{m^2}
\left(d_{aa,1}-2\, d_{aa,2}+m_{\perp}^{2}\, d_{aa,3}-2\,(m_{\perp}^{2}+B_{e})\, d_{aa,5}\right)\ ,\\[3mm]
\mathbf{J}_{a_{1}a_{1}}^{\rm mag} & = & \frac{B_e}{m^2}\left(d_{aa,1}-\,\frac{2\,m_{\perp}^{2}}{B_e}\, d_{aa,2}+
m_{\perp}^{2}\, d_{aa,3}-4\, m_{\perp}^{2}d_{aa,5}+\,\frac{2\,m^{2}}{B_e}\,b_{aa,1}^{ud,\,{\rm unreg}}\right)\ .
\end{eqnarray}

\end{document}